\documentclass[11pt,a4paper]{article}
\usepackage[utf8]{inputenc}
\pdfoutput=1
\usepackage{jheppub}
\usepackage{amsfonts}
\usepackage{amsmath}
\usepackage{color}
\usepackage[normalem]{ulem}
\usepackage{verbatim}
\usepackage{rotating, graphicx}
\usepackage{subcaption}
\usepackage{xcolor}
\usepackage{hyperref}
\usepackage{orcidlink}
\usepackage{etoolbox}

\usepackage{siunitx}

\newcommand{\beq}{\begin{equation}}
\newcommand{\eeq}{\end{equation}}
\newcommand{\bea}{\begin{eqnarray}}
\newcommand{\eea}{\end{eqnarray}}

\newcommand{\herwig}{\textsc{Herwig}{}}

\newcommand{\apprentice}{\textsc{Apprentice}{}}
\newcommand{\rivet}{\textsc{Rivet}{}}
\newcommand{\yoda}{\textsc{Yoda}{}}

\title{Matching Hadronization and Perturbative Evolution: The Cluster
  Model in Light of Infrared Shower Cutoff Dependence}

\preprint{
	\begin{flushright}
		UWThPh-2023-23\\
		MCnet-24-05
	\end{flushright}
}

\author[a]{Andr\'e H. Hoang\orcidlink{0000-0002-8424-9334}}
\author[a]{Oliver L. Jin,}
\author[b,a]{Simon Pl\"atzer\orcidlink{0000-0002-7616-4103},}
\author[c]{Daniel Samitz\orcidlink{0009-0006-6858-7049}}

\affiliation[a]{Faculty of Physics, University of Vienna, Boltzmanngasse 5, A-1090 Vienna, Austria}
\affiliation[b]{Institute of Physics, NAWI Graz, University of Graz, Universitätsplatz 5, A-8010 Graz, Austria}
\affiliation[c]{Stefan Meyer Institute for Subatomic Physics, Dominikanerbastei 16, A-1010 Vienna, Austria}

\emailAdd{andre.hoang@univie.ac.at}
\emailAdd{lizhi.oliver.jin@univie.ac.at}
\emailAdd{simon.plaetzer@uni-graz.at}
\emailAdd{daniel.samitz@oeaw.ac.at}

\abstract{In the context of Monte Carlo (MC) generators with parton
  showers that have next-to-leading-logarithmic (NLL)
  precision, the cutoff $Q_0$
  terminating the shower evolution should be viewed as an infrared
  factorization scale so that parameters or non-perturbative effects
  of the MC generator may have a field theoretic interpretation with a
  controllable scheme dependence. This implies that the generator's
  parton level should be carefully defined within QCD perturbation
  theory with subleading order precision.  Furthermore, it entails
  that the shower cut $Q_0$ is not treated as one of the generator's
  tuning parameters, but that the tuning can be carried out reliably
  for a range of $Q_0$ values and that the hadron level description is
  $Q_0$-invariant.  This in turn imposes non-trival constraints on the
  behavior of the generator's hadronization model, so that its
  parameters can adapt accordingly when the $Q_0$ value is changed. We
  investigate these features using the angular ordered parton shower
  and the cluster hadronization model implemented in the \herwig~7.2
  MC generator focusing in particular on the $e^+e^-$ 2-jettiness
  distribution, where the shower is known to be NLL precise and where
  QCD factorization imposes stringent constraints on the hadronization
  corrections. We show that the \herwig{} default cluster
  hadronization model does not exhibit these features or consistency
  with QCD factorization with a satisfying precision. We design a
  modification of the cluster hadronization model, where some
  dynamical parton shower aspects are added that are missing in the
  default model. For this novel dynamical cluster hadronization model
  these features and consistency with QCD factorization are realized
  much more accurately.}

\begin{document}
\maketitle
\flushbottom

\section{Introduction}
\label{sec:Introduction}

Multi-purpose Monte Carlo event generators (MCs) are indispensable
tools to describe realistic, fully detailed hadronic final states for
essentially all processes at collider experiments.  Their underlying
structure and components reflect the large hierarchies of energy
scales involved in these processes. These scale hierarchies are also
the basis of numerous factorization theorems in analytic QCD
approaches, which are partly also reflected in the MC components.
After the determination of hard scattering cross sections using matrix
elements obtained from fixed-order perturbation theory, all-order
leading contributions coming from the radiation of soft and collinear
partons are resummed by parton shower algorithms. The latter evolve
the incoming and outgoing energetic partons participating in the hard
scattering to lower scales tied to an ordering variable.  This
evolution terminates when the ordering variable reaches a cut-off
value, which we refer to as $Q_0$, and which defines a low scale kinematic
restriction on the gluon radiation and gluon branching into
quarks and gluons. The value of $Q_0$ is typically in the range between $0.5$ and
$2$~GeV. Phenomenological models of hadronization then describe how
hadrons are formed out of the partonic final states that have emerged
from the parton shower. The parameters of these hadronization models
are fixed through the tuning procedure which is based on a fit to a
reference data set.  In current state-of-the-art MCs it is common that
the shower cutoff $Q_0$ is also treated as a hadronization parameter. 
This means that the value of $Q_0$
is fixed by demanding best agreement with the reference
data.  In collisions of extended objects, such as protons, also
multi-parton interactions taking place in the initial stages of the
collision are simulated. The modelling of multi-parton interactions
is, however, not subject of the present paper.

One of the major recent advances in the development of MCs has been to
include fixed-order (QCD) corrections for the description of the hard
scattering process and the production of additional hard jets through
NLO matching and multijet merging algorithms.  These developments need
to be complemented by more accurate parton shower algorithms to
improve the description of the resummed higher order corrections
arising from the soft and collinear dynamics. Such improvements for a
large class of observables have only recently received more attention:
While, since a long time, parton shower approaches such as coherent
branching can be rigorously ({\it i.e.} analytically and numerically)
proven to be accurate at the next-to-leading logarithmic (NLL) level
for dijet-based $e^+e^-$ event-shape variables such as thrust or other
event-shapes related to jet masses in the peak
region~\cite{Catani:1992ua,Hoang:2018zrp,Bewick:2019rbu,Bewick:2021nhc},
similar progress for dipole parton showers, which are more convenient
for matching and merging, has only recently been achieved, see {\it
  e.g.}
\cite{Dasgupta:2020fwr,Forshaw:2020wrq,Nagy:2020rmk,Herren:2022jej,vanBeekveld:2022zhl,Hoche:2024dee}
for recent proposals and implementations. Going even
beyond this level of accuracy for the parton shower evolution is also
subject to an active area of research, but the implementation of these
developments in the framework of multi-purpose MCs applicable for
experimental analyses may likely still require significant
time.\footnote{A first study \cite{vanBeekveld:2024wws}
    has appeared just after the first version of this work has been
    submitted.}

In the context of these advancements much less work has been invested
in the developments of the MC hadronization models. However, the
presence and impact of hadronization is an essential ingredient for a
realistic description of infrared sensitive observables alongside with
improved matching and merging scheme and NLL accurate parton shower
algorithms. This is particularly important when MCs are employed in
the context of the determination of scheme-dependent Lagrangian QCD
parameters, where a systematic separation of perturbative and
non-perturbative effects is crucial.  This has for example been
stressed since a long time for strong coupling determination analyses from
$e^+e^-$ event-shapes, when non-perturbative hadronization corrections
are determined from MC simulations~\cite{Huston:2023ofk} and the
associated uncertainties are determined from simulations with
different MC settings.  The properties and quality of the MC
hadronization models is also essential for the interpretation of the
event generator's top quark mass parameter $m_t^{\rm MC}$ determined
in direct top mass measurements~\cite{Hoang:2020iah,CMS:2024irj},
since they directly affect the top quark mass sensitive aspects of the
observables used for the measurements~\cite{Hoang:2018zrp}.

Inspired by earlier studies of some of the authors in
Ref.~\cite{Proceedings:2018jsb}, we argue that accurate and improved
parton shower algorithms to be employed for MC predictions should not
be considered independently of the MC hadronization models. In
principle this is already obvious at the purely practical level in the
context of the tuning procedure of an event generator to a set of
reference data, since the tuning traditionally considers a combined
fit of parton shower and hadronization model parameters. However, what
we mean is, that eventually one should go beyond this practical level
by {\it demanding in addition} that the combined matrix-element and
parton shower partonic description, on one side, and the effects of
the hadronization model on the other, by themselves have a
well-defined field theoretic meaning in the context of QCD with a
systematic and controllable scheme-dependence.  It is obvious that
this is particularly important when combining hadronization
corrections extracted from MCs with high precision perturbative QCD
predictions for the determination of hadron level eventshape
distribution, or for the interpretation of $m_t^{\rm MC}$ obtained
from the direct measurements. This implies that {\it (i) the field
theoretic meaning of the parton level description and the
hadronization effects should not depend on any parameter subject to
tuning hadronization effects and (ii) in particular that the parton
shower infrared cutoff $Q_0$ should be considered as a factorization
scale.} 

This view has also been advocated in
Ref.~\cite{Platzer:2022jny}, where the same reasoning has been applied
for the construction of an algorithm for parton branching at the
amplitude level~\cite{Forshaw:2019ver} based on an underlying
factorization of jet cross sections.  The way how the shower cutoff 
is implemented (in terms of an infrared restriction 
related to an evolution variable)
defines a particular scheme of this factorization, and the
factorization property ensures that the hadron level description is
scheme-invariant. This in particular implies that, at least within
some limited $Q_0$ interval, the combined partonic and hadron-model
prediction for infrared sensitive quantities should be independent of
the $Q_0$ value and the hadronization model should be capable to describe
the perturbative evolution in $Q_0$. The essential merrit of a model satisfying
this property would be that its genuine hadronization effects related to $\Lambda_{\rm QCD}$ 
would become universal and thus well-defined within QCD. 
In contrast a model not having the property (or a model where this property has never been examined 
in a dedicated way), can yield hadronization effects 
that are not consistent with QCD.  As it is probably too ambitious
to demand exact $Q_0$-independence, in practice at least a sizeable
systematic cancellation between the dominant linear
$Q_0$-dependencies of the parton level predictions and of the
associated hadronization corrections (at the precision level of the 
parton shower) should be realized. This level of
$Q_0$-insensitivity with respect to linear $Q_0$ contributions is
what we refer to as '$Q_0$-independence' in this article.

Within such a framework the analytic properties of a (NLO matched and)
NLL accurate parton shower could be transferred in a controllable
way to the hadron level description provided by the MC generator and
new avenues to scrutinize the combined action of parton level and
hadron level descriptions provided by the MCs would be made possible. This
level of control is also mandatory to study the impact of the shower
cutoff $Q_0$ dependent top quark mass parameter $m_Q^{\rm CB}(Q_0)$,
which was recently proven to emerge from using the coherent branching
algorithm for massive quark initiated $e^+e^-$ event-shape
distributions~\cite{Hoang:2018zrp}, in the context of MC simulations
for boosted top pair production.

For state-of-the-art MCs it is largely unknown to which extent their
hadronization models satisfy the factorization criterium formulated
above. It is obvious that the paradigm of
parton-shower cutoff scale independence of the hadron level
simulations imposes additional nontrivial constraints on the 
hadronization models. At least within some limited range for $Q_0$, 
they need the
flexibility to match with the corresponding evolution of the parton
shower description.  Furthermore, the parton shower cutoff $Q_0$
should not be considered as a non-perturbative parameter, but always
in a hierarchy $\Lambda_{\rm QCD} \ll Q_0 \ll Q$, where $Q$ is the
hard scale of the process of interest, and $\Lambda_{\rm QCD}$ is the
intrinsic scale of QCD. Since typical parton shower cutoff values are
around the charm mass scale, this interpretation is also practically
feasible for current state-of-the-art MCs.

In this article we further promote the idea of the shower cutoff $Q_0$
being a factorization scale by providing an actual implementation of a
hadronization model acting in this direction. 
This entails that the
(linear) evolution in $Q_0$ of the parton level description and the
corresponding hadronization effects {\it individually} follow the concrete
predictions of QCD perturbation theory with good precision, yielding hadron level 
descriptions that are less $Q_0$-dependent in a controllable manner. 
To
be definite, we carry out our considerations in the context of the
angular ordered parton shower and cluster hadronization model as
implemented in the \herwig~7 MC generator
\cite{Bellm:2015jjp,Bellm:2019zci,Bewick:2023tfi}. Our new model will become
available with an upcoming \herwig~7.4 release. We focus in particular
on the $e^+e^-$ dijet event-shape distribution 2-jettiness. For the
latter it is known since a long time
ago~\cite{Catani:1992ua,Contopanagos:1996nh,Almeida:2014uva} that the
coherent branching algorithm, which is the basis of the angular
ordered \herwig~parton shower, is NLL precise. For the 2-jettiness
distribution the dominant linear shower cut $Q_0$-dependence of the angular ordered
parton shower description in the dijet limit has also been determined
analytically at NLO (${\cal O}(\alpha_s)$), and the \herwig~7.2 parton
shower has been shown to be fully consistent with these analytic
results also for quasi-collinear massive quarks in
Ref.~\cite{Hoang:2018zrp}.\footnote{ 
The results obtained in Ref.~\cite{Hoang:2018zrp} were possible due to the
global character of jet mass based event-shapes such as 2-jettiness (or thrust), and because
of the knowledge on their factorization at the hadron level.
At this time results along the same lines have not been determined
for other types of observables.}  
Using the 	analytic results for this linear
$Q_0$-dependence and the association of the generator's
parton-to-hadron level migration matrix $T$ with the dijet shape
function $S_{\rm had}$ appearing in QCD factorization and
soft-collinear effective theory (SCET) for 2-jettiness in the dijet
region~\cite{Lee:2006nr}, we can derive a QCD constraint concerning
the migration matrix $T$, which can then be tested in detail. This
constraint involves the linear $Q_0$-dependence of the first moment of
the transfer matrix and demands in addition independence concerning
the c.m.\ energy $Q$.

We find that the default cluster hadronization model of \herwig~7.2
does not satisfy this QCD constraint in a consistent manner.\footnote{
  In this respect, \herwig~7.3 \cite{Bewick:2023tfi}, which has been
  released recently, behaves similar to \herwig~7.2
  \cite{Bellm:2019zci}.}  This motivates the construction of a novel
and improved cluster hadronization model, which we call the
``dynamical cluster model". The dynamical cluster model satisfies the
QCD constraints more accurately and leads to a significantly improved
shower cut $Q_0$-independence of the hadron level description in the
physically important shower cutoff interval
$1~\mbox{GeV} < Q_0 < 2$~GeV.  While even the behavior of the new
dynamic hadronization model with respect to $Q_0$-variations is not
yet perfect, the results of this article provide first important steps
towards giving the hadronization effects provided by MC simulations a
well-defined QCD meaning with a controllable scheme-dependence, and a
stepping stone to achieve hadronization models fully consistent with
QCD factorization, as outlined in \cite{Platzer:2022jny}. Eventually
variations of the shower cut may also be used as an instrument to
quantify the theoretical uncertainty of event generator predictions in
analogy to the canonical renormalization or factorization scale
variations in analytic calculations.  In this article we focus on
hadronization in the presence of light quarks, which is a vital step
towards understanding the interplay of the infrared cutoff,
hadronization and the shower cutoff dependence of heavy quark mass
parameters in an upcoming paper~\cite{ourMCtopmasspaper}.

The content of this article is a follows: In
Sec.~\ref{sec:generatorgeneral} we briefly review the basic conceptual
components of the coherent branching algorithm implemented in the
\herwig{} MC. As a novelty, we describe the access to the parton level
involving quarks with current masses and massless gluons. This ``true
parton level'' (which was already used in Ref.~\cite{Hoang:2018zrp})
has not been directly accessible in earlier \herwig{} releases, but it
is an essential prerequisite for the analyses in this work, and
can also be compared to other approaches as those outlined in
  \cite{Reichelt:2021svh}.  In Sec.~\ref{sec:expectations} we then
recall the general aspects of the factorization between parton level
and hadronization effect for MC simulations from
Ref.~\cite{Platzer:2022jny} and the concrete resulting constraints on
their (linear) shower cutoff $Q_0$-dependence for the 2-jettiness
distribution in the dijet region obtained from the \herwig{} coherent
branching implementation. These constaints, which are based on QCD
factorization and Soft-Collinear Effective Theory (SCET), can be
formulated in the form of RG evolution equations for distribution
cumulants and moments and have been determined previously in
Ref.~\cite{Hoang:2018zrp}. Here we also discuss the concrete relation
of the non-perturbative shape function appearing in QCD factorization
and the parton-to-hadron level migration matrix that can be extracted
from the MC simulations which plays an essential role for our
phenomenological analyses.  The structure of the default cluster
hadronization model available in the current \herwig{} release is
explained in detail in Sec.~\ref{sec:Clusters} as a prequisite for the
improved dynamical aspects of our novel cluster hadronization model
introduced in Sec.~\ref{sec:NewFission}. In this article, the
consistency with respect to QCD factorization of the default and the
novel dynamical hadronization models is tested through a number of
analyses of the hadronization effects based on tunes for different
fixed shower cutoff $Q_0$ values. These tuning analyses are described
in detail in Sec.~\ref{sec:tuningphenosetup}. Here we also address how
to estimate uncertainties in the determination of the hadronization
model tuning parameters, as this becomes relevant in the context of
their cutoff scheme-dependence.  Finally, in
Sec.~\ref{sec:NewFissionpheno} we discuss the results of the
$Q_0$-dependent tuning analyses from the phenomenological perspective,
demonstrating a substantially better performance of the novel
dynamical hadronization model with respect to the constraints imposed
by QCD factorization.  In Sec.~\ref{sec:Conclusions} we conclude.

The reader not interested in the details concerning the \herwig{} 
hadronization models and our tuning analyses, 
may skip Secs.~\ref{sec:Clusters} and \ref{sec:NewFission} and most
of Sec.~\ref{sec:tuningphenosetup}. However, we recommend to read
Sec.~\ref{sec:extractmigration} as it contains useful information 
for the understanding of the following phenomenological discussion.

\section{Coherent Branching}
\label{sec:generatorgeneral}

\subsection{General remarks}
\label{sec:generalremarks}

Coherent branching is a parton shower algorithm which is based on
$1\to 2$ parton splitting processes incorporating QCD coherence
through the ordering in an emission angular variable $\tilde{q}$, see
e.g.\ the classic articles~\cite{Catani:1990rr,Catani:1992ua}.  In
this work we use the implementation of coherent branching in the
\herwig{} event generator, as previously described and analyzed in
\cite{Gieseke:2003rz,Hoang:2018zrp}.  For global observables in simple
2-jet processes such as event-shapes distributions in
$e^+e^-\to$~hadrons, coherent branching is NLL precise.  Here we do
not discuss details of \herwig's coherent branching
algorithm\footnote{We refer the reader to Ref.~\cite{Hoang:2018zrp}
  for a detailed discussion of features in \herwig's coherent
  branching parton shower relevant for the shower cutoff analyses in
  this article.}  but focus on its main features relevant for the
kinematic properties of the emerging partonic final state and its
characteristics relevant for the onset of the simulated hadronization
dynamics taking place after the parton shower has terminated.

The first important aspect related to hadronization is that the
coherent branching algorithm generates colour structures which are
largely compatible with the space-time structure of hadronization in
the sense that large-angle soft gluons will be emitted first in the
parton shower evolution. Upon hadronization this effectively isolates
the colors of the outgoing hard partons in the form of jets at the
expense of forming a few, soft hadrons transverse to the hard jet
momenta. The latter feature is also the basis of the analytic
factorization methods for large-angle soft and collinear radiation
dynamics such as in Soft-Collinear Effective Theory
(SCET)~\cite{Bauer:2002nz}.  The second important aspect is the
preconfining property of the coherent QCD evolution. This property
requires that, in the large-$N_c$ limit, color singlets (i.e.\ colour
connected quark-antiquark pairs which form at the end of the shower
through additional $g\to q \bar{q}$ branchings to be discussed in more
detail in Sec.~\ref{sec:Clusters} and \ref{sec:NewFission}) acquire a
universal invariant mass spectrum that is peaked at scales similar to
the infrared cutoff $Q_0$ of the shower evolution and falls off
exponentially for larger invariant masses. These 'clusters',
expressing the fact that color correlations are very local in phase
space, can be interpreted as excited hadronic states and form the
basic degree of freedom of the cluster hadronization model.
The way how the coherent evolution builds up colour
structure in the large-$N_c$ limit also reflect the time scales
involed in the build-up of jets and the way colour charges of the
inital $q\bar{q}$ pair are isolated from each other, predicting few
soft hadrons transverse to the hard jet directions. Coherent
branching and the cluster model, together with global observables
which can reliably be predicted using coherence, thus fit very well
to study the interface between shower and
hadronization.\footnote{Non-global effects might not be predicted
in an entirely reliable way from coherent branching, however hadronization
dynamics in this case are also far from well-understood.}

\subsection{Kinematic Reconstruction, Reshuffling and True Parton Level}
\label{sec:KinematicReconstruction}

The coherent branching algorithm proceeds by generating, for each hard
jet progenitor produced in the hard scattering or a decay process, a
sequence of evolution variables: the angular scales $\tilde{q}$,
momentum fractions $z$ and azimuthal orientations $\phi$ of the
emissions. The values of these evolution variables are distributed
according to the Sudakov densities and the splitting functions
describing the individual branchings respecting the angular ordering
restrictions already mentioned before~\cite{Catani:1990rr}.  The full
kinematics of the emerging partonic final state is, however, not yet
determined during this process. Rather, it is inferred at the end of
the shower evolution, when the sequence of these variables is
terminated through an infrared resolution criterion. For the coherent
branching algorithm this criterion is based on the transverse momentum
of the branching (which is a function of $\tilde{q}$ and $z$) being
larger than a cutoff value $Q_0$. Once the transverse momentum drops
below $Q_0$, the parton shower evolution terminates. At this point a
process of {\it kinematic reconstruction} takes place to determine a
concrete physical final state with partons having four-momenta
satisfying on-shell conditions and overall momentum conservation.
This proceeds in two steps.

First, the momenta of the final state partons emerging from each
progenitor are calculated from the sequence of evolution variables
$\tilde{q},z, \phi$ according to the progenitor (forward) and partner
(backward) direction associated to the evolving jet. At this point the
four-momenta of all partons do not yet satisfy overall energy-momentum
conservation since the total four-momenta of the partons emerging from
each progenitor acquire invariant masses larger than the original
progenitor masses. We call these invariant masses also progenitor
virtualities below.  Then, as the second step, a {\it reshuffling
  algorithm} is employed to balance the resulting jets against each
other, maintaining overall energy-momentum conservation.  In this
algorithm the spatial momenta of all emitted final state partons in
their common center-of-mass frame are rescaled by a global factor such
that overall energy-momentum conservation is satisfied.

The (on-shell) masses of the final state partons emerging from the
kinematic reconstruction procedure are in principle free parameters.
In the context of processes involving only massless quarks and gluons,
one would expect them to be zero.  This is what we refer to as the
{\it true parton level}, which can also be related to massless parton
final states in standard computations in perturbative QCD. In general
the true parton level is associated to the quarks having {\it current
quark masses} $\hat m_i$.\footnote{This true final state parton
level has been employed and analyzed in detail already in our previous
work~\cite{Hoang:2018zrp}. In practice, for a light quarks a mass of
$10$~keV is adopted. In that previous work we have also
analyzed in depth how the parton shower algorithm affects the
interpretation of the mass parameter for heavy quarks in relation to
mass renormalization schemes.}  However, the initial steps to
interface the parton level to the cluster hadronization model requires
(much larger) {\it constituent quark masses} $m_i$ for all final state
quarks.  These constituent quark masses are parameters of the
hadronization model and are constrained such that a cluster is
kinematically allowed to decay at least into a pair of the lightest
hadrons.  Furthermore, the cluster hadronization model involves a
branching of each gluon into $q\bar{q}$ pairs as the basis to form the
initial clusters. The model thus assigns the gluons a mass $m_g$, such
that this $g\to q\bar{q}$ splitting process is kinematically allowed
at least for the lightest quarks {\it with respect to their
  constituent masses}.\footnote{Below we also refer to the gluon mass
  $m_g$ as a constituent mass for simplicity.}

For practical purposes, the
previous default implementations of the parton shower in the Herwig
event generator have only been providing a 'constituent' parton
  level with the previously mentioned constituent quark and gluon
  masses. So the kinematic reconstruction and reshuffling procedures
  have been directly accounting for these masses, so that {\it there
    has not been any direct access to the true parton level}.  In
  other words, the Herwig's default parton level has already
  incorporated some aspects of hadronization.  As long as the true
  parton level is of no relevance in the simulation, for example when
  the main focus is to describe experimental data, this is not an
  issue. However, for analyses (such as those carried out in this
  work) where the effects of hadronization need to be cleanly and
  accurately distinguished from the perturbative dynamics, it is. In
  particular, for events with large gluon multiplicities, the mass
  related kinematic effects can be quite significant given that tunes
  to data yield gluon mass values typically of the order of $1$~GeV.

In order to have access to the true parton level so that the effects
of the hadronization model can be cleanly separated and quantified
on an event-by-event basis, we have therefore extended
\herwig's functionality to perform the kinematic reconstruction and
reshuffling (as described above) based on massless gluons and current
quark masses $\hat m_i$.  Subsequently, as the first part in the
implementation of the hadronization model an {\it additional
  reshuffling procedure} is carried out which changes current quark
masses $\hat m_i$ to constituent masses $m_i$ and provides the mass
$m_g$ to the gluon.  This reshuffling procedure is quite similar to
the reshuffling carried out for the true parton level. First all final
state partons are assigned their new mass ($m_i$ or $m_g$) and then
again a global spatial momentum rescaling is carried out in the final
state partons' center-of-mass frame to ensure yet again overall
energy-momentum conservation. The resulting partonic final state with
quark constituent and gluon masses is not identical with, but very
close to the constituent parton level of the default
implementation. This procedure ensures that the process where the
partons acquire quark constituent and gluon masses is separated
completely from the shower evolution, so that it can be cleanly
considered as a part of the hadronization model.  This is vital to
extract event-by-event migration matrices which describe how a parton
level observable value correlates with a hadron level observable in a
single event and enables us to calculate binned migration matrices for
any observable by reading out two-dimensional histograms, see
Sec.~\ref{sec:extractmigration}.

\section{Expectations and Constraints on Hadronization}
\label{sec:expectations}

The default cluster hadronization model has predominantly 
been motivated by the
preconfinement property of coherent QCD evolution, and was otherwise
driven by minimal assumptions {\it e.g.} using only information on phase space 
and available quantum numbers as well as simple power laws for the dynamics
within the model.\footnote{A detailed description of the cluster model
  will follow in Sec.~\ref{sec:Clusters}.} The parton shower evolution
provides kinematic and color connection information to the
hadronization model that depends on the value of the scale $Q_0$,
where the partonic evolution terminates. The value of the shower cut
$Q_0$ has then typically been inferred through the {\it tuning
  procedure}, where also all parameters of the hadronization model
(including also the quark constituent and gluon masses) are fixed in a
fit to a set of reference data. This has assigned the shower cut
effectively the role of an additional hadronization parameter even
though its value affects the properties of the final states that
emerge from the parton shower.  Different hadronization models, while
technically inter-operable among different types of parton showers and
applicable for different values of the shower cut, have thus always
shown an implicit dependence on $Q_0$ through the tuning procedure.
However, this dependence has not been systemtically studied, neither
to design and improve hadronization models, nor for systematic
investigations of the uncertainty in the hadronization modeling or the
size of the hadronization corrections.  Nevertheless, it has been
common practice to adopt hadronization effects extracted from MC
simulations, through parton-to-hadron level migration matrices or by
taking hadron-parton level ratios, as estimates for hadronization
corrections for high-precision and potentially resummed perturbative
QCD calculations, where for the latter the limit of a vanishing
infrared regulator has been applied. A popular application of this
kind is constituted by the previously mentioned strong coupling
determinations from $e^+e^-$ event-shape distributions or jet
rates~\cite{ParticleDataGroup:2022pth}.

In recent works~\cite{Hoang:2018zrp,Platzer:2022jny}, however, we have
been pointing out that the role of the parton shower cutoff $Q_0$
should be understood in the sense of an infrared factorization scale.
This implies that the value of $Q_0$ and the way how the shower cut is
implemented define a particular scheme how the partonic and the
non-perturbative dynamics are separated and that the hadron level
descriptions should exhibit some invariance under variations of
$Q_0$. This factorization scale invariance would ensure that the
hadronization model does not modify the infrared structure provided by
perturbative QCD in an uncontrolled manner.  In other words the shower
evolution and the hadronization model should match at this scale, in
the sense that the partonic final state provided by the parton shower
at the scale $Q_0$ provides the starting point of the evolution of the
hadronization model towards even lower scales where the
non-perturbative dynamics at the scales of individual hadrons sets
in. Thus at least within some limited range, the evolution of the
hadronization model should be driven by the perturbative dynamics
encoded also in the parton shower evolution. In
Ref.~\cite{Hoang:2018zrp} we studied the parton shower 
	evolution with $Q_0$ for \herwig's coherent branching
parton shower in detail for massless and massive quark $e^+e^-$ event
shape distributions. We showed that the evolution is dominated by
effects linear in $Q_0$ which can be quantified accurately through
RG-evolution equations that can be calculated at NLO either from the
coherent branching algorithm itself or from a common diagrammatic
computation. This $R$-evolution equation is reviewed below in
Sec.~\ref{sec:observable}.

\subsection{General Formulation}
\label{sec:general}

Schematically, our starting point can be summarized either as a
factorisation at the level of the cross section or at the level of the
colour density operator \cite{Platzer:2022jny}, by studying a
convolution of partonic and hadronic cross sections as 
\begin{multline}
 \label{eq:egfactorization}
  \frac{{\rm d}\sigma_H}{{\rm d}w}
  = \sum_{n,m,c} \int \int {\rm d}\phi_m(p_1,...,p_m|Q)\\
  {\rm d}\sigma_{P,n,c}(q_1,..,q_n|Q;Q_0) S_{mn,c}(p_1,...,p_m|q_1,...,q_n;Q_0)
  \delta\left(w - W(p_1,...,p_m)\right)\,.
\end{multline}
Here ${\rm d}\phi_m(p_1,...,p_m|Q)$ is the integration measure of the
final-state hadron momenta and the term
${\rm d}\sigma_{P,n,c}(q_1,..,q_n|Q;Q_0)$ stands for the partonic cross
section, including the partonic phase space integration of total
momentum $Q$ and a certain colour-flow $c$. The term
$S_{mn,c}(p_1,...,p_m|q_1,...,q_n;Q_0)$ represents the action of the
hadronization model in converting $n$ partons into $m$ hadrons subject
to a given model and momentum mapping inherent to it. $W(p_1...,p_m)$
in turn is the observable's definition at hadron level. In practice
for an event generator we have $m>n$. We also stress the fact that
such a {\it probabilistic} factorization is merely possible in
presence of the large-$N$ limit and more generally would involve the
presence of a colour flow in the parton level amplitude and its
conjugate, see \cite{Platzer:2022jny} for more details.  In the above
form it is clear that the demand of $Q_0$-independence implies sets of
evolution equations for both the partonic and hadronic factor, which
mix different partonic multiplicities, a generic feature which a
hadronization model consistent with $Q_0$ independence must
respect. To this end, we can use the evolution of the parton shower,
which we can schematically write for a gluon exchange or emission with
momentum $q_n$ in terms of a virtual $V_c(q_1,..,q_n;Q_0)$ and real
emission contributon $R_{c,c'}(q_1,..,q_n|p_1,...,p_{n-1};Q_0)$ as
\begin{multline}
  \label{eq:psevolution}
  Q_0\frac{\partial}{\partial Q_0} {\rm d}\sigma_{P,n,c}(q_1,...,q_n|Q;Q_0) =
    V_c(q_1,..,q_n;Q_0) {\rm d}\sigma_{P,n,c}(q_1,...,q_n|Q;Q_0)  \\  -
    \sum_{c'}\int  R_{c,c'}(q_1,..,q_n|p_1,...,p_{n-1};Q_0){\rm d}\sigma_{P,n-1,c'}(p_1,...,p_{n-1}|Q;Q_0)
    {\rm d}^d p_1\cdots {\rm d}^d p_n  \ ,
\end{multline}
where $V$ and $R$ include the definition of some momentum mapping, and
can eventually be expressed in terms of splitting functions.
Demanding that the hadronic cross section is independent of $Q_0$, one
can then derive an evolution equation for
$S_{mn}$~\cite{Platzer:2022jny}. This evolution
  demands that the hadronization model should be able to mirror the dynamics of the
  parton shower in evolving towards a low-scale process of producing
  hadrons. For the present purpose we note that
Eq.~(\ref{eq:egfactorization}) is an accurate analytic model of an
event generator, and Eq.~(\ref{eq:psevolution}) one of parton shower
evolution. We can also ask the question how such a factorization then
gives rise to observable-specific hadronization corrections, which can
be extracted in a MC sense by snapshots of events at the parton level
and at the level of the final hadronic states.  For $e^+e^-$
event-shapes this is discussed in the next subsection. Furthermore, we
are naturally led to conclude that $Q_0$ should not be a tuning
parameter and that we should investigate the $Q_0$ dependence of
$S_{mn}$, for example, by studying the dependence of the tuned
hadronization model parameters on variations of $Q_0$.  As already
mentioned in the introduction, the least we need to demand from an
improved $S_{mn}$ is a smooth change across multiplicities in the
sense that higher partonic multiplicities at lower $Q_0$ lead to
comparable effect as smaller multiplicities at higher $Q_0$ do: the
dynamics governing $S_{mn}$ near the infrared cutoff need to be the
same as that of the shower close to the infrared cutoff. This puts
severe constraints on the onset of the hadronization model, to be
discussed in Sec.~\ref{sec:Clusters} and
Sec.~\ref{sec:NewFission}. Using the evolution equations implied for
$S_{mn}$ for the construction of an evolving model is subject of
additional ongoing work.

\subsection{Thrust Distribution, Factorization and $R$-Evolution}
\label{sec:observable}

The concrete observable we consider in most detail 
in this work is the {\it 2-jettiness} $\tau$
event-shape variable in $e^+e^-$ collisions, defined by
\begin{equation}
\label{eq:tau2def}
\tau = \frac{1}{Q} \min_{\vec{n_t}}\sum_i(E_i - |\vec{n_t} \cdot \vec{p_i}|)\,,
\end{equation}
where $Q$ is the $e^+e^-$ c.m.\ energy, and the sum runs over all
final state particles with momenta $\vec{p_i}$. The maximum defines
the thrust axis $\vec{n_t}$. In the limit of vanishing hadron masses
$\tau$ is identical to thrust. The difference is very small and does
not play any role in the context of our studies. We therefore call
$\tau$ also thrust sometimes in the rest of this article. 
Thrust is an IR safe
and global shape variable and in the limit of small $\tau$,
referred to as the dijet region, the events are 
characterized by two energetic
back-to-back jets along the thrust axis. In this section we render
factorization aspects of the coherent branching shower cutoff $Q_0$
and the resulting RG equation explicit for the thrust distribution in
the dijet region. This RG equation serves as the basis of the concrete
numerical studies we carry out in Secs.~\ref{sec:NewFissionpheno} for
the default and our novel dynamical hadronization model.
As already indicated in the introduction,
results along these lines are currently only available for jet mass related
event-shapes such as 2-jettiness. 

Using the notations of Ref.~\cite{Abbate:2010xh} the hadron level
thrust distribution in the dijet region can be written in the
factorized form
\begin{align}
\label{eq:thrustmassless1}
\frac{\mathrm{d}\sigma}{\mathrm{d}\tau}(\tau,Q) \,=\, &
\int\limits_0^{Q\tau}\!\mathrm{d}\ell\; 
\frac{\mathrm{d}\hat\sigma}{\mathrm{d}\tau}\Big(\tau-\frac{\ell}{Q},Q\Big)\,\,S_{\rm had}(\ell) \\ \nonumber
\, =\, & \int\limits_0^{\tau}\!\mathrm{d}\hat\tau\; 
 \frac{\mathrm{d}\hat\sigma}{\mathrm{d}\hat\tau}(\hat\tau,Q)\,\,Q\,S_{\rm had}(Q(\tau-\hat\tau))
\end{align}
where $\mathrm{d}\hat\sigma/\mathrm{d}\hat\tau(\hat\tau,Q)$ is the 
parton level distribution containing the resummed partonic QCD
corrections. In the standard analytic QCD approach these computations
are carried out in the limit of a vanishing IR cutoff, so that these
perturbative QCD corrections encode terms of the form
$\alpha_s^n\delta(\hat\tau)$ and plus-distributions of the form
$\alpha_s^n [\ln^k(\hat\tau)/\hat\tau]_+$) to all orders of perturbation
theory and potentially additional fixed-order corrections to improve
the descriptions when $\tau$ increases.
In this context $\mathrm{d}\hat\sigma/\mathrm{d}\hat\tau(\hat\tau,Q)$
has been determined up to N$^3$LL$+{\cal O}(\alpha_s^3)$
order~\cite{Becher:2008cf,Abbate:2010xh}.  The exact form of the
parton level distribution, which is based on an additional
perturbative factorization between large-angle soft and energetic
collinear radiation, is not relevant for the studies in this article.
The relevant aspect is that the factorized form of
Eq.~(\ref{eq:thrustmassless1}) applies for any scheme to regulate the IR
momenta in the partonic distribution $\mathrm{d}\hat\sigma/\mathrm{d}\hat\tau(\hat\tau,Q)$.

The function $S_{\rm had}(\ell)$ is the shape function that
describes the leading hadronization effects. For thrust in the 
dijet region it arises from the non-perturbative dynamics of the 
large-angle soft radiation  in the vicinity of the hemisphere plane 
perpendicular to the thrust axis. Hadronization effects also exist 
for the energetic collinear radiation, but these are strongly 
suppressed and negligible.\footnote{While large-angle soft 
dynamics is linearly sensitive to non-perturbative scales, 
the collinear dynamics is only quadratically sensitive and 
furthermore is associated to momentum fluctuations at higher 
scales of order $Q\sqrt{\tau}$ compared to the soft scales 
that are of order $Q\tau$.}
The shape function satisfies the normalization condition
\begin{align}
\label{eq:Omega0}
\int \!\mathrm{d}\ell\, S_{\rm had}(\ell) \, = \, 1\,.
\end{align} 
It has an unambiguous definition related to vacuum-to-hadrons matrix
elements of soft gluon Wilson lines. The scale $\ell$ is the non-perturbative 
light-cone momentum associated to coherent large-angle soft radiation 
respecting the thrust hemisphere constraint related to the thrust
axis~\cite{Korchemsky:1998ev,Bauer:2008dt}. The shape function is known 
to be universal for many event-shape distributions associated to 
the thrust axis hemisphere definition~\cite{Lee:2006fn}. In the 
canonical approach for analytic (and numerical) perturbative 
QCD computations the partonic distribution  
$\mathrm{d}\hat\sigma/\mathrm{d}\hat\tau(\hat\tau,Q)$ 
and thus also the shape function are defined in the scheme of a 
vanishing IR cutoff. The shape function $S_{\rm had}(\ell)$
exhibits a peaked behavior for $\ell$ values around $1$~GeV and is
strongly falling to zero for larger $\ell$. Due to the smearing 
effects induced by convolution over the shape function 
in Eq.~(\ref{eq:thrustmassless1}) the peak of the hadron level $\tau$
distribution is shifted from the parton level threshold at
$\hat\tau=0$ to positive values by an amount of order
$\Lambda/Q$, where $\Lambda$ is in the range of $1$ to $2$~GeV. 
It is this peak region, which we focus on in our studies. 

The point essential for our study is that the form
of Eq.~(\ref{eq:thrustmassless1}), which factorizes parton level and 
hadronization effects, is an umambiguous property of QCD
independent of the scheme that is adopted to regulate IR momenta in
the partonic distribution.
In Ref.~\cite{Hoang:2018zrp} the NLL partonic thrust distribution was
analyzed from the perspective of using the transverse momentum cut
$Q_0$ that is employed in \herwig's angular ordered parton shower to
regulate IR momenta and keeping track of the dominant linear
dependence on $Q_0$.\footnote{In Ref.~\cite{Hoang:2018zrp} the
analysis was carried out for the coherent branching algorithm as
well in soft-collinear effective theory (SCET). The exact relation
of NLL precision for the angular ordered parton shower and terms in
the NLL$+{\cal O}(\alpha_s)$ order counting in SCET was specified 
as well.} It was found that 
(a) to achieve NLL for the solution of the coherent branching algorithm 
only the splitting
function $P_{qq}$ for the radiation of a gluon off a quark needs to be accounted for\footnote{
This feature was already discussed a long time ago in Ref.~\cite{Catani:1992ua}. }
and that (b) 
only the 
large-angle soft radiation can cause a linear $Q_0$ sensitivity. 
The partonic distribution
$\mathrm{d}\hat\sigma/\mathrm{d}\hat\tau(\hat\tau,Q,Q_0)$ in the presence of a
finite value of $Q_0$ can be written as
\begin{align}
\label{eq:partonicthrustQ0}
\frac{\mathrm{d}\hat\sigma}{\mathrm{d}\hat\tau}(\hat\tau,Q,Q_0)\,=\, &
\frac{\mathrm{d}\hat\sigma}{\mathrm{d}\hat\tau}\left(\hat\tau,Q\right) 
+ \frac{1}{Q}\Delta_{\rm soft}(Q_0)\, \frac{\mathrm{d}^2\hat\sigma}{\mathrm{d}\hat\tau^2}\left(\hat\tau,Q\right) \\ \nonumber
\,=\, &
\frac{\mathrm{d}\hat\sigma}{\mathrm{d}\hat\tau}\left(\hat\tau+\frac{1}{Q}\Delta_{\rm soft}(Q_0),Q\right)
\end{align} 
with the gap function 
\begin{align}
\label{eq:Deltasoftv1}
\Delta_{\rm soft}(Q_0) \, = & \,
16\, Q_0\,\frac{\alpha_s(Q_0)C_F}{4\pi}\,+\,{\cal O}(\alpha_s^2(Q_0))
\end{align}
and where $\mathrm{d}\hat\sigma/\mathrm{d}\hat\tau(\hat\tau,Q)$ is the 
partonic thrust distribution for a vanishing IR cutoff shown in 
Eq.~(\ref{eq:thrustmassless1}).
Note that the simple form of Eq.~(\ref{eq:Deltasoftv1}) arises from 
a multipole expansion keeping
the dominant linear dependence on $Q_0$ of the full result 
(which is also given in
Ref.~\cite{Hoang:2018zrp}). Upon convolution with the shape 
function this multipole expansion 
provides an excellent approximation to the full result.
The strong coupling in the gap function $\Delta_{\rm soft}(Q_0)$ is
evaluated at the renormalization scale $\mu=Q_0$ since the gap function 
quantifies the
effects of the unresolved (large-angle) soft radiation below the 
scale $Q_0$, which can only depend on the scale $Q_0$ in perturbation theory. 
The NLL precision of the coherent branching algorithm ensures 
that this $Q_0$ dependence is also realized by the \herwig{} 
simulation parton level results.

The ${\cal O}(\alpha_s^2)$ term indicated in
Eq.~(\ref{eq:Deltasoftv1}) can only be specified once the cutoff
prescription in the context of a more precise NNLL order shower
evolution has been defined. Such a prescription is currently unknown and 
we therefore drop 
these higher order contributions from now on. The result for the gap 
function $\Delta_{\rm soft}(Q_0)$ implies it
satisfies the renormalization group equation
\begin{align}
\label{eq:gapRGE}
R\,\frac{\mathrm{d}}{\mathrm{d}R}\,\Delta_{\rm soft}(R)\, = \, 16\,R\,\frac{\alpha_s(R)C_F}{4\pi} \,.
\end{align}
This evolution equation describes a linear scale dependence and has
been called $R$-evolution in
Refs.~\cite{Hoang:2008fs,Hoang:2008yj,Hoang:2009yr}.\footnote{
The point of the $R$-evolution is that the perturbation series for 
$\Delta_{\rm soft}(R)$ contains an ${\cal O}(\Lambda_{\rm QCD})$ IR 
renormalon, but that Eqs.~(\ref{eq:gapRGE}) and 
(\ref{eq:partonicthrustQ0primeQ0}) are renormalon-free.
}
The NLL partonic thrust distribution at two different shower cutoff values
$Q_0$ and $Q_0^\prime$ are therefore related by the equality
\begin{align}
\label{eq:partonicthrustQ0primeQ0}
\frac{\mathrm{d}\hat\sigma}{\mathrm{d}\hat\tau}(\hat\tau,Q,Q_0)\,=\, 
\frac{\mathrm{d}\hat\sigma}{\mathrm{d}\hat\tau}\left(\hat\tau + \frac{1}{Q}\Delta_{\rm soft}(Q_0,Q_0^\prime)  ,Q,Q_0^\prime\right)\,,
\end{align} 
where 
\begin{align}
\label{eq:Deltasoftv2}
\Delta_{\rm soft}(Q_0,Q_0^\prime) \, = \,
16\,\int\limits_{Q_0^\prime}^{Q_0}\mathrm{d}R\,\,\Bigl[\,\frac{\alpha_s(R)C_F}{4\pi}\,\Bigr]\,.
\end{align}
We emphasize again that the NLL precision of the parton shower (for the
thrust distribution) is essential, since otherwise
Eqs.~(\ref{eq:partonicthrustQ0}) and
(\ref{eq:partonicthrustQ0primeQ0}) and the evolution equation in
Eq.~(\ref{eq:Deltasoftv2}), which can be computed in a straightforward way in analytic QCD computations,
may not be realized by the parton level MC simulations.

Since factorization of the partonic thrust distribution and the
non-perturbative shape function also applies in the context of a finite
transverse momentum cutoff $Q_0$,
\begin{align}
\label{eq:thrustmassless2}
	\frac{\mathrm{d}\sigma}{\mathrm{d}\tau}(\tau,Q) \,=\, &
	\int\limits_0^{\tau}\!\mathrm{d}\hat\tau\; 
	\frac{\mathrm{d}\hat\sigma}{\mathrm{d}\hat\tau}(\hat\tau,Q,Q_0)\,\,Q\,S_{\rm had}(Q(\tau-\hat\tau),Q_0)\,,
\end{align}
we can also derive the relation of the shape functions for two
different cutoff values $Q_0$ and $Q_0^\prime$:
\begin{align}
\label{eq:ShadQ0primeQ0}
S_{\rm had}(\ell,Q_0^\prime) \,=\,  S_{\rm had}(\ell -\Delta_{\rm soft}(Q_0^\prime,Q_0),Q_0)\,.
\end{align} 
Here we remind the reader that this relation is exact at the level of
terms linear in the cutoff
$Q_0$. Equalities~(\ref{eq:partonicthrustQ0primeQ0}) and
(\ref{eq:ShadQ0primeQ0}) are concrete realizations of the more general
and generic relations quoted in Sec.~\ref{sec:general}.

The quantity that conveniently quantifies the shape function's 
linear dependence on $Q_0$ is the shape functions first
moment\footnote{The weight factor $1/2$ in the definition of the first
moment is motivated by the fact that 2-jettiness is related to the
sum of the two squared hemisphere (jet) masses in the small $\tau$
dijet region. The factor normalizes back to quantify the
hadronization effects of a single jet.}
\begin{align}
\label{eq:Omega1Q0}
\Omega_1(Q_0) \, \equiv\,
\frac{1}{2}	\int \!\mathrm{d}\ell\,\ell\, S_{\rm had}(\ell,Q_0)\,,
\end{align} 
which leads to the relation
\begin{align} 
\label{eq:Omega1Q0primeO0}
\Omega_1(Q_0\prime) \, =\, \frac{1}{2}\,\Delta_{\rm soft}(Q_0^\prime,Q_0) +\Omega_1(Q_0)\,.
\end{align} 
All other higher order cumulant moments do not have any linear cutoff
dependence. The relation of the shape function's first moment for different 
parton shower cutoff values shown in Eq.~(\ref{eq:Omega1Q0primeO0}) is 
an essential prediction of QCD and plays an
important role in our subsequent analysis of \herwig's hadronization models. 

\subsection{Shower Cutoff Dependence for Thrust and Migration Matrix}
\label{sec:numericalQ0}

At this point it is highly instructive to cross check the level of
validity of relation~(\ref{eq:partonicthrustQ0primeQ0}) for the
implementation of the angular-ordered parton shower in \herwig{}. As
we already mentioned, \herwig{}'s angular-ordered parton shower is NLL
precise for $e^+e^-$ event-shapes in the dijet limit.  Even though the
NLL precision is guaranteed conceptually, it is certainly useful to
examine it for the practical implementation of the \herwig{} event
generator. Note that a similar analysis 
has already been carried out in Ref.~\cite{Hoang:2018zrp}.
To this end, we consider the shower-cutoff dependence of
the 2-jettiness cumulant
\begin{align}
\label{eq:cumiulantdef}
\hat\Sigma(\hat\tau,Q,Q_0) \equiv \int\limits_0^{\hat{\tau}} \mathrm{d}\bar\tau\,\,
\frac{\mathrm{d}\hat\sigma}{\mathrm{d}\bar\tau}(\bar\tau,Q,Q_0)\,.
\end{align}
Using Eq.~(\ref{eq:partonicthrustQ0primeQ0}) it is straightforward 
to derive the partonic cumulant difference relation
\begin{align}
\label{eq:cumiulantdiffference}
Q\,\frac{\hat\Sigma(\hat\tau,Q,Q_0)-\hat\Sigma(\hat\tau,Q,Q_0^\prime)}{\frac{\mathrm{d}\hat\sigma}{\mathrm{d}\hat\tau}(\hat\tau,Q,Q_0^\prime)} 
= \Delta_{\rm soft}(Q_0,Q_0^\prime)\,,
\end{align}
where the normalization condition 
$\int\mathrm{d}\ell S_{\rm had}(\ell,Q_0) = 1$ applies.
We make the important observation that the RHS of 
Eq.~(\ref{eq:cumiulantdiffference}) is independent of
$\hat{\tau}$ and the hard scale $Q$.  As long as the soft-collinear
approximations associated to the dijet region are valid, 
which are the basis of
the factorization theorem in Eq.~(\ref{eq:thrustmassless1}) and the
shift relation in Eq.~(\ref{eq:partonicthrustQ0}), this
universality should hold to a good approximation. The range of thrust
values where relation (\ref{eq:cumiulantdiffference}) is realized for
\herwig's angular ordered parton shower also indicates the expected 
range of validity of the factorization
formula~(\ref{eq:thrustmassless2}) and the relation of the shape 
function's first moment for different cutoff values shown in 
Eq.~(\ref{eq:Omega1Q0primeO0}) that tests consistency with QCD.

\begin{figure}[h]
	\center
	\includegraphics[width=\textwidth]{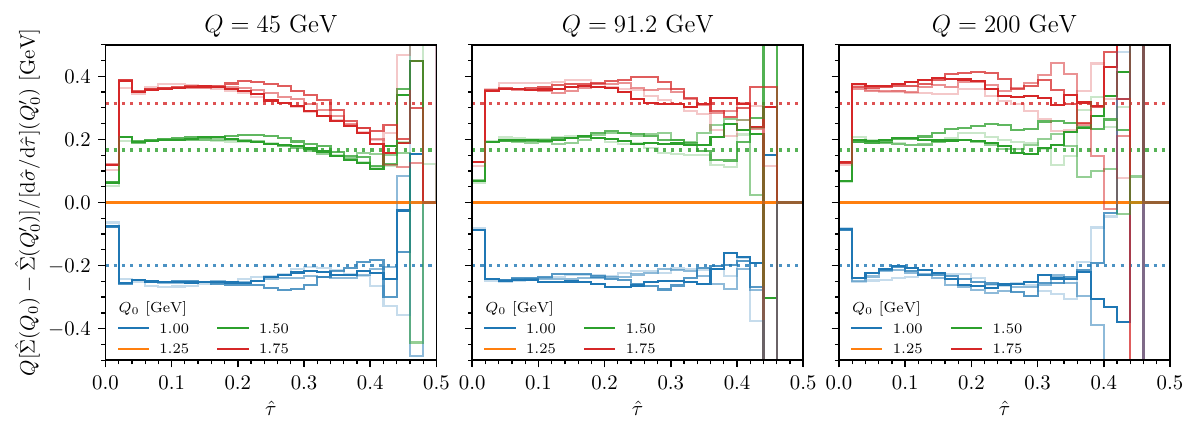}
	\caption{Binned cumulant differences of the parton level thrust 
    distribution defined in Eq.~(\ref{eq:cumiulantdiffference}) 
    generated by the \herwig~7.2 angular ordered parton
    shower for $Q=45$~GeV (left panel), $91.2$~GeV (middle panel) and 
    $200$~GeV (right panel) and shower cutoff values $Q_0=1$~GeV (blue)
    $1.5$~GeV (green) and $1.75$~GeV (red) with respect to the 
    reference cutoff $Q_{0,{\rm ref}}=1.25$~GeV. The dashed lines 
    show the results expected from QCD factorization employing 
    the $\overline{\rm MS}$ strong coupling extracted from \herwig{}.
}
    \label{fig:cumulantdifference}
\end{figure}
In Fig.~\ref{fig:cumulantdifference} we have displayed the cumulant
difference of Eq.~(\ref{eq:cumiulantdiffference}) for $Q_0=1.0$
(blue), $1.25$ (orange), $1.5$ (green) and $1.75$~GeV (red) with
respect to the reference result for $Q_0^\prime=1.25$~GeV for the hard
scales $Q=45$, $91.2$ and $200$~GeV. We have employed bins of size
$\Delta\hat\tau=0.02$, and each bin value is determined from the
average of the cumulant at the respective upper and lower bin
boundaries divided by the differential cross section of the bin. To
visualize the impact of matching corrections which affect the
2-jettiness distribution outside the dijet region, four different
\herwig{} matching settings have been used.  The four different
intensities for the same color correspond to (from darkest to the
brightest color): (i) leading-order matrix element with matrix-element
(ME) correction for hard QCD radiation (which is \herwig's default);
(ii) the same without ME correction; (iii) NLO matrix elements with
multiplicative (POWHEG-type) matching and (iv) additive (MC@NLO-type)
NLO matching, as available from the Matchbox framework
  \cite{Platzer:2011bc}.  The dotted horizontal lines correspond to
$\Delta_{\rm soft}(Q_0,Q_0^\prime)$, where we use the strong coupling
in the $\overline{\rm MS}$ scheme extracted from \herwig{} to solve
the RG equation of Eq.~(\ref{eq:Deltasoftv2}). As expected, the impact
of the matching and matrix element corrections increases for larger
$\hat\tau$ values since the soft-collinear approximations in the
coherent branching algorithms work more efficiently in the dijet
region where $\hat\tau$ is small. We also see that the region in
$\hat\tau$, where the matching corrections become sizeable, increases
toward smaller $\hat\tau$ ranges for increasing c.m.\ energy $Q$. This
indicates that the dijet 2-jettiness region around $\hat\tau=0$
decreases when the hard scattering scale gets larger and is consistent
with the fact that the peak location and peak width scale with
$\Lambda/Q$.

We see that, except for the very first $\hat\tau$ bin, where the
no-radiation events and the implementation dependent details of the
shower-cutoff dependence are still being resolved (and therefore not
relevant for testing Eq.~(\ref{eq:cumiulantdiffference})), the partonic
cumulant differences obtained from \herwig's angular ordered shower
are nicely compatible with the QCD value of
$\Delta_{\rm soft}(Q_0,Q_0^\prime)$ for $\hat{\tau}$ up to about
$0.2$. 
The level of agreement is as expected from a comparison between the analytic result
of Eq.~(\ref{eq:Deltasoftv2}) (which was obtained from an anomalous dimension at 
${\cal O}(\alpha_s)$ for the dominant linear cutoff scale effects in Ref.~\cite{Hoang:2018zrp})
and the parton shower result (where higher power terms in $\alpha_s$ and $Q_0$ are 
generated as well). 
The observed deviations at the level of $10$ to $15\%$ are consistent with the expected size of the higher order NNLL QCD and quadratic $Q_0$ contributions generated in the parton shower simulation, but not contained in the analytic NLL QCD result for $\Delta_{\rm soft}$ in Eq.~(\ref{eq:Deltasoftv2}). We also refer to Ref.[2] for analytic solutions of the coherent branching algorithm for thrust with NLL precision.
Overall, we can
conclude that the factorization formula~(\ref{eq:thrustmassless1})
should be applicable for 2-jettiness values $\hat\tau$ up to around
$0.2$ which well includes the peak region and a large fraction of
2-jettiness distribution tail.

Given that the angular ordered parton shower of \herwig{} exhibits the
correct NLL shower cutoff $Q_0$ dependence, the hadronization model
must have at an associated inverse $Q_0$ dependence so that the
hadron level description is $Q_0$ independent and the shower cut $Q_0$
can be interpreted as an IR factorization scale. For the thrust distribution the effect of the
hadronization in the MC generator appears in terms of a
parton-to-hadron level migration matrix function $T$ connecting bins in
the partonic thrust $\hat\tau$ to the hadron level thrust
$\tau$. Writing the expression for binned distributions for simplicity
in integral form, the generator's hadron level thrust distribution can
be written as
\begin{align}
\label{eq:thrustmassless3}
\frac{\mathrm{d}\sigma}{\mathrm{d}\tau}(\tau,Q) \,=\, &
\int\!\mathrm{d}\hat\tau\; 
\frac{\mathrm{d}\hat\sigma}{\mathrm{d}\hat\tau}(\hat\tau,Q,Q_0)\,\,T(\tau,\hat\tau,\{Q,Q_0\}))\,.
\end{align}
The transfer matrix $T$ is determined by reading out, for each event,
the thrust values at the true parton level ($\hat{\tau}$) and at the hadron
level ($\tau$) and by determining the resulting 2-dimensional
histogram. Unitarity ensures that
$\int \mathrm{d}\tau \, T(\tau,\hat\tau,\{Q,Q_0\})=1$ for each parton
level $\hat{\tau}$.  Comparing to the factorization analogue in
Eq.~(\ref{eq:thrustmassless2}) we see the close relation between the
shape function $Q S_{\rm had}( Q(\tau-\hat\tau),Q_0)$, which satisfies
the analogous normalization condition, and the migration
matrix $T(\tau,\hat \tau,\{Q,Q_0\})$.  Changing variables to $k=Q\tau$
and $\hat k=Q\hat{\tau}$ we define the rescaled MC parton-to-hadron level migration matrix function
\begin{align}
\label{eq:tildeSMCTrelation}
\tilde S^{\rm MC}(k,\hat k,\{Q,Q_0\}) \equiv \frac{1}{Q} T\bigg(\frac{k}{Q},\frac{\hat k}{Q},\{Q,Q_0\}\bigg)\,.
\end{align}
which is precisely the MC analogue to the shape function $S_{\rm had}(k-\hat k,Q_0)$. 

To visualize the correspondence to the shape function 
$S_{\rm had}(\ell,Q_0)$ more directly we
can also consider the rescaled MC parton-to-hadron level migration
function as a function of $\ell=k-\hat k$ for given values of
$\hat k$, $Q$ and $Q_0$. We therefore define a shifted version of the
rescaled migration matrix as a function of $\ell$,
\begin{align}
\label{eq:SMCdef}
	S^{\rm MC}(\ell,\{\hat k,Q,Q_0\}) \equiv \tilde S^{\rm MC}(\ell+\hat k,\hat k,\{Q,Q_0\})\,.
\end{align}
Its first moment, defined in analogy to Eq.~(\ref{eq:Omega1Q0}), 
should thus also satisfy the QCD constraint~(\ref{eq:Omega1Q0primeO0}).
Apart from the dependence on the shower cutoff scale $Q_0$, it is the
additional potential dependence on the partonic momentum variable
$\hat k$ and the hard scattering scale $Q$, which is not contained in
the shape function $S_{\rm had}(\ell,Q_0)$, that we will discuss in
the phenomenological analyses in the later sections of this article.
Details on how the hadronization model migration matrix function
$\tilde S^{\rm MC}$ is extracted in \herwig{} are provided in
Sec.~\ref{sec:extractmigration}. Examples for the migration matrix
functions $\tilde S^{\rm MC}$ (left panels) and $S^{\rm MC}$ (right
panels) are displayed in Fig.~\ref{fig:transfermatrix3Ddefault} for
the default \herwig{} hadronization model and in
Fig.~\ref{fig:transfermatrix3Ddynamic} for the novel dynamical model.

The reader mostly interested in the phenomenology of the default and
the novel dynamic hadronization models may now directly jump to
Sec.~\ref{sec:extractmigration}.

\section{The Default Cluster Hadronization Model}
\label{sec:Clusters}

The cluster hadronization model is motivated by the preconfinement
property of coherent QCD cascades. As we already briefly mentioned in
Sec.~\ref{sec:generatorgeneral}, in the first step, gluons are split
into quark-antiquark pairs such that (in the large-$N_c$ limit),
colour-neutral $q\bar{q}$ systems, the 'clusters', emerge. These
clusters are interpreted as highly excited hadronic systems, and
successively fission into lighter clusters, which, once below a
certain threshold, decay into pairs of hadrons. Within our novel
dynamical hadronization model the implementation of this final hadron
decay process is the same as for the default model. So here we are
mainly concerned with the gluon splitting, cluster formation and
cluster fission processes, which we briefly review in this section
focusing on the default \herwig{} implementation.  These three
processes are the important steps relevant for the matching to the
$Q_0$-dependent infrared regime of the parton shower. For many of the
details including a comprehensive description of the default
implementation, we refer the reader to Ref.~\cite{Bahr:2008pv}.

\subsection{Low-scale Gluon Splitting and Cluster Formation}
\label{sec:ForcedGluonSplitting}

After the parton shower has terminated, the final state consists of
quarks and gluons.  All gluons present thus need to undergo a
branching into quark-antiquark pairs such that the colour neutral
(mesonic quark-antiquark) clusters can be determined by the colour
connections which the parton shower has produced.  In the \herwig{}
default hadronization model every gluon is assigned the {\it same
  fixed} constituent mass $m_g$, while for the novel dynamical model
this gluon mass is generated dynamically, as explained in
Sec.~\ref{sec:dynamicgluonmass}.  As already mentioned in
Sec.~\ref{sec:KinematicReconstruction}, in the existing default
hadronization model implementation the gluon mass $m_g$ and the
constituent quark masses $m_i$ have been implemented by a reshuffling
procedure directly after the kinematic reconstruction, so that the
true parton level (with current quarks and massless gluons) has not
been accessible.  For the hadronization model implementations used in
this article we added access to the true parton level by first
implementing current
quarks and massless gluons (within the kinematic reconstruction and
  reshuffling at the end of the parton shower) and a separate
subsequent reshuffling to constituent quarks and
massive gluons. The latter 'constituent' parton level constitutes the
first step of the cluster hadronization model.

The gluon constituent mass $m_g$ of the default hadronization model is
one of its parameters subject to the tuning procedure.  In principle
the same applies to the quark constituent masses $m_i$. However, for
the cluster model they are highly constrained such that enough energy
is available for a cluster to produce at least the lightest pairs of
hadrons in its final decay. In practice the quark constituent quark
masses are therefore fixed parameters of the default model, see
Ref.~\cite{Bahr:2008pv}, as well as for our novel dynamical model.

After the reshuffling to the constituent parton level has been
performed, each (now massive) gluon is forced to split into a light
quark-antiquark pair. This decay is isotropic in the gluon rest frame,
and the flavor of the emerging light quark-antiquark pairs are
assigned randomly, see Sec.~7.1 in Ref.~\cite{Bahr:2008pv} for
details. The associated probabilities are also tuning parameters, but
they are not expected to carry any shower cutoff dependence and
therefore fixed to the default in all our following analyses.

\subsection{Cluster Fission}
\label{sec:ClusterFission}

After the forced gluon splitting, the final state consists only of
quarks and antiquarks having constituent quark masses $m_i$. The color
connected quark-antiquark pairs are now combined into the
clusters. For each cluster we have full information about its flavor
content and the 4-momenta of its two constituents, which define the
cluster's mass $M$. The step that now follows in the hadronization is
the cluster fission.  Each cluster that fulfills the relation
\begin{equation}
\label{eq:IsHeavy}
 {M}^{\rm{Cl}_{\rm{pow}}}\geq {\rm{Cl}_{\rm{max}}}^{\rm{Cl}_{\rm{pow}}}+(m_1+m_2)^{\rm{Cl}_{\rm{pow}}}\,,
\end{equation}
where the $m_i$ are the masses of the clusters's constituents, is
considered ``heavy'' and will undergo fission. Otherwise it is called
``light''. Cluster fission is a $1\to 2$ process, where one parent
cluster is split into two daughter clusters. To do so a $q\bar{q}$
pair is popped from the vacuum, and together with the already two
existing constituent quarks form two new color singlet clusters. If a
daughter cluster is again ``heavy'' according to
Eq.~\eqref{eq:IsHeavy}, it will itself undergo another fission, and so
on, until all clusters are ``light''. These final light clusters then
decay into the pair of hadrons we already mentioned in
Sec.~\ref{sec:KinematicReconstruction}.  Depending on the type of
cluster fission implementation (see below) and on the values of the
parameters $\rm{Cl}_{\rm{max}}$ and $\rm{Cl}_{\rm{pow}}$, it can
happen that also a cluster that is considered ``heavy'' according to
Eq.~\eqref{eq:IsHeavy} cannot undergo fission anymore because it is
impossible to produce two physical daughter clusters
which would be able to decay into hadrons
  individually. Also these clusters then decay
  directly into hadrons.  The parameters $\rm{Cl}_{\rm{max}}$ and
  $\rm{Cl}_{\rm{pow}}$ are tuning parameters that govern how long the
  cluster fission proceeds and how heavy the clusters can be when they
  finally decay into hadrons. There are separate $\rm{Cl}_{\rm{max}}$
  and $\rm{Cl}_{\rm{pow}}$ parameters for clusters containing charm
  and bottom quarks, but we have identified them in our analysis. The
  cluster fission condition~(\ref{eq:IsHeavy}) and the parameters
  $\rm{Cl}_{\rm{max}}$ and $\rm{Cl}_{\rm{pow}}$ are implemented for
  the default and our novel dynamic hadronization model.  The
  difference is the dynamics of how the $q\bar{q}$ pair is produced
  from the vacuum.

In the default cluster fission the process is entirely one-dimensional
in the sense that the momenta of the produced $q\bar{q}$ pair are
directed
along the axis defined by the cluster's constituent 3-momenta in the
cluster rest frame prior to the fission and that the direction of the
3-momenta of the original constituents remains unchanged.
The only free parameters in this simplistic process are the masses of
the two daughter clusters, $M_1$ and $M_2$. Only light flavored
$q\bar{q}$ pairs can be popped from the vacuum and their flavor is
picked randomly, with constant probabilities which are tuning
parameters.  
The daughter cluster masses $M_1$ and
$M_2$ are picked from a probability distribution that is a function of
the parent's mass $M$, the masses of the original constituents $m_1$
and $m_2$, the mass $m_q$
of the quarks popped from the vacuum and one
or more tuning parameters $\vec{\lambda}$:
\begin{equation}
\label{eq:ProbM1M2}
 \frac{\mathrm{d}^2P}{\mathrm{d}M_{1}\mathrm{d}M_{2}}=f\bigl(M,m_1,m_2,m_q,\vec{\lambda}\bigr)\,.
\end{equation}
These general properties of the cluster fission are very similar for
the default and the novel dynamical model. They differ in the way 
how the probability distribution in
Eq.~\eqref{eq:ProbM1M2} is obtained. We note that these probability
distributions are independent components and in principle not tied to
how the forced gluon splitting discussed in
Sec.~\ref{sec:ForcedGluonSplitting} is handled.  In order to better
understand the novel aspects in the cluster fission of the dynamical
model, we now describe briefly how the probability function is
determined for the default cluster fission.

For the default fission the daughter cluster masses $M_i$ for each fission process 
are generated from the equation
\begin{equation}
\label{eq:Gendefault1}
 M_{i}=m_i+(M-m_i-m_q)\times r_i^{1/\rm{PSplit}}\,\qquad i=1,2\,.
\end{equation}
Here, $r_i$ is a random variable drawn from a uniform distribution
$\rm{unif}(0,1)$. Additionally, it is also required that the kinematic
constraints
\begin{equation}
\label{eq:Gendefault2}
M_{i}\geq m_i+m_q\,,\qquad M_1+M_2\leq M\,.
\end{equation}
are fulfilled. The parameter $\mathtt{PSplit}$ is the only tuning
parameter for the cluster mass distribution of the default fission,
i.e.\ $\vec{\lambda}=(\mathtt{PSplit})$. 
There are separate parameters for clusters containing charm and bottom
quarks. 
The resulting double $M_1\leftrightarrow M_2$ symmetric differential
mass distribution generated from Eqs.~\eqref{eq:Gendefault1}
and~\eqref{eq:Gendefault2} reads
\begin{equation}
  \frac{\mathrm{d}^2P}{\mathrm{d}M_1\mathrm{d}M_2}\propto\Theta\bigl(M_0-M_1-M_2\bigr)\times g(M_1)\times g(M_2)\,,
\end{equation}
with
\begin{equation}
  g(M_i)=\frac{\Theta\bigl(m_i+m_q<M_i<M_0-m_q\bigr)}{M_0-m_1-m_q}\biggl[\frac{M_i-m_i}{M_0-m_q-m_i}\biggr]^{-1+\rm{PSplit}}\,,
\end{equation}
and where we suppressed the normalization factor. Integrating let's say over mass $M_2$ we can also write down the single differential mass distribution of the default cluster fission:
\begin{align}
\label{eq:defaultsinglediff}
  \frac{\mathrm{d}P}{\mathrm{d}M_1}\propto & \frac{\Theta\bigl(m_1+m_q<M_1<M_0-m_2 -m_q\bigr)}{M_1-m_1}\Biggl[\frac{M_1-m_1}{\Bigl(M_0-m_1-m_q\Bigr)\Bigl(M_0-m_2-mq\Bigr)}\Biggr]^{\rm{PSplit}}\notag \\
  &\qquad\times\biggl[\Bigl(M_0-M_1-m_2\Bigr)^{\rm{PSplit}}-m_q^{\rm{PSplit}}\biggr]\,,
\end{align}
The smallest possible mass that a cluster generated in the fission can have is $m_i+m_q$. Therefore the lower bound on the mass of a cluster that can still undergo fission (i.e.\ being classified as ``heavy'' by Eq.~\eqref{eq:IsHeavy}) is
\begin{equation}
 M_{\rm{min}}=m_1+m_2+2m_q\,.
\end{equation}
Neglecting for simplicity the quark constituent masses, the single
differential cluster mass distribution in
Eq.~\eqref{eq:defaultsinglediff} reduces to the expression
\begin{equation}
\label{eq:dPM1distribution}
 \frac{\mathrm{d}P}{\mathrm{d}M_i}\approx 
 \frac{\Theta\bigl(M_0-M_i\bigr)}{M_0}\Bigl(\frac{M_i}{M_0}\Bigr)^{\rm{PSplit}-1}
 \Bigl(1-\frac{M_i}{M_0}\Bigr)^{\rm{PSplit}}\,.
\end{equation}
We remind the reader that the result in
Eq.~(\ref{eq:dPM1distribution}) provides the daughter cluster mass
distribution for a single cluster fission process. The final resulting cluster
mass distribution, at the point when the cluster fission processes of a single
event have terminated,
is more involved and also depends on the tuning parameters
${\rm{Cl}_{\rm{pow}}}$ and ${\rm{Cl}_{\rm{max}}}$ due to the heavy
cluster condition in Eq.~(\ref{eq:IsHeavy}).

\section{Dynamical Cluster Hadronization Model}
\label{sec:NewFission}

Within \herwig's cluster hadronization model implementation the two
essential features that influence to which extend the hadronization
model dynamics can properly match the infrared features of the parton
shower are the gluon mass $m_g$, the kinematics of the subsequent
forced gluon splitting in the first step of the hadronization process,
and the dynamics behind the cluster fission. For a fixed gluon mass
value the problematic aspect is that the gluon splitting process
carried out by the parton shower entails in contrast a nontrivial
distribution of invariant masses of the $g\to q\bar q$ process. This
distribution depends on the value of shower cut $Q_0$. But it is also
clear that a fixed gluon mass value can only mimic some averaged
features of the parton shower splitting, and this prohibits an exact
matching to the parton shower.  For the cluster fission process, where
an additional light quark-anti-quark pair is produced from the vacuum,
the parton shower analogue is a radiated gluon which afterwards
branches into the light quark-antiquark pair. For the default cluster
fission process, the dynamical aspects encoded in these parton shower
processes are missing. It can therefore not be expected that fixing
the parameters $m_g$, ${\rm{Cl}}_{\rm{max}}$, ${\rm{Cl}}_{\rm{pow}}$
and $\rm{PSplit}$ from the tuning procedure should result in an exact
matching to the parton shower.\footnote{
    Reference~\cite{Chahal:2022rid} also discusses a dynamical
    splitting of gluons depending on their colour connections to other
    perturbatively produced partons in the context of an
    implementation of a cluster hadronization model in the SHERPA
    event generator.  From the algorithmic structure this model may be
    able to also relate the underlying splitting function to the
    shower evolution.}

The motivation for the construction of the dynamical cluster
hadronization model is to add these parton shower aspects back to the
model implementation such that the ultraviolet aspects of the
hadronization modelling have an improved compatibility to the infrared
behavior of the parton shower. The essential novel aspects are (i) a
dynamically generated distribution for the gluon mass $m_g$ and
parton-shower-like kinematics for the forced splitting in the initial
stage of the hadronization and (ii) the implementation of a
parton-shower-like dynamics behind the cluster fission process.  
In particular, for (i) we supplement the splitting function $P_{g\to q\bar{q}}$ for the process $g\to q\bar{q}$ and 
base the cluster fission dynamics (ii) on the process $q\to qg \to qq'\bar{q}'$ governed by the splitting functions
$P_{q\to qg}$ and $P_{g\to q\bar{q}}$. As the clusters are (currently) 
based only on quark-antiquark constituents,  the model has no analogue to the gluon splitting $g\to gg$. 
These implementations
are explained in the following two subsections.

\subsection{Dynamic Gluon Mass Distribution}
\label{sec:dynamicgluonmass}

The idea of the dynamic gluon mass distribution is to adopt
essential features of the perturbative parton shower gluon splitting
also for the non-perturbative gluon splitting in the hadronization
model. In the parton shower, if a gluon branches into a quark-antiquark
pair that will then be part of the final state (i.e.\ these quarks do not
split any further and their momenta are set on-shell with current
quark masses in the kinematic reconstruction process), the gluon
adopts a finite virtuality associated to the quarks' 4-momenta. The
probability distribution of this gluon virtuality follows from the
form of the splitting function and the implementation of the splitting
algorithm.

For the construction a dynamical non-perturbative gluon splitting at
scales lower than those probed by the parton shower, we implement
important elements of this partonic branching by using the same
splitting function with some modifications.  The first obvious
modification is that the splitting function implementation 
does not have the infrared
$Q_0$-dependent cutoff of the parton shower and the second
modification is to adopt a constituent mass $m_q$ for the produced
quark pair (which automatically regulates the splitting function by
the kinematic constraint $p_g^2>4m_q^2$). Furthermore, since we cannot
use the perturbative QCD strong coupling $\alpha_s(\mu)$ for
renormalization scales $\mu$ well below $1$~GeV, we adopt the frozen
strong coupling value $\alpha_s(\mu_0)$ at the scale $\mu_0=1$~GeV for
scales $\mu<1$~GeV.  As an additional feature we also account for the
Sudakov form factor $\Delta(\tilde{Q}_g^2,\tilde{q}^2)$ to quantify a
non-splitting probability between some scale $\tilde{Q}_g$ from which
we start our ``non-pert. shower'' to the scale $\tilde{q}$ where the
gluon splitting takes place.  The scale $\tilde{Q}_g$ is a new tuning
parameter of this gluon mass model (replacing the fixed gluon
constituent mass $m_g$ appearing for the default version).

\begin{figure}
	\begin{center}
		\includegraphics[width=0.75\textwidth]{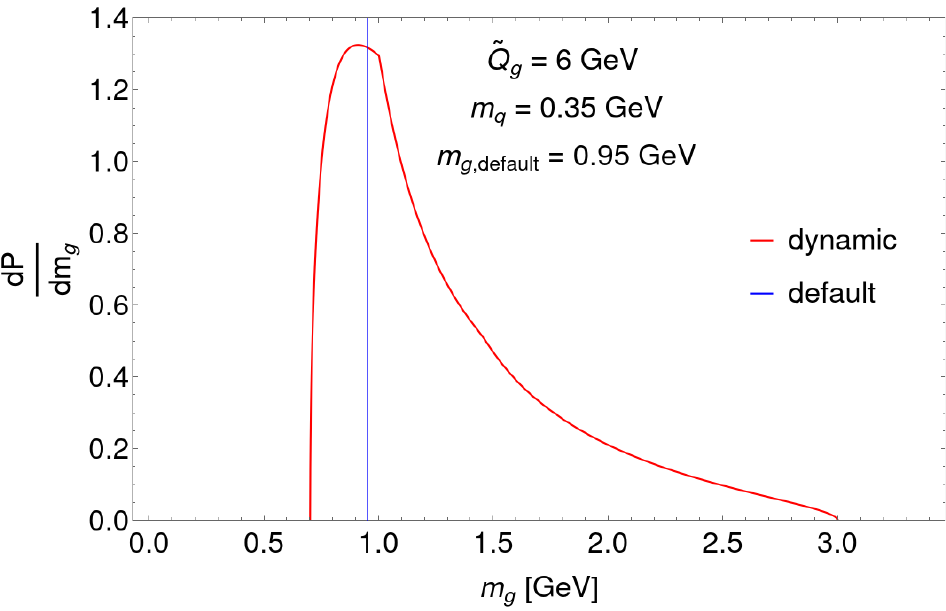}
		\caption{Dynamic gluon mass distribution for given
        $\tilde{Q}_g=6$~GeV and $m_q=350$~MeV (red curve) compared to 
        the fixed default model gluon mass $m_{g,\rm{default}}=950$~MeV 
        (blue vertical line).}\label{fig:gluonmass}
	\end{center}
\end{figure}

To keep things analytically trackable we approximate the 
Sudakov form factor with a Theta-function, 
because the scale hierarchies considered here are not large:
\begin{align}
\label{eq:sudakovapprox}
\Delta(\tilde{Q}_g^2,\tilde{q}^2)\approx \Theta(\tilde{Q}_g^2-\tilde{q}^2)\,.
\end{align}
A similar approximation can also been used for the analytic Laplace 
space solution of the coherent branching algorithm 
in Refs.~\cite{Catani:1992ua,Hoang:2018zrp}, 
see e.g.\ Eq.~(4.11) in Ref.~\cite{Hoang:2018zrp}.
Substituting $m_g^2=z(1-z)\tilde{q}^2$ for the gluon's virtuality
generated by the splitting, which we identify now with the gluon's
constituent mass $m_g$, the splitting probability for linear momentum
fraction $z$ and gluon constituent mass $m_g$ for a single quark species
is given by the expression 
\begin{align}
\label{eq:Probmg1}
\mathrm{d}P&\propto\Delta\Bigl(\tilde{Q}_g^2,\frac{m_g^2}{z(1-z)}\Bigr)\,
\mathrm{d}P_{g\to q\bar{q}}\Bigl(\frac{m_g^2}{z(1-z)},z,Q_0=0\Bigr) \nonumber \\
&=\frac{\mathrm{d}m_g^2}{m_g^2}\,\mathrm{d}z\,
\frac{\alpha_s(m_g^2)T_F}{2\pi}\Bigl(1-2z(1-z)+\frac{2m_q^2}{m_g^2}\Bigr) \nonumber \\
& \hspace{1cm} 
\times \Theta(4m_q^2<m_g^2<z(1-z)\tilde{Q}_g^2)\,
\Theta\Bigl(z_-\bigl(\frac{m_q}{m_g}\bigr)<z<z_+\bigl(\frac{m_q}{m_g}\bigr)\Bigr) 
\end{align}
with
\begin{equation}
z_{\pm}(x)\equiv\frac{1}{2}\Bigl(1\pm\sqrt{1-4x^2}\Bigr)\,.
\end{equation}
The restrictions on $z$ in the second line arise from the kinematics
of the splitting process:
\begin{equation}
0<p_{\perp}^2=z(1-z)m_g^2-m_q^2 \quad \Rightarrow\quad 2m_q<m_g
\,\,\mbox{and}
\,\,
z_-\bigl(\frac{m_g}{m_q}\bigr)<z<z_+\bigl(\frac{m_g}{m_q}\bigr)\,.
\end{equation}
Integrating over $z$ we can then obtain the probability distribution for the
dynamic gluon mass,
\begin{align}
\label{eq:Probmg}
\frac{\mathrm{d}P}{\mathrm{d}m_g}\propto \frac{\alpha_s(m_g^2)}{m_g}\,\biggl[&\Theta\Bigl(2m_q<m_g<\sqrt{m_q\tilde{Q}_g}\Bigr)
\sqrt{1-\frac{4m_q^2}{m_g^2}}\Bigl(1+\frac{2m_q^2}{m_g^2}\Bigr)\notag \\
&+\Theta\Bigl(\sqrt{m_q\tilde{Q}_g}<m_g<\frac{\tilde{Q}_g}{2}\Bigr)
\sqrt{1-\frac{4m_g^2}{\tilde{Q}_g^2}}\Bigl(1+\frac{3m_q^2}{m_g^2}-\frac{m_g^2}{\tilde{Q}_g^2}\Bigr)\biggr]\,.
\end{align}
The resulting dynamic gluon mass distribution is shown as the red curve
in Fig.~\ref{fig:gluonmass} for the values
$\tilde{Q}_g=6$~GeV (which is the typical value we obtain from the
tuning analyses discussed later in Sec.~\ref{sec:tuningphenosetup})
and the constituent quark mass $m_q=350$~MeV for the quark generated
from the splitting. The gluon mass value for the standard tune of
\herwig's default forced gluon splitting,
$m_{g,\rm{default}}=950$~MeV, is also indicated by the blue
vertical line.

For the implementation of the dynamic forced gluon splitting a value
$m_g$ is drawn from the probability distribution in
Eq.~\eqref{eq:Probmg}, that is then used for this gluon for the
(second) reshuffling to constituent masses.  After this reshuffling,
each gluon with its dynamic gluon mass is now split into a light
quark-antiquark pair in order to allow for clusters to be
formed. While for the default implementation an isotropic
quark-antiquark pair production process in the gluon rest frame has
been used, for the novel dynamical model a process more closely
resembling the parton shower dynamics is adopted. So the kinematics of
the dynamical forced gluon splitting depends on the longitudinal
momentum fraction $z$ and the azimuthal angle $\phi$ of the
emission. While the latter azimuthal angle is uniformly distributed,
the value $z$ is drawn from the distribution
\begin{align}
\label{eq:Probz}
\frac{\mathrm{d}P}{\mathrm{d}z}&\propto
\Bigl(1-2z(1-z)+\frac{2m_q^2}{m_g^2}\Bigr)\biggl[\Theta\bigl(2m_q<m_g<\sqrt{m_q\tilde{Q}_g}\bigr)
\Theta\Bigl(z_-\bigl(\frac{m_q}{m_g}\bigr)<z<z_+\bigl(\frac{m_q}{m_g}\bigr)\Bigr)\notag \\
&\hspace{3.0cm}+\Theta\Bigl(\sqrt{m_q\tilde{Q}_g}<m_g<\frac{\tilde{Q}_g}{2}\Bigr)
\Theta\Bigl(z_-\bigl(\frac{m_g}{\tilde{Q}_g}\bigr)<z<z_+\bigl(\frac{m_g}{\tilde{Q}_g}\bigr)\Bigr)\biggr]\,,
\end{align}
which is just the $z$-dependent part of Eq.~(\ref{eq:Probmg1}).

Using the standard decomposition into forward, backward and transverse
momenta, the concrete expression for the quark momentum reads
\begin{equation}
P_q^\mu=zP_g^{\mu}+\frac{m_q^2-(zP_g+q_\perp)^2}{2z\,P_g\cdot\bar{n}}\,\bar{n}^{\mu}+q_\perp^\mu\,.
\end{equation}
The dependence on the azimuthal angle $\phi$ is contained in the
transverse momentum component $q_\perp$ with respect to the axis
$\bar{n}$, which has the concrete form
\begin{equation}
q_\perp^\mu=\sqrt{-q_\perp^2}\Bigl(\cos(\phi)\,n_{\perp,1}^{\mu}+\sin(\phi)\,n_{\perp,2}^\mu\Bigr)\,,
\end{equation}
where
\begin{equation}
P_g \cdot   n_{\perp,i}=0\,, 
\quad\,
\bar{n}\cdot n_{\perp,i}=0\,,
\quad\,
n_{\perp,1}\cdot n_{\perp,2}=0\,,
\quad\,
n_{\perp,i}^2=-1\,.
\end{equation}
Using in addition the relations
\begin{align}
P_g^2=m_g^2\,,\qquad -q_{\perp}^2=z(1-z)m_g^2-m_q^2\,,
\end{align}
we finally arrive at
\begin{equation}\label{eq:QuarkMom}
P_q^\mu=zP_g^{\mu}+\frac{m_g^2(1-2z)-2P_g\cdot q_\perp}{2\,P_g\cdot\bar{n}}\,\bar{n}^{\mu}+q_\perp^\mu\,.
\end{equation}
For the antiquark momentum we have $P_{\bar{q}}^\mu=P_g^\mu-P_q^\mu$, which just
corresponds to the replacements $z\to(1-z)$ and
$q_\perp^\mu\to-q_\perp^\mu$. 

At this point we still need to provide a concrete expression for the
backwards light-like direction $\bar{n}$.  In the parton shower the
direction $\bar{n}$ is uniquely defined as the backwards direction to
the momentum of the progenitor of the branching tree. However, for the
forced gluon splitting this information is not available any more, and
in principle there is no unique ``correct'' way assigning the
direction of $\bar{n}$ so that we need to decide on a
prescription. However, all choices should be close (or collinear) with
the progenitor's backward direction.  To determine $\bar{n}$ we define
a ``new progenitor'' momentum $P$, from which we obtain the light-like
backwards direction as
\begin{equation}
\bar{n}^\mu=\begin{pmatrix}1\\ \frac{-\vec{P}}{|\vec{P}|}\end{pmatrix}\,.
\end{equation}
To do this we first identify the momenta $P_i$ of the two
(large-$N_c$) color connected {\it final state} partons of the splitted
gluon, which either are a quark and antiquark, a quark and a gluon or
an antiquark and a gluon. For our implementation we adopt one of these
partons as the ``new progenitor'', choosing the one whose direction of
motion leads to a smaller transverse momentum of the gluon. With this
choice for $\bar{n}$ together with the values for $m_g$, $z$ and
$\phi$ we have now fully determined the quark momentum in
Eq.~\eqref{eq:QuarkMom}.

Since the gluon splitting function is symmetric in $z$ around $z=1/2$,
the quarks and antiquarks generated in the splitting have the same
probability for both going in the direction of their respective color
partner (with that they will then form a cluster) as for both going in
the opposite direction of their color partner. There is in principle
nothing wrong with that since this also happens in the parton
shower. However, since one can argue that the non-perturbative
splitting considered here is already the first step of the cluster
formation, it is more ``natural'' that the quarks are predominantly
emitted in the direction of their color partner.\footnote{In
fact, colour reconnection models~\cite{Gieseke:2017clv,Gieseke:2018gff} 
would prefer to align the
colour connections in such a way as to minimize the cluster
masses.} We therefore restrict the range of $z$ for the emitted
quark to $z<1/2$ when the progenitor is the color connected quark and
to $z>1/2$ when the progenitor is the antiquark.

So far we have discussed only one quark flavor of mass $m_q$ being
produced in the splitting process. The gluon splitting, however,
generates all three light flavors. In the default model implementation
there have been fixed (tuned) probabilities $p_i$ for the three
possible flavors $i=u,d,s$ to be chosen when the gluon is split. In
the dynamic gluon splitting we go a different way. Writing the (not
normalized) gluon mass distribution for light flavor $q$ in
Eq.~\eqref{eq:Probmg} as $\mathrm{d}P(m_g,\tilde{Q}_g,m_q)$, the full
gluon mass distribution $\mathrm{d}P(m_g,\tilde{Q}_g)$ is obtained
from the sum over all light quark flavors
\begin{equation}
\frac{\mathrm{d}P(m_g,\tilde{Q}_g)}{\mathrm{d}m_g}
\propto\sum_{i=u,d,s}\frac{\mathrm{d}P(m_g,\tilde{Q}_g,m_i)}{\mathrm{d}m_g}\,.
\end{equation}
We draw the gluon masses from this distribution and then do the
reshuffling from the true to the constituent parton level. When coming
do the point where we have to split the gluon, we have to decide for
one light flavor. This is done randomly where the gluon mass dependent
probability for flavor $i$ reads
\begin{equation}
p_i=\frac{\mathrm{d}P(m_g,\tilde{Q}_g,m_i)}{\mathrm{d}m_g}
\times\biggl(\sum_{j=u,d,s}\frac{\mathrm{d}P(m_g,\tilde{Q}_g,m_j)}{\mathrm{d}m_g}\biggr)^{-1}\,.
\end{equation}
This ensures that $\sum_{i=u,d,s}p_i=1$ and that $p_i=0$ if
$2m_i>m_g$. We note that this implementation provides an
exact treatment of the flavor-dependent gluon mass distribution
$\mathrm{d}P(m_g,\tilde{Q}_g,m_q)$ adapted to the basic setup of the
cluster hadronization where the gluon splitting in the light
quark-antiquark pairs takes place after the reshuffling to the
constituent parton level.

\subsection{Embedding Gluon Branching into Cluster Fission}
\label{sec:dynamicalclusterfission}

\begin{figure}
	\begin{center}
		\includegraphics[width=1.0\textwidth]{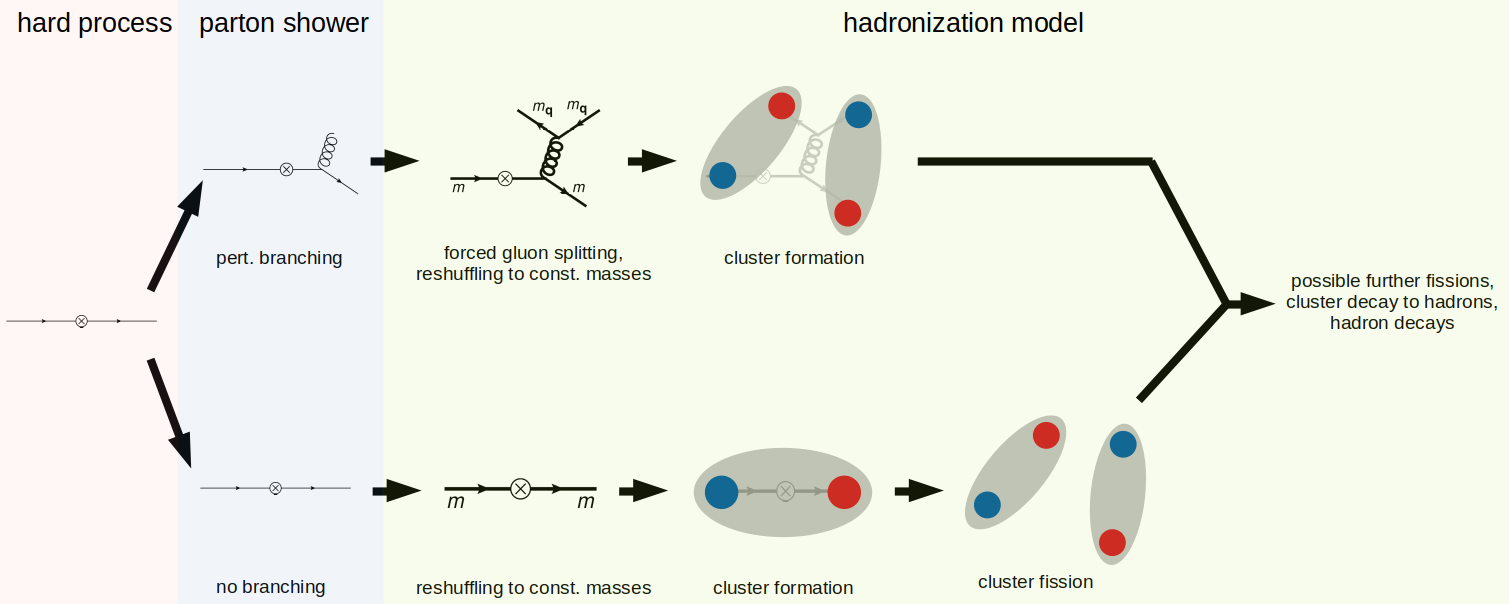}
		\caption{Cluster formation and fission for the simple case of a
               quark-antiquark final state system produced by the hard 
               scattering process. The upper path shows the
               process in the presence of a soft gluon that was just 
               barely radiated perturbatively in the parton shower, which
               then splits into a quark-antiquark pair. The lower path
               shows the same system without any perturbative
               branching. After the formation of the first two
               primary clusters in the upper path, and the fission
               of the primary cluster into two secondary clusters
               in the lower path, the final state consists in both
               cases of two clusters. The further hadronization
               steps (more fission processes, cluster decay, hadron
               decays) are identical for the two
               paths. }\label{fig:clusterfission}
	\end{center}
\end{figure}

The idea behind the novel dynamical cluster fission is, similar as for
the dynamic gluon mass distribution, to implement a fission process
that mimics important aspects of the parton shower dynamics.  To
illustrate our implementation let us have a look on the generic
aspects of the process of cluster formation and cluster fission for
the case of a simple quark-antiquark system, once in the presence of a
(soft) gluon that was radiated in the parton shower and once without
any perturbative radiation, see Fig.~\ref{fig:clusterfission}.  If a
gluon has been radiated by the parton shower (upper path in
Fig.~\ref{fig:clusterfission}), in the first step of the hadronization
model all particle momenta are reshuffled to their constituent masses
and the gluon is forced to split into a quark-antiquark pair as we
have discussed in Secs.~\ref{sec:ForcedGluonSplitting} and
\ref{sec:dynamicgluonmass}. In the next step the color-connected
quarks and antiquarks are combined into two clusters. At this point
the final state consists of two clusters.  If instead no gluon has
been radiated in the parton shower (lower path in
Fig.~\ref{fig:clusterfission}), we start the hadronization with only
the quark-antiquark pair. In the first step of the hadronization, the
two quarks' momenta are reshuffled to their constituent masses, and
subsequently combined into one single primary cluster. If this cluster
is heavy, it undergoes cluster fission and splits into two lighter
clusters. At this point the final state consists of two clusters as
well.

Let us now assume that the perturbative radiation of the gluon in the
shower in the upper path of Fig.~\ref{fig:clusterfission} happens at a
very low sale just slightly above the cutoff scale $Q_0$. So the upper
path in Fig.~\ref{fig:clusterfission} (perturbative gluon radiation,
formation of two primary clusters) and the lower path in
Fig.~\ref{fig:clusterfission} (no perturbative radiation, with one
primary cluster that undergoes cluster fission) are only separated by
an very small shift of the cutoff $Q_0$ that either allows or vetoes
the soft gluon radiation in the shower. In both cases one ends up with
two separate clusters in the final state (that can then either decay
directly to hadrons if they are light, or undergo further fission
processes if they are heavy, and eventually decay into hadrons). We
see that a smooth transition between the perturbative and the
non-perturbative dynamics at the cutoff scale requires that the
cluster fission in the lower path mimics the dynamics of the
parton shower's gluon radiation and gluon splitting.  The idea behind
the novel dynamical cluster fission model is therefore a generalization
of the parton-shower-like gluon splitting dynamics we have adopted for
the forced gluon splitting and dynamic gluon mass distribution described
in Sec.~\ref{sec:dynamicgluonmass} to a constituent quark $q\to qg$
(or antiquark $\bar q\to \bar qg$) branching as the basis of the
cluster fission.  In the following we will explain some technical
details of the implementation.

The dynamical cluster fission starts from a primary cluster made of a
color-connected quark-antiquark pair emerging from the forced gluon
splitting of Sec.~\ref{sec:dynamicgluonmass}. We consider the
cluster's rest frame, where the quark and antiquark, which we call
constituents, are back-to-back. It is now randomly chosen (with equal
probability) from which of the two constituents the intermediate gluon
is being radiated. Let us call the momentum of the constituent from
which the gluon is radiated $P_1^\mu$, and the momentum of the other
constituent $P_2^\mu$. Both momenta are on-shell with respect to their
constituents' masses, i.e. $P_i^2=m_i^2$.  We now define the
like-light backwards direction $\bar{n}^\mu$ to parametrize the
branching process from the backward direction of radiating
constituent, i.e.\
\begin{align}
\label{eq:nbardef}
\bar{n}^\mu\equiv (1,-\vec{P}_1/|\vec{P}_1|)\,.
\end{align}
The momenta $k^\mu$ of the (anti)quark and $g^\mu$ of the gluon that
emerge from the branching are
\begin{align}
k^\mu &= zP_1^\mu+\frac{m_1^2+p_{\perp}^2-z^2m_1^2}{z(\bar{n}\cdot P_1)}\times\frac{\bar{n}^\mu}{2}+q_{\perp}^\mu\,,\\
g^\mu &= (1-z)P_1^\mu+\frac{m_g^2+p_{\perp}^2-(1-z)^2m_1^2}{(1-z)(\bar{n}\cdot P_1)}\times\frac{\bar{n}^\mu}{2}-q_{\perp}^\mu\,,
\end{align}
where $p_\perp^2=-q_\perp^2$. The gluon mass $m_g$ is the virtuality
the gluon acquires from its splitting into a $q^\prime\bar{q}^\prime$
pair following the algorithm already described in
Sec.~\ref{sec:dynamicgluonmass}.  The invariant mass the radiating
constituent acquires due to this 'mini-shower', i.e.\ the $q\to qg$
(or $\bar q\to \bar qg$) and the $g\to q^\prime\bar{q}^\prime$
branching, reads
\begin{align}
M_1^2=(k^\mu+g^\mu)^2=\frac{m_1^2}{z}+\frac{m_g^2}{1-z}+\frac{p_\perp^2}{z(1-z)}\,.
\end{align}
Expressing the perp momentum in terms of the evolution variable for
the gluon emission branching
\begin{align}
p_\perp^2=(1-z)^2(z^2\tilde{q}^2-m_1^2)\,,
\end{align}
the invariant mass $M_1$ can also written as
\begin{align}
M_1^2=m_1^2+\frac{m_g^2}{1-z}+z(1-z)\tilde{q}^2\,.
\end{align}

At this point we need to restore energy-momentum conservation and
modify the 3-momentum of constituents $1$ and $2$. We thus solve the
equation
\begin{align}
M_{\rm{cl}} = \sqrt{|\tilde{p}|^2+M_1^2}+\sqrt{|\tilde{p}|^2+m_2^2}\,,
\end{align}
which yields
\begin{align}
	\tilde{P}_1^\mu &= (\sqrt{|\tilde{p}|^2+M_1^2},\,|\tilde{p}|\times\vec{P}_1/|\vec{P}_1|) \,,\\
	\tilde{P}_2^\mu &= (\sqrt{|\tilde{p}|^2+m_2^2},\,-|\tilde{p}|\times\vec{P}_1/|\vec{P}_1|)\,.
\end{align}
with
$|\tilde{p}|=\frac{1}{2 M_{\rm{cl}}} \lambda^{1/2}(
M_{\rm{cl}}^2,M_1^2,m_2^2)$ for the new 3-momentum of the two
constituents. The concrete expressions for the momenta of the quark
(or antiquark) and the gluon emerging from the branching of constituent~1
read
\begin{align}
\tilde{k}^\mu &= z\tilde{P}_1^\mu+\frac{m_1^2+p_{\perp}^2-z^2M_1^2}{z(\bar{n}\cdot \tilde{P}_1)}\times\frac{\bar{n}^\mu}{2}+q_{\perp}^\mu \notag \\
&= z\tilde{P}_1^\mu+\biggl(2(1-z)m_1^2-\frac{z}{1-z}m_g^2+(1-z)z(1-2z)\tilde{q}^2\biggr)\times\frac{\bar{n}^\mu}{2(\bar{n}\cdot \tilde{P}_1)}+q_{\perp}^\mu\,,\\
\tilde{g}^\mu &= (1-z)\tilde{P}_1^\mu+\frac{m_g^2+p_{\perp}^2-(1-z)^2M_1^2}{(1-z)(\bar{n}\cdot \tilde{P}_1)}\times\frac{\bar{n}^\mu}{2}-q_{\perp}^\mu \notag \\
&= (1-z)\tilde{P}_1^\mu-\biggl(2(1-z)m_1^2-\frac{z}{1-z}m_g^2+(1-z)z(1-2z)\tilde{q}^2\biggr)\times\frac{\bar{n}^\mu}{2(\bar{n}\cdot \tilde{P}_1)}-q_{\perp}^\mu\,.
\end{align}

The momenta $\tilde{k}^\mu$, $\tilde{g}^\mu$ and the momentum
$\tilde{P}_2^\mu$ of constituent~2 are now parametrized in terms of
the splitting variables $z$ and $\tilde{q}$ that are determined from
the $P_{q\to qg}$ splitting function, and the gluon mass $m_g$ that we
have to draw from the dynamic gluon mass distribution as described in
Sec.~\ref{sec:dynamicgluonmass}.  The splitting variables $\tilde{q}$
and $z$ are drawn from the probability distribution given by the
$q\to qg$ splitting function (including a flat distribution of
azimuthal angles)
\begin{align}
\mathrm{d}P_{q\to qg}\propto \frac{\mathrm{d}\tilde{q}^2}{\tilde{q}^2}\frac{\mathrm{d}z}{1-z}\alpha_s\Bigl(z^2(1-z)^2\tilde{q}^2\Bigr)\biggl[1+z^2-\frac{m_1^2}{z\tilde{q}^2}\biggr]\Theta(z^2\tilde{q}^2-m_1^2)\,,
\end{align}
where the quark mass $m_1$ should always be understood to be the
constituent mass of the branching quark, times the Sudakov form
factor, that we approximate again via a Theta-function,
\begin{align}
\Delta(\tilde{Q}_q^2,\tilde{q}^2)\approx \Theta(\tilde{Q}_q^2-\tilde{q}^2) \,.
\end{align}
The scale $\tilde{Q}_q$, which is the quark analogue of the scale
$\tilde{Q}_g$ for the gluon splitting in Eq.~(\ref{eq:sudakovapprox}),
is one of the new tuning parameters of the dynamic model, and sets the
starting scale of the shower evolution for the non-perturbative
(anti)quark branching in the cluster fission process.

The gluon with momentum $\tilde{g}^\mu$ now splits into a
$q^\prime\bar{q}^\prime$ pair following the description of the forced
gluon splitting in Sec.~\ref{sec:dynamicgluonmass}. However, the
backwards direction $\bar n^\mu$ is given in Eq.~(\ref{eq:nbardef}).
Furthermore, the analogue of the scale $\tilde{Q}_g$ (shown in
Eq.~(\ref{eq:sudakovapprox})) which here we denote by
$\tilde{Q}_g^{(f)}$ is related to the splitting variables $z$ and
$\tilde{q}$ by the angular ordering relation\footnote{ We note that
  there is also the option to use $\tilde{Q}_{g}^{(f)}$ as an
  additional tuning parameter that sets the scale of the gluon
  splitting in the fission independent of the (anti)quark
  branching. We have, however, not used this option in our analyses.}
\begin{align}
\tilde{Q}_g^{(f)} = (1-z)\tilde{q}\,.
\end{align}
We emphasize that this means that the gluon splitting scale
$\tilde{Q}_g$ in the forced gluon splitting and $\tilde{Q}_g^{(f)}$ in
the cluster fission are unrelated, hence the
superscript $(f)$ for 'fission'.
Finally, the color-connected quarks and antiquarks (one being a
constituent and the other being either $q^\prime$ or $\bar{q}^\prime$)
are paired into the two new clusters.

\begin{figure}
	\begin{center}
		\includegraphics[width=0.49\textwidth]{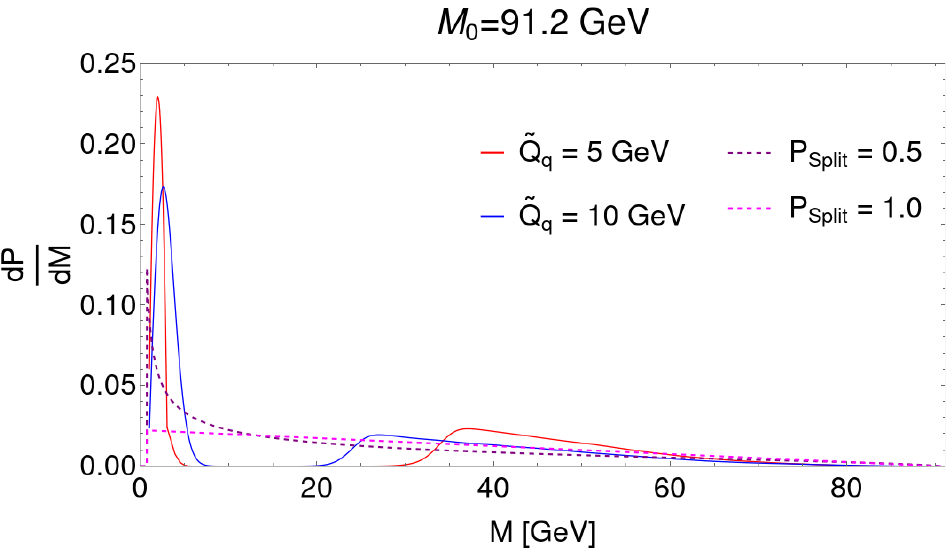}
		\includegraphics[width=0.49\textwidth]{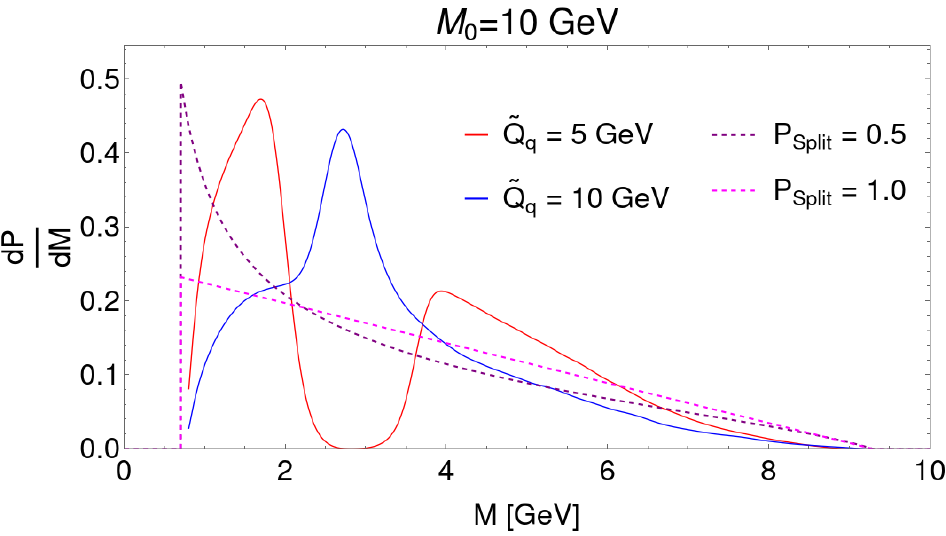}
		\caption{Distribution of the daughter cluster masses
			$M$ in the fission for a parent cluster with mass
			$M_0=91.2$~GeV (left panel) and $M_0=10$~GeV
			(right panel). The solid lines show the distribution for 
			the dynamical cluster fission for 
			$\tilde{Q}_q=5~\mbox{GeV}$ (red)
			and $10$~GeV (blue). The dashed lines 
			show the default mass distribution for 
			$\mathtt{PSplit}=0.5$ (red) and $1$ (blue)..}\label{fig:dynamic_cluster_masses}
	\end{center}
\end{figure}

The resulting distribution for the daughter cluster masses $M$ are
shown in Fig.~\ref{fig:dynamic_cluster_masses} for the examples of a
parent cluster mass $M_0=91.2$~GeV (left panel) and $M_0=10$~GeV
(right panel).  The solid lines show the novel dynamical mass
distribution for $\tilde{Q}_q=5~\mbox{GeV}$ (red) and $10$~GeV
(blue). The dashed colored lines show the mass distribution for the
default cluster fission for $\mathtt{PSplit}=0.5$ (red) and $1$
(blue). The values for $\tilde{Q}$ and $\mathtt{PSplit}$ are within
the typical ranges obtained in our tuning analyses.  All constituent
quark masses are set to $m_q=0.35\,\rm{GeV}$. We see that the novel
dynamical cluster fission yields a substantially richer structure of
the daughter cluster mass distribution than the default model, which
is merely assuming a power law behavior supplemented by constraints
from the phase space limits.  In contrast, the dynamical model
naturally implements phase space constraints through the kinematics of
the fission process and therefore incorporates physical thresholds
where both light, but also very asymmetric heavy/light cluster
configurations are allowed.  This can be seen from the peak structures
exhibited by the solid lines.  It is typical that two peaks
appear. The lighter peak corresponds to the cluster that is formed
from the constituent quark of the parent cluster from which the gluon
was radiated and the color connected quark generated from the gluon
splitting. Since the gluon and the quark pair emerging from its
splitting are radiated predominantly into the direction of that
constituent quark, this cluster is typically light. The heavier peak
then corresponds to the cluster that contains the other constituent
quark. Furthermore, in the dynamical model there is always a smooth
and fast suppression in the limit of vanishing cluster masses which is
more physical than the sharp cutoff resulting from the simply power
law mass distribution of the default model.

\section{Description of the Shower Cutoff Dependent Tuning Analyses}
\label{sec:tuningphenosetup}

\subsection{Tuning Procedure and Reference Tune}
\label{sec:referencetune}

The goal of our numerical analysis is to investigate the $Q_0$-dependence 
of hadronization tuning parameters and observables
generated by \herwig. The hadronization parameters are determined by
tuning \herwig{} to reference data. To obtain this {\it reference
  data} we employ observables generated by \herwig{} itself for the
shower cut $Q_{0,\mathrm{ref}}$. The corresponding tune, which we call
the \emph{reference tune}, is obtained from a regular tune to
experimental $e^+e^-$ data obtained at the $Z$-pole for
$Q=\SI{91.2}{\GeV}$ from the different LEP collaborations, which
include event-shapes, particle multiplicities and jet rates at
$Q=\SI{91.2}{\GeV}$. This amounts to 3180 observable bins for which
official \rivet{}~\cite{Bierlich:2019rhm} analysis code
implementations are available.

The tuning procedure is performed with the software library
\apprentice~\cite{Krishnamoorthy:2021nwv}. For the determination of
the reference tune we convert the experimental data bins, which are
provided in form of \yoda{} files, to a reference data file by using
the \apprentice{} Python script ``app-datadirtojson''.  The second
input required by \apprentice{} are \herwig{} samples for the
corresponding observable bins generated from the \rivet{} analysis
code related to the experimental data evaluated at different values of
the tuning parameters within the parameter ranges. The \apprentice{}
script ``app-build'' determines a polynomial interpolation from these
samples for both height $f_\mathrm{MC,k}(\{p_i\})$ and statistical
error\footnote{We have fixed a bug in \apprentice{} which led to wrong
  values in the error interpolation.}
$\Delta f_\mathrm{MC,k}(\{p_i\})$ of each bin $k$ separately. The tunes
are then produced by using the \apprentice{} script ``app-tune2''
which performs a weighted $\chi^2$ minimization by evaluating these
interpolations.  The goodness of fit function that \apprentice{}
minimizes is
\begin{equation}
\mathrm{GOF}(\{p_i\})=\sum_{k}w_k^2\frac{(f_\mathrm{MC,k}(\{p_i\}) - f_\mathrm{ref,k})^2}{(\Delta f_\mathrm{MC,k}(\{p_i\}))^2 + (\Delta f_\mathrm{ref,k})^2},
\end{equation}
where $f_\mathrm{ref,k}$ and $\Delta f_\mathrm{ref,k}$ are the height
and the corresponding statistical error of bin $k$ of the reference
histogram, respectively. In our analysis the weights $w_k$ are chosen
to be common for bins belonging to the same observable.

For the {\it reference data} of the $Q_0$-dependent tuning analysis we
include the 3180 observable bins related to LEP measurements at
$Q=\SI{91.2}{\GeV}$ which already enter the reference tune using the
official \rivet{} analysis code for the data generation at the
reference cutoff $Q_{0,\mathrm{ref}}$.  This means that for these
observables all bins from the entire spectra are
included. Additionally, we add 14 equidistant bins of the 2-jettiness
observable with $\SI{1.2}{\GeV}\le Q\tau < \SI{6.8}{\GeV}$. This range
covers the peak region and some part of the tail.  For the
implementation of 2-jettiness we employ a custom, in-house \rivet{}
analysis code.  This additional partial 2-jettiness distribution does
not contribute to the determination of the reference tune, but it is
included in the reference data of the $Q_0$-dependent tune analyses,
since it is the 2-jettiness distribution for which we analyze the
properties of the transfer function.  Note that the 2-jettiness and
the classic thrust observables are very similar at the c.m.\ energies
$Q$ we consider, and we have checked that the outcome for using either
2-jettiness or classic thrust for the additional partial distribution
are fully equivalent. For the GOF function the relative weights
$w_k^2$ from the 3180 observable bins related to the LEP measurements
at $Q=\SI{91.2}{\GeV}$ that are used for the determination of the
reference data remain unchanged and the overall contribution of the
additional 2-jettiness distribution in terms of total bin weights is
set to 8\%.  In all other aspects the $Q_0$-dependent tunes in our
analysis are produced in a way very similar to the out-of-the-box
\herwig{} tune, so that they also provide compatible realistic
simulations.

For our $Q_0$-dependent tuning analysis we create tunes at fixed
values of $Q_0$. To track the details of
$Q_0$-dependence we adopt $Q_0$ values between \SI{0.75}{\GeV} and
\SI{2.00}{\GeV} in steps of \SI{0.05}{\GeV}. This constitutes $26$ 
different $Q_0$ values. The lower and upper
bounds are motivated by $Q_0$ being at low-energy scale close to the
hadronization scale, but still within the realm of perturbation
theory. At $Q_0=0.75$~GeV we do not expect perturbation theory (and
the parton-shower description) to work very well, but we include
scales below $1$~GeV as a monitoring tool to visualize the expected
breakdown of the perturbative description. Scales above $2$~GeV are
not considered since we cannot expect that a hadronization model could
(and actually should) provide a description of parton branching at
high scales. So the level of agreement (or disagreement) of the
properties of the hadronization corrections with QCD factorization in
the range $1\,\mbox{GeV}< Q_0< 2$~GeV will be our analysis instrument
to quantify the quality and consistency of the hadronization model.

In the Herwig input files, which describe the settings for each sample
run, we choose the built-in leading order
$e^+e^- \rightarrow q\bar{q}$ matrix element (5 flavours
$d, u, s, c, b$) and we turn off QED radiation. We specify the
hadronization model (default or dynamical) and set the corresponding
tuning parameters. These are sampled uniformly within a
hyperrectangle. We include the shower cutoff parameter $Q_0$ as one of
the interpolation variables so that we can use the same interpolation
for the determination of the reference tune and the corresponding {\it
  reference data} at $Q_{0,\mathrm{ref}}$ as well as for the
subsequent $Q_0$-dependent tuning analyses.  We made sure that
the region around the minimal GOF value parametrized 
by $Q_0$ is fully
contained within the sampled hyperrectangle, while at the same time
keeping the hypervolume small enough to ensure a good \apprentice{}
interpolation quality. Note that we produce two reference tunes, one
for the default hadronization model and one for the 
novel dynamical hadronization model, so that
our tuning analyses are self-consistent. We pick
$Q_{0,\mathrm{ref}}=\SI{1.25}{\GeV}$ as the reference shower cut value
since it is close to the value obtained from the global minimum tune,
see also our discussion in Sec.~\ref{sec:tuningquality}. We emphasize 
that we have checked that the simulations based on the reference tunes 
for both the default and novel dynamical hadronization models 
are very close to the
\herwig{}'s out-of-the-box default $e^+e^-$ tune and thus provide the
same realistic overall data description. We refer the interested reader 
to the webpages
\href{https://herwig.hepforge.org/}{https://herwig.hepforge.org/} or
\href{http://mcplots.cern.ch/}{http://mcplots.cern.ch/}, where the 
data description of standard \herwig{} tunes (and for other MCs) are
collected.

In order to have a means to cross check that stability of
interpolation procedure we carry out two independent analyses for the
``app-build'' interpolations, one using cubic and one using quartic
polynomial orders (which yields 120 and 330 polynomial coefficients,
respectively, for seven parameters). We use the more precise quartic 
interpolation as
our default interpolation and the less precise cubic one as a 
reference for the stability checks. 
For the cubic and quartic interpolations we
generate $10^5$ and $10^6$ events per parameter space point,
respectively, to obtain the binned distributions. These leads to
statistical errors that are, for both interpolations, approximately of
the same size as our estimate for the interpolation uncertainties
which are described in more detail below. The latter can be kept small
by providing a sufficiently large oversampling factor compared to the
number of coefficients of the polynomial interpolation. The number of
sampled parameter space points are (hadronization model: interpolation
order, number): (dynamic: cubic, 493), (dynamic: quartic, 2257),
(default: cubic, 484), (default: quartic, 5380). This corresponds to
at least a four-fold oversampling.

The ``app-tune2'' minimization script is run with the following options:
The minimization algorithm is set to the \emph{truncated Newton} (TNC)
algorithm, where the starting point for the minimization is obtained
by taking the point with the minimum $\mathrm{GOF}$-function value out
of 100 randomly sampled points. The full minimization is then repeated
10 times to reliably find the global minimum.  After determining the
reference tune, the ``app-tune2'' script also automatically saves
prediction histograms for all observables in an output file called
``predictions\_tnc\_100\_10.yoda''\footnote{We have fixed a bug in
  \apprentice{} which led to wrong values in this prediction
  file.}. It obtains these predictions by evaluating the \apprentice{}
interpolation at the parameter values given by the reference tune.  We
convert this \yoda{} file to a \emph{reference data file} by using the
``app-datadirtojson'' script again. This standard reference file
provides the reference data for the $Q_0$-dependent analysis.

\subsection{Treatment of the Hadronization Model Parameters}
\label{sec:descriptiontuning}

\subsubsection*{Tuning Parameters}

The default and the novel dynamical cluster hadronization models 
both feature a number of tuning
parameters. In our tuning studies we do not consider all of them as
floating parameters to be determined by the tuning fits. Rather we
consider those as floating which are associated with the 
'hard scale' for the hadronization dynamics (i.e.\ they should 
show some correlation to the parton shower's IR cutoff $Q_0$) and 
on which the event-shapes, jet rates and
charged particle multiplicities that we account for in the reference
data depend in a significant way. Furthermore, we also include
parameter which are related to low-scale hadron-specific 
processes unrelated to
that 'hard scale' which, from the physical perspective, should be
rather $Q_0$-independent.
Overall, for the default and for
the novel dynamical hadronization models each, 6 parameters are
treated as floating parameters in the fits to the reference data. 
We also refer to them as $p_i$ ($i=1,\ldots 6$) below. In
this section we discuss the concrete features of these parameters 
in more detail. All
other parameters, which are common to both hadronization models, are
set to their default values.
 
The only hadron-specific hadronization model parameters we consider as
floating tuning parameters are $\mathtt{PwtSquark}$ and
$\mathtt{PwtDIquark}$.  They control for example the
  strangeness and Baryon production rates and thus directly impact
the charged particle multiplicities. Since hadron production takes
place after the cluster fission process, $\mathtt{PwtSquark}$ and
$\mathtt{PwtDIquark}$ appear for the default as well as for the novel
dynamical hadronization model. Among all the
hadron-specific parameters of the hadronization model it is important
to include at least these two as floating parameters in the tuning
analyses to maintain a realistic correlation between various
parameters that affect the charged particle multiplicities, but in the
end also have an impact on other observables. We also note that
several of the cluster fission parameters are flavour dependent, in
order to allow them to account for effects of the heavy quark
masses. Sensitivity to those can only be gained by explicitly
considering flavour dependent observables, but otherwise their role is
no different from their light-quark counterparts. Since we are not
specifically addressing heavy quark fragmentation in our analysis, we
choose to identify those heavy flavor specific parameters with their
light quark counterparts for the studies in this article.

Apart from the two hadron-specific parameters  $\mathtt{PwtSquark}$ 
and  $\mathtt{PwtDIquark}$
just explained, there are two additional hadronization model
parameters which we treat as floating tuning parameters and which are
common to both hadronization models.  These are the parameters
$\mathtt{Cl}_\mathtt{max}$ and ${\mathtt{Cl}_{\mathtt{pow}}}$. They 
appear in the
``heavy'' cluster condition of Eq.~(\ref{eq:IsHeavy}) which determines
whether the fission of a (heavy) cluster
into two lighter clusters takes place or whether that cluster is
already considered light and decays into hadrons, see
Sec.~\ref{sec:ClusterFission}. While $\mathtt{Cl}_\mathtt{max}$ (which has
dimension of mass) sets the overall scale of the heavy cluster fission
threshold, the dimension-less parameter ${\mathtt{Cl}_{\mathtt{pow}}}$ 
essentially quantifies a smearing of the threshold that also depends 
on the masses of the cluster's constituents.

The two remaining hadronization parameters we consider in our
$Q_0$-dependent tuning analyses differ for the two hadronization
models.  For the default hadronization model these are the gluon
constituent mass $m_g$ and the dimension-less parameter
$\mathtt{PSplit}$. The gluon constituent mass $m_g$ is important for the
kinematics of the forced gluon splitting taking place at the initial
states of the hadronization process, see
Sec.~\ref{sec:ForcedGluonSplitting}.  For the novel dynamical
hadronization model the fixed gluon mass $m_g$ is replaced by a gluon
mass distribution. The parameter $\mathtt{PSplit}$ governs the shape and
steepness of the daughter cluster mass distribution according to
Eq.~(\ref{eq:Gendefault1}) in the cluster fission algorithm described
in Sec.~\ref{sec:ClusterFission}.  It has no counterpart in the novel
dynamical model, where the daughter cluster mass distribution is
generated from a splitting process. We refer to our discussion 
of Fig.~\ref{fig:dynamic_cluster_masses}.
For the novel dynamical
hadronization model the two remaining hadronization parameters are
$\tilde Q_g$ and $\tilde Q_q$ both which have dimension of energy.
The scale $\tilde Q_g$ is the `hard energy scale' of the
$g\to q\bar q$ branching process that is the basis of the forced gluon
splitting in the dynamical model, see
Sec.~\ref{sec:dynamicgluonmass}. The scale $\tilde Q_q$ is the
`hard energy scale' of the $q\to qg$ (or $\bar q\to \bar qg$)
branching process that governs the cluster fission process in the
dynamical model, see Sec.~\ref{sec:dynamicalclusterfission}.

Since $\tilde Q_g$ and $\tilde Q_q$ represent the scales where the 
non-perturbative splitting and fission processes start we can 
expect some linear
dependence of their fitted values on the value of the parton shower
cutoff $Q_0$, if a proper matching of the dynamical hadronization to
the parton shower is realized through the tuning to the reference
data. Furthermore, in this case, the other four parameters of the
dynamical hadronization model should be rather insensitive to the 
shower cutoff value since they govern dynamical aspects that are only of
non-perturbative nature taking place at scales below the hard scales
$\tilde Q_g$ and $\tilde Q_q$. On the other hand, for the default
hadronization model, which is not designed to provide any systematic
matching and where $Q_0$ essentially plays the role of just another
hadronization parameter, such behavior cannot be naturally expected.  
To which extent these expectations are actually met by the outcome
of our tuning analyses is subject to our phenomenological
discussion in Sec.~\ref{sec:tuningquality}. 

\subsubsection*{Error Estimate}
\label{sec:tuninguncertainties}

The description up to this point is complete with regards to the
determination of the central values $p_\mathrm{i,cent}$ ($i=1,\ldots 6$) 
of the hadronization tuning parameters. Interestingly, at this time 
there is no general canonical approach to estimate the uncertainties 
of MC hadronization model tuning parameters. 
However, in the context of having hadronization effects being defined in
a particular scheme, it is also relevant to quantify an uncertainty on
its parameters. In the following we explain the prescription we adopt
for an uncertainty estimate of the tuning parameter we treat as floating
in our tuning fits. As there is no canonical approach,  
our prescription is to some extent ad-hoc and only provides a first step
towards are systematic treatment of the tuning parameter uncertainties. 
However, we believe that viewed together with the differences obtained 
from the cubic and quartic interpolations it provides 
a sufficiently fair treatment at this point.

There are two sources of uncertainties we consider. 
The first is the statistical uncertainty related to the number of
events in our MC simulations.  The second is an estimate for the
``app-build'' interpolation error.
The statistical error is determined from the inverse of the Hessian
matrix
$H_{ij}\equiv\partial^2 (\mathrm{GOF})/(\partial p_i \partial p_j) $
of the fit at the best fit point multiplied with a heuristic rescaling
factor associated to the effective degrees of freedom induced by the
weights $w_k$:
\begin{equation}
\Delta p_{i,\mathrm{stat}}(Q_0) = 
\sqrt{2*(H^{-1})_{i,i}(Q_0) * \left(\sum_k{w_k^2}\right)^2\left/\middle(\sum_k{w_k}\right)^2}\,.
\end{equation}
To estimate the interpolation uncertainty we analyze the difference
for the tuning parameter obtained from the reference data (that is
based on the \apprentice{} interpolation) versus data generated from a
full \herwig{} simulation run, which we call {\it exact \herwig{}
data}.  Given the central values for the hadronization parameters
${p_{i,\mathrm{cent}}(Q_0)}$ obtained at cutoff $Q_0$, we can carry
out a second tuning fit for the cutoff  
$Q_0$ with the reference data being replaced by exact data
obtained from a \herwig{} run using the tuning parameters
$p_{i,\mathrm{cent}}(Q_0)$. This yields a new set of best tuning
parameter values which we call $p_{i,\mathrm{refH}[Q_0]}(Q_0)$, 
where the subscript $\mathrm{refH}[Q_0]$ stands for the shower cutoff
where the exact data is generated and the argument $(Q_0)$ for the cutoff
where the tuning fit is carried out. 
The difference between these two sets of tuning parameters
$\Delta
p_i(Q_0)=p_{i,\mathrm{refH}[Q_0]}(Q_0)-p_{i,\mathrm{cent}}(Q_0)$ is an
estimate for the \apprentice{} interpolation errors at the cutoff
$Q_0$ since, for a perfect interpolation and in the absence of
statistical errors, we would have $\Delta p_i(Q_0)=0$. This approach
also allows to quantify the interpolation uncertainty at the reference
cutoff $Q_{0,\mathrm{ref}}$ itself.

To reduce computational cost of this uncertainty estimation method, we
carry out this procedure only for the six equidistant cutoff values
$Q_{0,m}^\textrm{H}\in \{0.75,1.00,...,2.00\}\,\si{\GeV}$. To obtain
an uncertainty estimate for all $Q_0$ values we consider, we apply the
following averaging procedure.  Since the minimization procedure
itself is actually not computationally expensive we can evaluate
$p_{i,\mathrm{refH}[Q_{0,m}^\textrm{H}]}(Q_0)$ for the $m=1,\ldots, 6$
exact Herwig data sets for all $Q_0$ values.  We then estimate the
final interpolation uncertainty of the hadronization parameters by
adopting a distance based average,
\begin{equation}
\Delta p_{i,\mathrm{inter}}(Q_0)=\sqrt{\sum_{m=1}^6  \sum_{n=1}^{26} w(Q_0, Q^\prime_{0,n}, Q^\textrm{H}_{0,m}) \Big[p_{i,\mathrm{refH}[Q_{0,m}^\textrm{H}]}(Q_{0,n}^\prime)-p_i(Q_0)\Big]^2}\,,
\end{equation}
where the weights are given by
\begin{equation}
w(Q_0, Q^\prime_{0,n}, Q^\textrm{H}_{0,m})\propto e^{-[\sigma_c^{-2}(Q^\textrm{H}_{0,m}-Q^\prime_{0,n})^2 +\sigma_s^{-2}(Q^\prime_{0,n}-Q_0)^2]/2}, 
\end{equation}
with the sum normalized to one
$ \sum_{m,n}w(Q_0, Q^\prime_{0,n}, Q^\textrm{H}_{0,m}) = 1$. The
correlation width $\sigma_c=1/32$~GeV in the first exponential ensures
that the exact Herwig data tunes at $Q_{0,m}^\textrm{H}$ closest to
$Q_{0}$ contribute most. The smoothing width $\sigma_s=1/16$~GeV in
the second exponential ensures that for $Q_0$ values in the middle
between two $Q^\textrm{H}_{0}$ values we obtain about the average of
the interpolation uncertainties at these $Q^\textrm{H}_{0}$ values.

\subsection{Extraction of Migration Matrix Functions}
\label{sec:extractmigration}

The technical implementation of the migration matrix extraction relies
on the true parton level state, in the form of the complete
description of all particles in that state, in the HepMC event record
for each event that \herwig{} produces. 
Before we continue the description we remind
the reader about the standard framework that is used to generate
observable histograms from MC generator runs: A MC generates each
event sequentially. For each event it produces an event record in the
HepMC format that also includes all particles in the final state,
which are marked by the ``final-state'' status code (which is just the
number 1). Each particle entry $[X, (p_E, p_x, p_y, p_z)]$ specifies
the particle species $X$ and its four-vector $p^\mu$. This HepMC event
record is read by the analysis framework \rivet. Analyses written for
\rivet{} first carry out a final-state \textit{projection} which gives
the set of all final-state particle four-momenta of the event,
$\{p^{\mu}_i\}$. This set is then converted to an observable value by
an \textit{observable projection} $\tau=\mathcal{O}(\{p^\mu_i\})$. For
each event this observable value is filled into a histogram to produce
the final binned distribution $\mathrm{d}{\sigma}/\mathrm{d}{\tau}$.

In the \herwig{} code that we use in our analyses we include in
addition the {\it true parton level state} as defined in
Sec.~\ref{sec:KinematicReconstruction} in the HepMC event record for
each event, with parton level particles marked by a fixed status code,
which lies in the value range that a MC can freely use for internal
purposes. This allows us to not only do a final-state, i.e.\ hadron
level, projection but also a true parton level one to obtain the
parton level four-vectors $\{\hat{p}^{\mu}_i\}$, denoted with a
hat. This functionality does not only allow for the evaluation of
parton level observables, such as
$\hat{\tau}=\mathcal{O}(\{\hat{p}^\mu_i\})$ and
$\mathrm{d}{\sigma}/\mathrm{d}\hat{\tau}$, analogous to the hadron
level ones, but also for the extraction of the combined set
$(\{\hat{p}^{\mu}_i\},\{{p}^{\nu}_j\})$ obtained for each event. This
provides access to a fully differential probability distribution
$P((\{\hat{p}^{\mu}_i\},\{ {p}^{\nu}_j\})$ in the combined
parton-hadron level space. We can therefore also extract any
parton-hadron level correlation function
$\mathrm{corr}[\mathcal{O}_1(\{\hat{p}^{\mu}_i\}),
\mathcal{O}_2(\{p^{\nu}_j\})]$ or any migration function (or matrix)
given by the conditional probability distribution
$P(\mathcal{O}(\{p^{\nu}_j\})\,|\,\mathcal{O}(\{\hat{p}^{\mu}_i\}))$.
This allows us to extract the probability of having a hadron level
observable value $\tau$ for a given parton level value $\hat{\tau}$.
In our phenomenological studies we analyze the behaviour of the
migration matrix function of the 2-jettiness observable
$(\hat{\tau},\tau)$ and the first moment of the probability
distribution $P(\tau\,|\,\hat\tau)$ in $\tau-\hat{\tau}$. In practice
these probability distributions are of course lists or matrices of
probabilities as we consider observable bins. There are two approaches
to saving the necessary data using \rivet. The first is to fill a very
finely binned 2-D histogram in the $(\hat{\tau},\tau)$ variables. The
second option is to save the tuple $(w,\hat{\tau},\tau)$ for each
event, i.e. unbinned data. The \emph{event weight} $w$ can in
  general differ from $1$, in particular for NLO-matched MC
simulations. We adopted the second method since it allows us to
generate any histograms with arbitrary binning specifications and to
calculate any average exactly without the potential need to rerun the
MC simulation.

Accounting for the additional dependence on the shower cutoff scale
$Q_0$ and the hard scattering scale $Q$ in our subsequent analyses, we
refer to the probability distribution $P(\tau\,|\,\hat\tau)$ as
$T(\tau,\hat\tau,\{Q,Q_0\}))$, which we already introduced in
Sec.~\ref{sec:numericalQ0}.  To be more specific, we concretely study
the quantitative behavior of the rescaled transfer functions
$\tilde S^{\rm MC}(k,\hat k,\{Q,Q_0\})$ and
$S^{\rm MC}(\ell,\{\hat k,Q,Q_0\})$ which are derived 
from $T(\tau,\hat\tau,\{Q,Q_0\}))$ in
Eqs.~(\ref{eq:tildeSMCTrelation}) and (\ref{eq:SMCdef}). We remind
the reader that $S^{\rm MC}(\ell,\{\hat k,Q,Q_0\})$ 
is the exact MC analogue of the shape function
$S_{\rm had}(\ell,Q_0)$ appearing in the analytic QCD factorization.

\section{Phenomenology of the New Model}
\label{sec:NewFissionpheno}

Finally, in this section we analyze quantitatively the properties 
of \herwig{}`s default and new dynamic hadronization models from the 
perspective of (i) the shower cut $Q_0$ taking the role of an IR 
factorization scale and (ii) the $Q_0$ scheme dependence  of the 
hadronization migration matrix function 
$S^{\rm MC}(\ell,\hat k,\{Q,Q_0\})$ in the 
2-jettiness dijet region demanded from the QCD factorization, where 
the hadronization effects are expressed in terms of a shape function. 
We remind the reader that unless stated otherwise we always use
the more precise quartic interpolations for the results that are 
discussed.

\subsection{Shower Cutoff Independence of Hadron Level Observables}

\begin{figure}
	\begin{center}
		\includegraphics[width=1.0\textwidth]{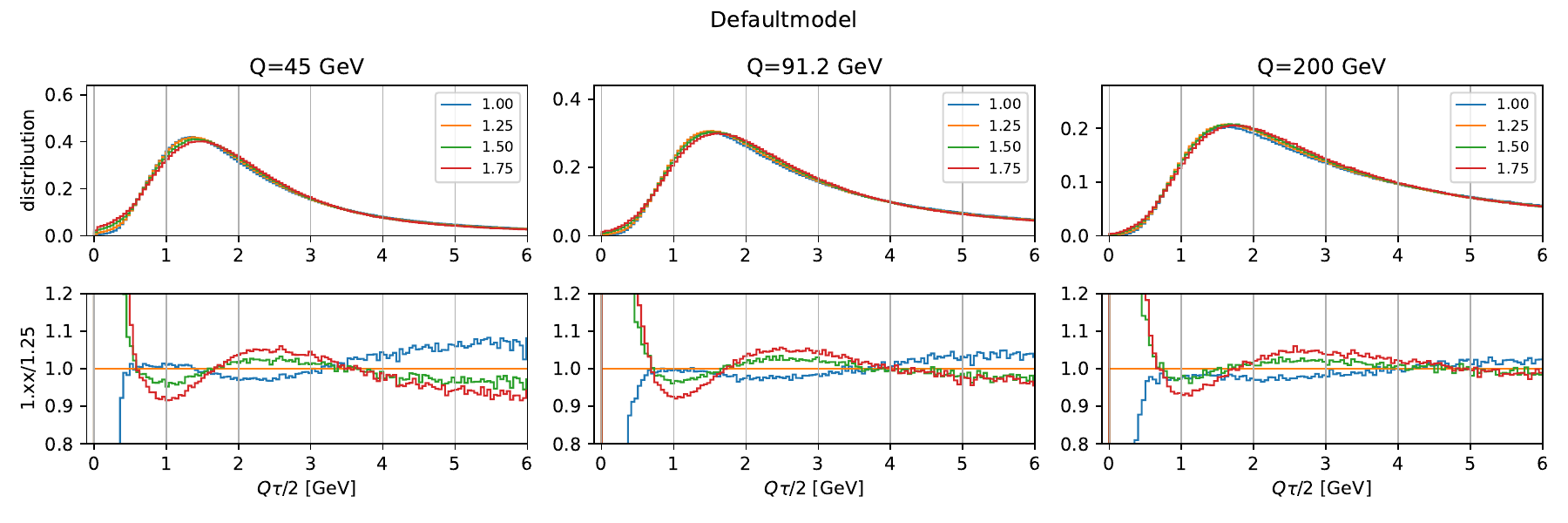}
		\caption{Upper panels: Shower cutoff $Q_0$-dependence of the 
	     \herwig{} hadron level 2-jettiness distribution 
		 in the peak region for the c.m.\ energies
		 $Q=45$ (left panels), $91.2$~GeV (middle panels) and 
		 $Q=200$~GeV (right panels) for the default
         hadronization model. Lower panels: ratio of the 
         2-jettiness distributions with respect to the reference 
         $Q_{0,{\rm ref}}=1.25$~GeV tune. 
         }\label{fig:2jettihadronleveldefault}
	\end{center}
\end{figure}

\begin{figure}
	\begin{center}
		\includegraphics[width=1.0\textwidth]{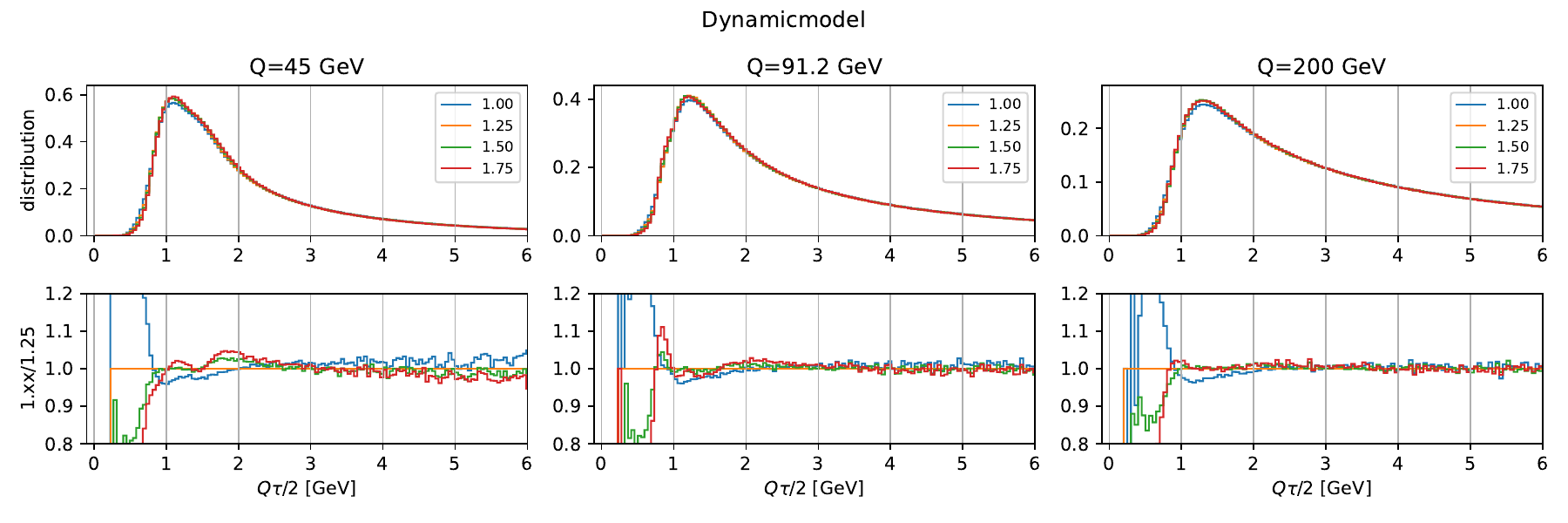}
		\caption{Upper panels: Shower cutoff $Q_0$-dependence of the \herwig{} 
			hadron level 2-jettiness distribution 
			in the peak region for the c.m.\ energies
			$Q=45$ (left panels), $91.2$~GeV (middle panels) and 
			$Q=200$~GeV (right panels) for the novel dynamical
			hadronization model. Lower panels: ratio of the 
			2-jettiness distributions with respect to the reference 
			$Q_{0,{\rm ref}}=1.25$~GeV tune. 
		    }\label{fig:2jettihadronleveldynamic}
	\end{center}
\end{figure}

We start by considering the simulation results for the hadron level
2-jettiness distribution for the different $Q_0$-dependent tunes in
the default and new dynamic hadronization model. In
Figs.~\ref{fig:2jettihadronleveldefault} and
\ref{fig:2jettihadronleveldynamic} the 2-jettiness $\tau$
distributions are shown for the default and the dynamic models,
respectively, for c.m.\ energies $Q=45$~GeV (left panels), $91.2$~GeV
(middle panels) and $200$~GeV (right panels) and for the shower cutoff
values $Q_0=1$ (blue), $1.25$ (orange), $1.5$ (green) and $1.75$~GeV
(red). The respective lower panels show the ratio of the differential
cross section values with respect to the reference scale
$Q_{0,\mathrm{ref}}=\SI{1.25}{\GeV}$. While the $Q_0$-dependence for
the default model yields variations in the simulated cross section in
the range between $5$ and $10$\% for the peak and the tail region, the
corresponding variations for the dynamic model are generally at the
level of a few percent, except for $Q=45$~GeV, where they can reach
$5$\% in the distribution tail for $\tau$ values above the
peak.\footnote{This behavior arises because the
  power-dependence on $Q_0$ related to the remaining uncanceled
  contributions comes with corresponding inverse powers of $Q$ for
  dimensional reasons. One should also note that the shower evolution
  needed to reproduce the expected resummation properties is typically
  over-populating the tails of event shapes, with larger variations in
  multiplicity due to the cutoff variations \cite{Bewick:2019rbu}.} We
can also observe a significantly smaller $Q_0$-dependence for the
dynamical hadronization model for $\tau$ values below the
peak. Overall, it is clearly visible that the $Q_0$-independence is
realized significantly better in the dynamical model than in the
default model.

It is highly instructive to see that this improvement does not only
happen for the simulation at $Q=91.2$~GeV, where the tuning procedure
is carried out, but also for other hard scattering energies. This is
highly rewarding as it shows that the design of the novel dynamical
hadronization model properly extrapolates the correct $Q_0$ dependence 
to other energies, that are not controlled directly through the tuning. 
This property is crucial for the interpretation of the shower cut being 
an IR factorization scale and the consistency of the hadronization
model's dynamical behavior in the context of QCD.

We have checked that these features are not only realized for
2-jettiness and event-shapes, on which most of our analytic insights
concerning the NLL precision of \herwig{}'s angular ordered parton
shower and of the partonic shower cut $Q_0$ dependence are based on,
but actually for all jet and event-shape related observables that have
been measured in $e^+e^-$ collisions in the past. In
Figs.~\ref{fig:Rivetdefaultotherdistributions} and
\ref{fig:Rivetdynamicotherdistributions} shown in
App.~\ref{app:otherfigures} this is demonstrated displaying the
simulations results obtained for the default and the novel dynamical
hadronization model, respectively, again for the shower cutoff values
$Q_0=1$, $1.25$, $1.5$ and $1.75$~GeV for a number of selected
observables for in comparison with actual LEP data gathered at $Q=44$,
$Q=91$ and $Q=133$~GeV.  We have displayed the results for a number of
other event-shapes (thrust, heavy jet mass), jet resolution rates as
well as charged particle multiplicities measured at the JADE, ALEPH,
DELPHI and OPAL
experiments~\cite{MovillaFernandez:1997fr,JADE:1999zar,ALEPH:2003obs,
  ALEPH:1996oqp,ALEPH:1991ldi,DELPHI:1996sen} The respective Rivet
analyses codes used for the generation of each of the histograms is
quoted at the bottom of each panel.  The lower sections of the
individual panels show ratios with respect to the MC reference
simulation results for $Q_{0,\rm{ref}}=1.25$~GeV. The improved shower
cutoff $Q_0$-independence for the dynamical model in comparison to the
default model for all observables is clearly visible. We emphasize
again that we have checked that the results shown in
Figs.~\ref{fig:Rivetdefaultotherdistributions} and
\ref{fig:Rivetdynamicotherdistributions} are representative for all
available $e^+e^-$ Rivet analyses. We also stress that, overall, the
quality of the description of the data provided by all the tunes we
obtained for the novel dynamical hadronization model is the same as
that of the standard \herwig{} release tune.\footnote{The fact that
  this description is not the best description of the current data is
  a known feature of the angular ordered shower in the default,
  analytically controllable and logarithmically accurate setup we
  employ here. This has been partially addressed in
  \cite{Bewick:2019rbu,Bewick:2021nhc} and improvements are subject to
  ongoing work. We do not expect, however, that this will change any
  of our findings on the hadronization since the data has merely
  served as a reference for a realistic tuning campaign.}

\subsection{Tuning Quality and Cutoff Dependence of the Tuning Parameters}
\label{sec:tuningquality}

\begin{figure}
	\begin{center}
		\includegraphics[width=1.0\textwidth]{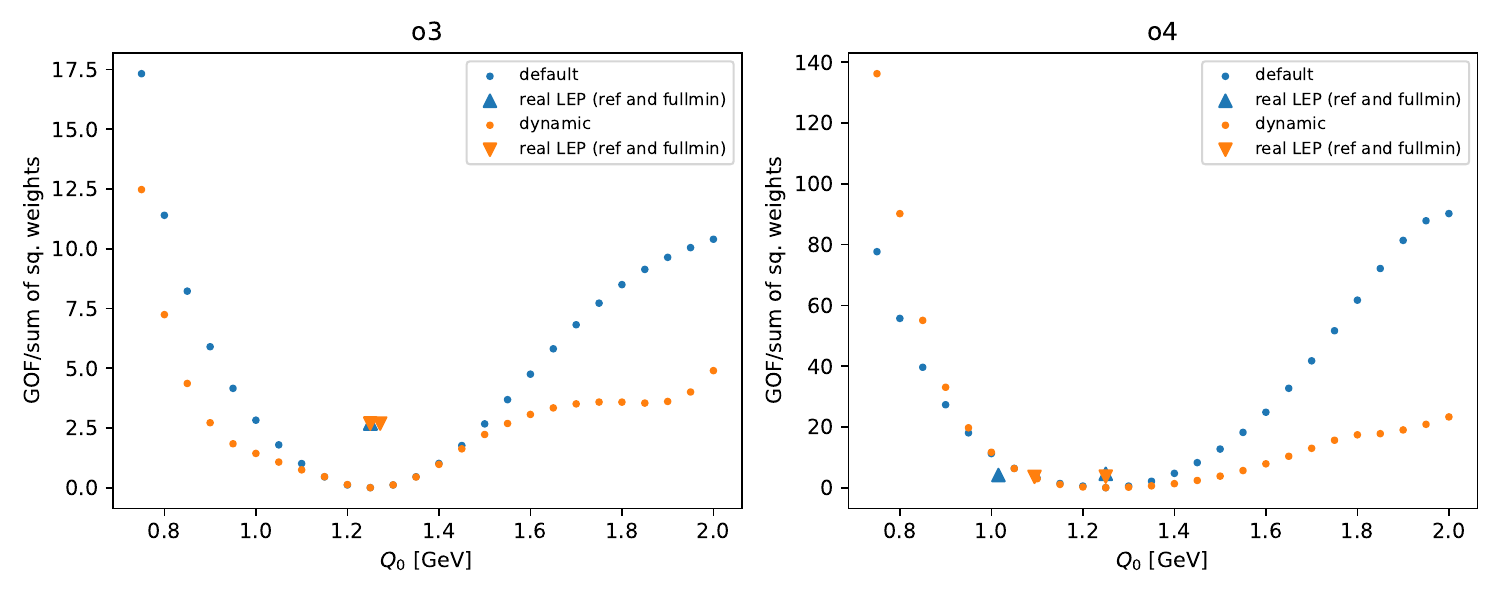}
		\caption{Minimal GOF values normalized to the sum of 
	    squared weights $(\sum_k{w_k^2})$
		obtained from the tuning procedures for the default (blue dots) 
		and the novel dynamical hadronization model (orange dots)
	    as a function of the shower cutoff $Q_0$. The left panel
	    shows the results for the cubic (o3) and right panel the 
	    results for the quartic (o4) interpolations. The colored 
	    triangles represent the normalized minimal GOF values obtained
	    from using actual LEP data as the reference date, once for
	    treating $Q_0$ as a floating fit parameter and once for
	    using the reference value $Q_{0,\rm{ref}}=1.25$~GeV. 
        }\label{fig:tuningchi2}
	\end{center}
\end{figure}

Let us now have a closer look on the quality of the fitting procedure
from which the cutoff dependent tuning parameters
${p_{i,\mathrm{cent}}(Q_0)}$ emerge. In Fig.~\ref{fig:tuningchi2} the
minimal GOF function value of the tune (normalized to the sum of squared bin
weights $\sum_i{w_i^2}$) is displayed
as a function of $Q_0$ in the range between $0.75$ and $2$~GeV for the
default (blue dots) and the dynamic model (orange dots). The left
panels show the results for the cubic (o3) interpolations and the
right panels those for the more accurate quartic (o4) interpolations.
For the reference shower cutoff $Q_{0,\mathrm{ref}}=\SI{1.25}{\GeV}$,
the GOF values are zero for both hadronization models as required by
consistency. Furthermore, both hadronization models yield very similar
small GOF values for $Q_0$ close to $Q_{0,\mathrm{ref}}$ in the range
between about $1$ and $1.4$~GeV, which is natural due to the small
difference to the reference cutoff value. Since cutoff values
$Q_0<1$~GeV are not feasible for the interpretation as an infrared
factorization scale, it is the GOF values for the other shower cutoff
values above $1.4$~GeV that we have to compare for the two
hadronization models. We can clearly see that the default
hadronization model (blue) yields significantly larger minimal GOF
values than the dynamical model (orange). The overall larger size of
the GOF values for the quartic interpolation arises due to the higher
statistics used when generating the interpolation. Thus, since the 
choice for $Q_{0,\mathrm{ref}}$ is as a matter of principle arbitrary,
we can conclude that the quality of the fits
is significantly better for the dynamical model than for the default
one in the physically relevant $Q_0$ interval between $1$ and
$2$~GeV. In the ideal case the minimal GOF value would stay close to
zero for all $Q_0$ values. In this respect the dynamical model does
much better than the default model, but the current version of the
dynamical model cannot yet achieve this challenging goal.

At this point an interesting aspect to be discussed is to which extent
the reference data used for the tuning analyses, which are generated
by the MC for the cutoff value $Q_{0,\mathrm{ref}}=1.25$~GeV,
represent a good proxy for the real LEP data. Furthermore, it should
be also clarified whether the experimental data indeed prefers shower cut
values in the range where perturbation theory is still valid, i.e.\ in
the range between $1$ and $2$~GeV that we consider in our discussion.
To address these questions we carry out two additional tuning analyses
where instead of the MC generated reference data the LEP data is used
that also enters the regular \herwig{} release tunes. In one tuning
analysis $Q_0$ is treated as a freely floating tuning parameter and in
the other the shower cutoff is fixed to $Q_{0,\mathrm{ref}}=1.25$~GeV.
The outcome is shown as the colored triangles for the cubic as well as
for the quartic interpolations.  The minimal GOF function values for
these tuning fits are of course not zero since the real data always
differs from simulations.  For the cubic interpolation both tunes
yield practically identical values, with minimal GOF values compatible
with the smaller dynamical model GOF values for $Q_0>1.4$~GeV.  For
the quartic interpolation the minimal GOF values for the tune with a
freely floating shower cutoff yields $Q_0$ values close to unity, and
the minimal GOF values are in the range from $4.2$ to $4.6$ for the default
and from $3.57$ to $3.64$ for the dynamical model.
The results confirm that the MC reference data of our
shower cutoff dependent tuning analysis is sufficiently close to the
experimental LEP data and, more importantly, that the interval of $1$
to $2$~GeV, where the shower cut $Q_0$ can be considered as an
infrared factorization scale, is well compatible with the experimental
data.

\begin{figure}
	\begin{center}
		\includegraphics[width=1.0\textwidth]{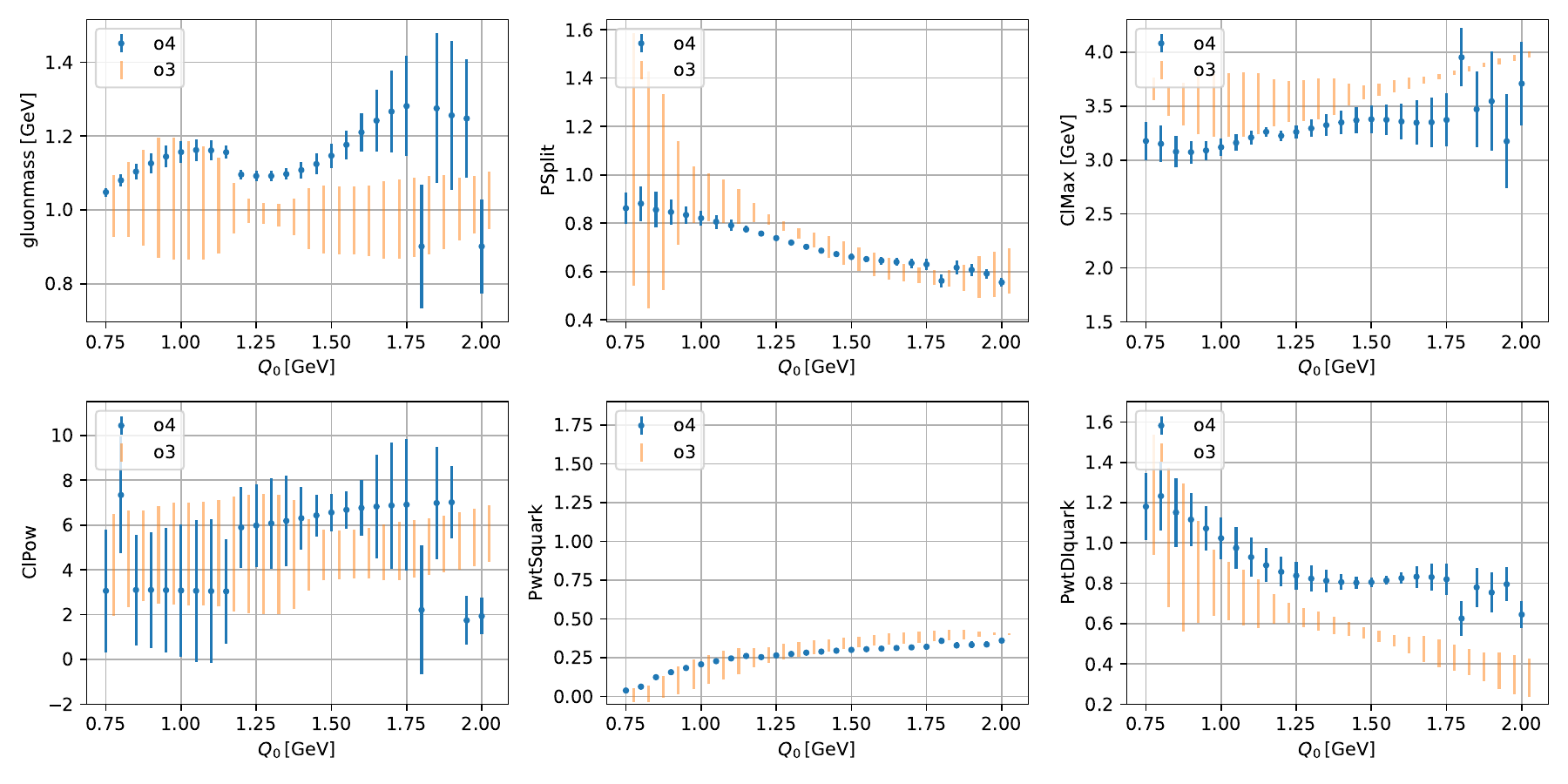}
		\caption{Dependence of the tuned parameters on the
               shower cutoff $Q_0$ for the default hadronization model. 
               The dots represent the central values and the vertical lines
               the combines statistical and interpolation uncertainty. 
               The orange results are based on cubic and the blue results on
               the quartic interpolations. 
               }\label{fig:tuningparadefault}
	\end{center}
\end{figure}

\begin{figure}
	\begin{center}
		\includegraphics[width=1.0\textwidth]{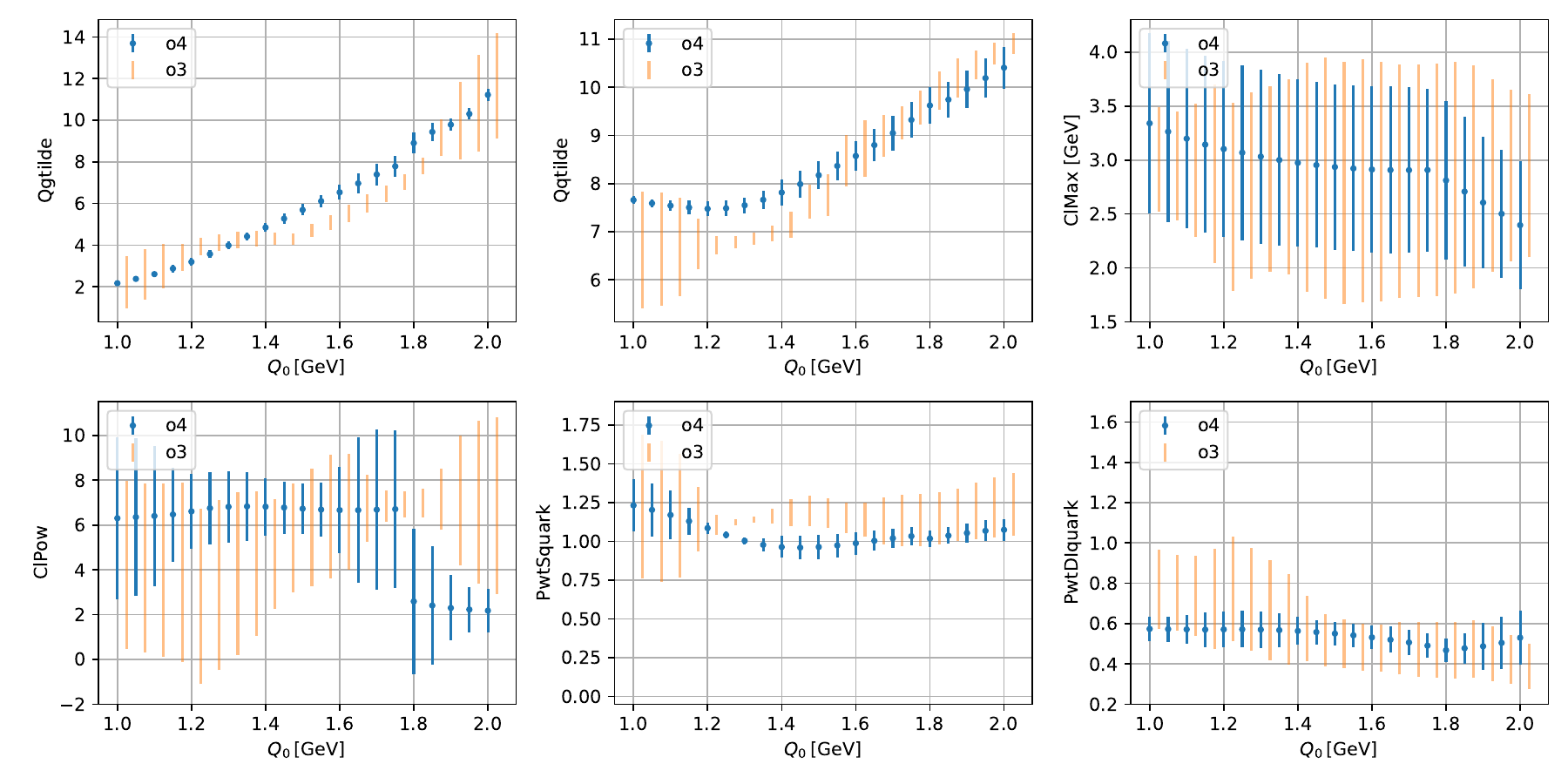}
		\caption{Dependence of the tuned parameters on the
			shower cutoff $Q_0$ for the novel dynamical hadronization model. 
			The dots represent the central values and the vertical lines
			the combines statistical and interpolation uncertainty. 
			The orange results are based on cubic and the blue results on
			the quartic interpolations. 
            }\label{fig:tuningparadynamic}
	\end{center}
\end{figure}

Let is now have a look at the $Q_0$-dependence of the six hadronization
model parameters which are treated as floating parameters in our
tuning analyses in each of the hadronization models. As we have
already mentioned in Sec.~\ref{sec:descriptiontuning}, for the
dynamical model a proper matching to the parton shower (which allows
for the shower cutoff $Q_0$ to be interpreted as an infrared
factorization scale) implies a linear $Q_0$-dependence for the
``hard'' starting scales $\tilde Q_g$ and $\tilde Q_q$ of the
non-perturbative branching processes and an insensitivity to $Q_0$ for
the other hadronization parameters. On the other hand, for the default
hadronization model (where the shower cutoff is merely another
hadronization parameter) similar implications can in principle not 
necessarily be
expected. In Figs.~\ref{fig:tuningparadefault} and
\ref{fig:tuningparadynamic} we show the $Q_0$-dependence of the six
tuning parameters which were treated as floating parameters for the
default and the dynamic models, respectively, 
for $Q_0$ values between $1$ and $2$~GeV. We show the results
based on the cubic (orange) and the quartic (blue) interpolations. The
dots show the central values ${p_{i,\mathrm{cent}}(Q_0)}$ and the
vertical lines stand for the uncertainties which are obtained from the
quadratic sum of the statistical und interpolation uncertainties
$\Delta p_i=\sqrt{(\Delta p_{i,\mathrm{stat}}(Q_0))^2 + (\Delta
  p_{i,\mathrm{inter}}(Q_0))^2}$, as described in
Sec.~\ref{sec:tuninguncertainties}. Overall we see that tuning
results based on the cubic and quartic interpolations are nicely
compatible, even though the error bars do not always overlap.

In Fig.~\ref{fig:tuningparadynamic} we see that, indeed, the
parameters $\tilde Q_g$ and $\tilde Q_q$ exhibit a $Q_0$-dependence
that is quite close to linear particularly for $\tilde Q_g$, which 
governs the gluon splitting dynamics of the dynamical model. For
$\tilde Q_q$ the expected behavior only sets in for $Q_0>1.3$~GeV,
which is likely related to the fact that it governs the cluster fission
dynamics that takes place at a later stage and may not have sufficient
room to adapt for low $Q_0$ values. 
At the same time, in comparison the
other four hadronization parameters (${\mathtt{Cl}_{\mathtt{max}}}$,
${\mathtt{Cl}_{\mathtt{pow}}}$, $\mathtt{PwtSquark}$ and 
$\mathtt{PwtDIquark}$) are, indeed, rather
$Q_0$-insensitive. The behavior is again not perfect, with 
${\mathtt{Cl}_{\mathtt{max}}}$ and ${\mathtt{Cl}_{\mathtt{pow}}}$
exhibiting rather large uncertainties, but it can
nevertheless be observed rather clearly. In contrast, we can see a
somewhat larger $Q_0$-dependence of the same four parameters for the
default model in Fig.~\ref{fig:tuningparadefault}. This is visible
most prominently for $\mathtt{PwtDIquark}$. 
Since these parameters directly affect the
Baryon production and thus the charge particle multiplicities
generated in the simulation, we can conclude that in the default model
the shower cutoff value $Q_0$ affects these multiplicities so that
the tuned value of $\mathtt{PwtDIquark}$ needs to compensate in
a significant way.
In the novel dynamical hadronization model this rather unnatural
feature is not visible. We also note that the not completely
smooth behavior visible for some of the model parameters as a function
of $Q_0$ can be partly attributed to mass, threshold as well as
some exceptional kinematic effects that can take place
in the cluster model when $Q_0$ is varied.
Overall, the results shown in
Figs.~\ref{fig:tuningparadefault} and \ref{fig:tuningparadynamic}
again confirm that the novel dynamical hadronization model achieves a
better separation of the low-scale hadronization processes from the
parton level description provided by the parton shower. This
separation is essentially the MC simulation analogue of factorization
in analytic QCD studies and an important prerequisite for the parton
shower cutoff to be interpreted as a factorization scale.

\subsection{Rescaled Migration Matrix Function}
\label{sec:migrationvsshape}

\begin{figure}
	\centering
	\begin{minipage}{.5\textwidth}
		\centering
		\includegraphics[width=\textwidth]{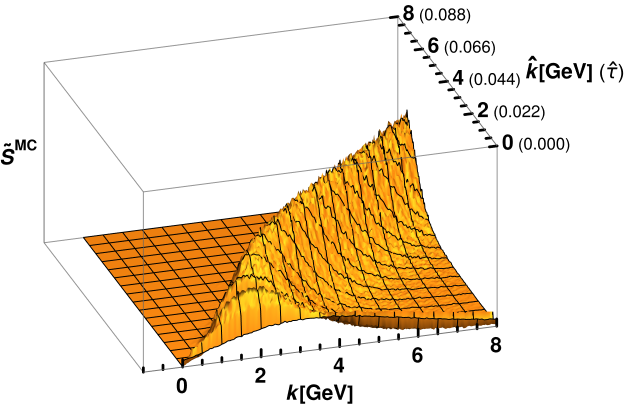}
	\end{minipage}%
	\begin{minipage}{.5\textwidth}
		\centering
		\includegraphics[width=\textwidth]{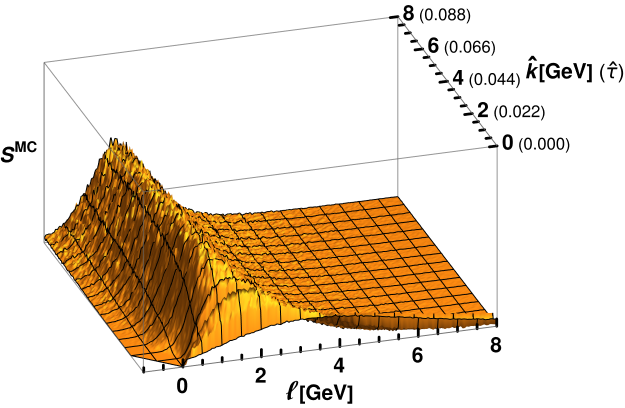}
	\end{minipage}
	\caption{Rescaled parton-to-hadron migration matrix funtions for the default 
		hadronization model for the 2-jettiness distribution for the shower cutoff
		$Q_0=1.25$~GeV extracted for c.m.\ energy $Q=91.2$~GeV. The left panel
		shows $\tilde S^{\rm MC}(k,\hat k,\{Q,Q_0\})$ and the right panel
		$S^{\rm MC}(\ell,\{\hat k,Q,Q_0\})$, which is analogous to the shape 
		function.
	    }
	\label{fig:transfermatrix3Ddefault}
\end{figure}

\begin{figure}
	\centering
	\begin{minipage}{.5\textwidth}
		\centering
		\includegraphics[width=\textwidth]{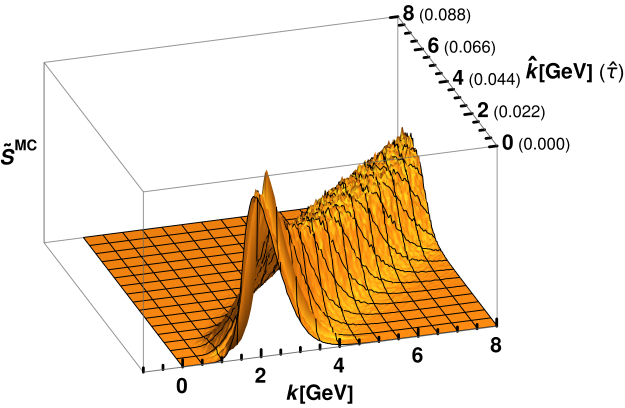}
	\end{minipage}%
	\begin{minipage}{.5\textwidth}
		\centering
		\includegraphics[width=\textwidth]{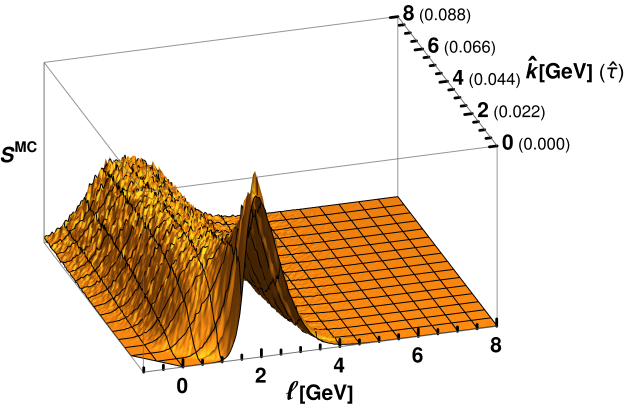}
	\end{minipage}
	\caption{Rescaled parton-to-hadron migration matrix funtions for the novel 
		dynamical hadronization model for the 2-jettiness distribution for the 
		shower cutoff
		$Q_0=1.25$~GeV extracted for c.m.\ energy $Q=91.2$~GeV. The left panel
		shows $\tilde S^{\rm MC}(k,\hat k,\{Q,Q_0\})$ and the right panel
		$S^{\rm MC}(\ell,\{\hat k,Q,Q_0\})$, which is analogous to the shape 
		function.}
	\label{fig:transfermatrix3Ddynamic}
\end{figure}

Let us now start the discussion on the MC results for the rescaled
migration matrix functions $\tilde S^{\rm MC}(k,\hat k,\{Q,Q_0\})$ and
$S^{\rm MC}(\ell,\{\hat k,Q,Q_0\})$ defined in
Eqs.~(\ref{eq:tildeSMCTrelation}) and (\ref{eq:SMCdef}), respectively,
for the default and the dynamic hadronization models obtained from the
2-jettiness $\tau$ distributions. For a given (true) parton level soft
momentum $\hat k=Q\hat\tau$ and shower cutoff $Q_0$, the migration
function $\tilde S^{\rm MC}(k,\hat k,\{Q,Q_0\})$ gives the
distribution of hadron level momenta $k=Q\tau$, while
$S^{\rm MC}(\ell,\{\hat k,Q,Q_0\})$ gives the distribution of the
non-perturbative soft momenta $\ell=k-\hat k$ that the hadronization
adds to the parton level configuration. We remind the reader that
$S^{\rm MC}(\ell,\{\hat k,Q,Q_0\})$ is the MC analogue of the infrared
cutoff-dependent shape function $S_{\rm had}(\ell,Q_0)$ in the dijet
2-jettiness QCD factorization formula discussed in
Sec.~\ref{sec:observable}.  Details concerning the extraction of the
migration matrix functions from the \herwig{} simulations have been
explained in Sec.~\ref{sec:extractmigration}.
 
In the left panel of Fig.~\ref{fig:transfermatrix3Ddefault} the MC
migration function $\tilde S^{\rm MC}(k,\hat k,\{Q,Q_0\})$ is shown
for the \herwig{} default cluster hadronization model. It has been
extracted at the c.m.\ energy $Q=91.2$~GeV for the reference
$Q_0=1.25$~GeV tune (explained in more detail in
Sec.~\ref{sec:referencetune}) in the range $0\le k,\hat k\le
8$~GeV. For $Q=91.2$~GeV this corresponds to thrust values between $0$
and $0.088$.  The corresponding shifted migration function
$S^{\rm MC}(\ell,\{\hat k,Q,Q_0\})$ is shown in the right panel.  In
Fig.~\ref{fig:transfermatrix3Ddynamic} the analogous migration
functions are shown for the novel dynamical model.  The shape and
structures shown in both figures are representative for the migration
matrix functions for all cases we obtain in our analyses
 
We see that for $\hat k>2$~GeV the shape of the migration matrix
functions for both hadronization models are very well 
consistent with the expectations from a shape
function, which states that the migration function
$S^{\rm MC}(\ell,\{\hat k,Q,Q_0\})$ should be independent of the
partonic momentum $\hat k$.  We can spot deviations from this
expectation for $\hat k<2$~GeV for the default as well as for the
novel dynamical hadronization model.  While for the default
hadronization model the migration function
$S^{\rm MC}(\ell,\{\hat k,Q,Q_0\})$ for $\hat k<2$~GeV is considerably
flatter than for $\hat k>2$~GeV, for the dynamic hadronization model
it is much peakier. These features play an essential role in the more
important quantitative analyses we carry out in
Sec.~\ref{sec:migrationquantitative}.

Before moving on, let us comment on a general feature of the migration
functions. It concerns that $S^{\rm MC}(\ell,\{\hat k,Q,Q_0\})$ is
finite for $\ell$ values down to $-1$~GeV. We can see
this feature clearly in the right panels of
Figs.~\ref{fig:transfermatrix3Ddefault} and
Fig.~\ref{fig:transfermatrix3Ddynamic} for partonic values
$\hat k > 1$~GeV.  Even though the major portion of
$S^{\rm MC}(\ell,\{\hat k,Q,Q_0\})$ is located at positive $\ell$,
which means that the bulk of hadronization corrections shift the
thrust distribution towards larger thrust values, this negative tail
shows that hadronization can sometimes also decrease the thrust
value. Since this feature arises from the tuning to actual LEP data,
it means that at least for the \herwig{} simulation implementation
this particular feature of the hadronization corrections is demanded 
by data. For $\hat k \ge 1.5$~GeV this behaviour arises consistently 
and becomes
independent of the value of $\hat k$ also for other $Q$ and $Q_0$.
This feature is per se not at all problematic and also consistent with
a shape function within QCD factorization. However, for partonic
$\hat k$ values below $1.5$~GeV the migration function is not capable to
build up such a negative tail, simply because the physical thrust
values are restricted to be positive. This entails that the first
$\ell$-moment of the rescaled migration function
$S^{\rm MC}(\ell,\{\hat k,Q,Q_0\})$ (see Eq.~(\ref{eq:Omega1Q0})) for
$\hat k$ values below $1.5$~GeV is always larger than for values larger
than $2$~GeV, where the first moment stabilizes. This feature is
absent in the shape function $S_{\rm had}(\ell,Q_0)$, which is
strictly $\hat k$-independent. We come back to this feature in our
discussion below.

\subsection{Shower Cutoff and Energy Dependence of the Migration Matrix Function}
\label{sec:migrationquantitative}

\begin{figure}
	\center
	\includegraphics[width=\textwidth]{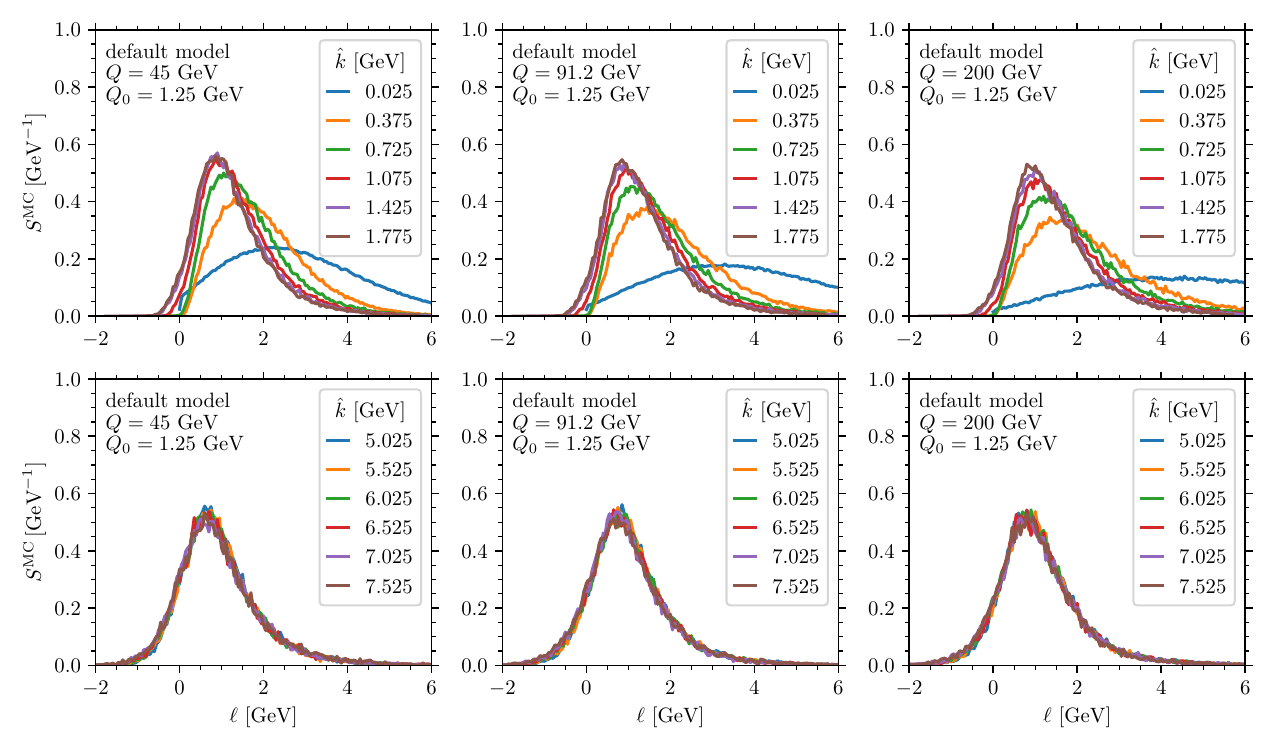}
	\caption{Rescaled parton-to-hadron migration matrix function
		$S^{\rm MC}(\ell,\{\hat k,Q,Q_0\})$ as a function of $\ell$ 
		for the default hadronization model for $Q_0=1.25$ and a 
		number of different 
		partonic $\hat k=\hat{\tau}/Q$ soft momenta. The upper panels 
		show the results for some $\hat k<2$~GeV and the lower panels
		for larger $\hat k$ values. The left, middle and right panels
		have been extracted from the 2-jettiness distributions for
		c.m.\ energies $Q=45$, $91.2$ and $200$~GeV, respectively.
	    }
	\label{fig:softfunctions_default}
\end{figure}

\begin{figure}
	\center
	\includegraphics[width=\textwidth]{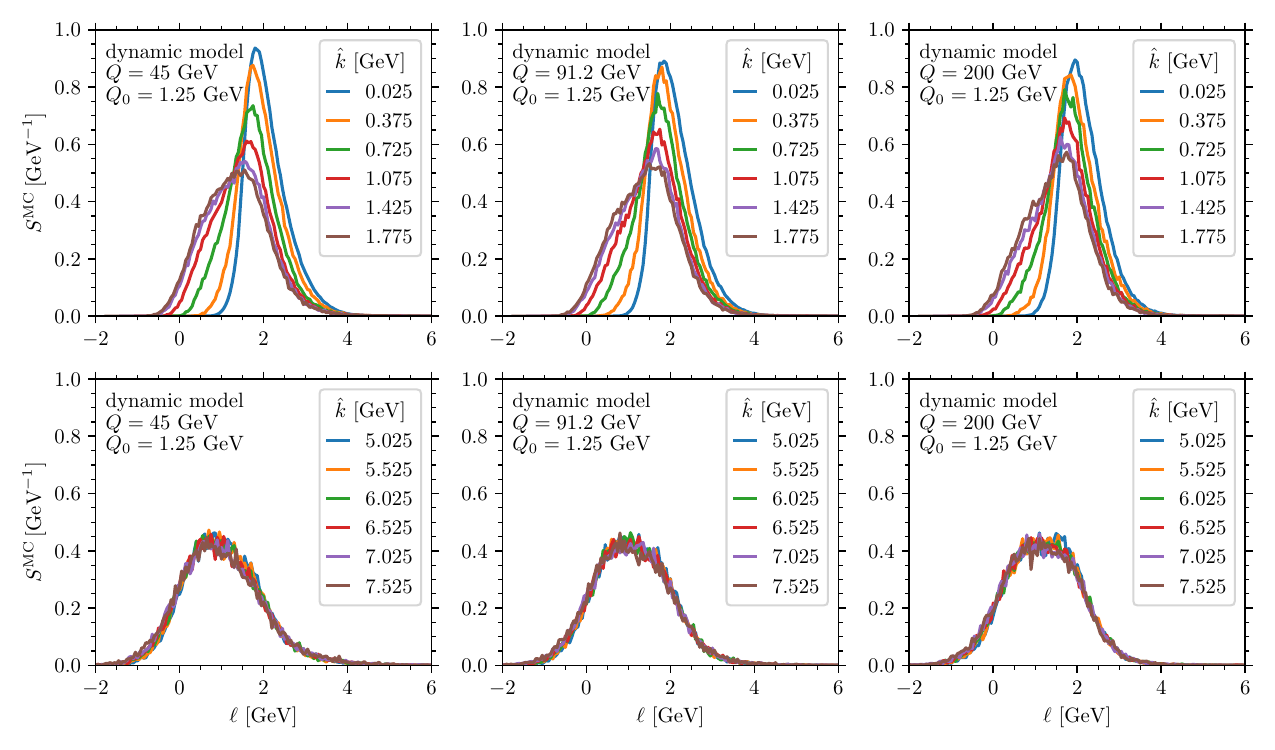}
	\caption{Rescaled parton-to-hadron migration matrix function
		$S^{\rm MC}(\ell,\{\hat k,Q,Q_0\})$ as a function of $\ell$ 
		for the novel dynamical hadronization model for $Q_0=1.25$ 
		and a number of different 
		partonic $\hat k=\hat{\tau}/Q$ soft momenta. The upper panels 
		show the results for some $\hat k<2$~GeV and the lower panels
		for larger $\hat k$ values. The left, middle and right panels
		have been extracted from the 2-jettiness distributions for
		c.m.\ energies $Q=45$, $91.2$ and $200$~GeV, respectively.
	}
	\label{fig:softfunctions_dynamic}
\end{figure}

Finally, let us now discuss at the more quantitative level the
properties of the migration matrix functions
$S^{\rm MC}(\ell,\{\hat k,Q,Q_0\})$ we have obtained from our
$Q_0$-dependent tuning analyses for the \herwig{} default and the
novel dynamical hadronization model.

In Fig.~\ref{fig:softfunctions_default} we show
$S^{\rm MC}(\ell,\{\hat k,Q,Q_0\})$ obtained from the default
hadronization model over $\ell$ for different $\hat k$ for the
reference shower cut scale $Q_{0,\mathrm{ref}}=\SI{1.25}{\GeV}$ for
the hard scales $Q=45$~GeV (left panels), $91.2$~GeV (middle panels)
and $200$~GeV (right panels). In Fig.~\ref{fig:softfunctions_dynamic}
the analogous results are displayed obtained from the dynamical
hadronization model.  The upper panels show
$S^{\rm MC}(\ell,\{\hat k,Q,Q_0\})$ for several small $\hat k$ values
below $2$~GeV, which is the range where $S^{\rm MC}$ is still strongly
depending on $\hat k$. The lower panels show
$S^{\rm MC}(\ell,\{\hat k,Q,Q_0\})$ for larger $\hat k$ values between
$5$ and $8$~GeV where it is rather $\hat k$-independent. We remind the
reader that $\hat k=8$~GeV corresponds to partonic 2-jettiness values
of $\hat{\tau}=(0.178,0.088,0.040)$ for $Q=\{45,91.2,200\}$\,GeV, so
all rescaled transfer functions which are shown are well within the
dijet region, see our discssion of Fig.~\ref{fig:cumulantdifference}.
It is one of the predictions of QCD factorization in the dijet region
that the shape function $S_{\rm had}(\ell,Q_0)$ is independent of the
hard scale $Q$ and the partonic $\hat k$ value. As we can see in the
lower panels of Figs.~\ref{fig:softfunctions_default} and
\ref{fig:softfunctions_dynamic}, this is indeed realized quite well
for $S^{\rm MC}(\ell,\{\hat k,Q,Q_0\})$ obtained from both
hadronization models at large $\hat k$ values. For the dynamic model
the shape of
$S^{\rm MC}(\ell,\{\hat k,Q,Q_0\})$ is somewhat broader than for the
default model but for both models the results are very stable
concerning the values of $Q$ and $\hat k$. For small $\hat k$ the
situation is, however, quite different. We see that for the default
model the $S^{\rm MC}(\ell,\{\hat k,Q,Q_0\})$ functions broaden
considerably for decreasing $\hat k$. This behavior is also visible in
the 3D plots of Fig.~\ref{fig:transfermatrix3Ddefault} and furthermore
depends significantly on the c.m.\ energy $Q$, as can be seen most
clearly from the blue curves for $\hat k=0.025$~GeV in the upper panels 
of Fig.~\ref{fig:softfunctions_default}. This feature of
the default hadronization model causes very large positive
hadronization corrections for events with no partonic branching or
small partonic $\hat{\tau}$ values. Since the no-branching events
still constitute a considerable fraction of the events
($(16.1, 6.2, 1.9)\%$ for
$Q=(45,91.2,200)$~GeV and $Q_0=\SI{1.25}{\GeV}$ for the \herwig's
shower), this broadening effect affects the hadron level $\tau$
distribution in a notable way. The important aspect of this effect
is, that it is incompatible with QCD factorization. For the dynamical
model there is still a visible dependence on $\hat k$, but it is
substantially milder. In particular there is no broadening for
decreasing values of $\hat k$ and the $Q$ dependence is significantly 
smaller as well, which can be seen by again comparing the blue curves 
in the upper panel of Fig.~\ref{fig:softfunctions_dynamic} for
$\hat k=0.025$~GeV and the three $Q$ values.  
Instead, decreasing $\hat k$ the shape of
$S^{\rm MC}(\ell)$ in the dynamical hadronization model becomes more
peaky. This behavior is an attempt of the dynamical
hadronization model to compensate for the principle inability to
describe the negative hadronization corrections for $\hat k\le 1.5$~GeV,
which we already mentioned in the previous subsection, while at the 
same time avoiding the QCD-incompatible broadening effects of the
default model.

\begin{figure}
	\begin{center}
		\includegraphics[width=1.0\textwidth]{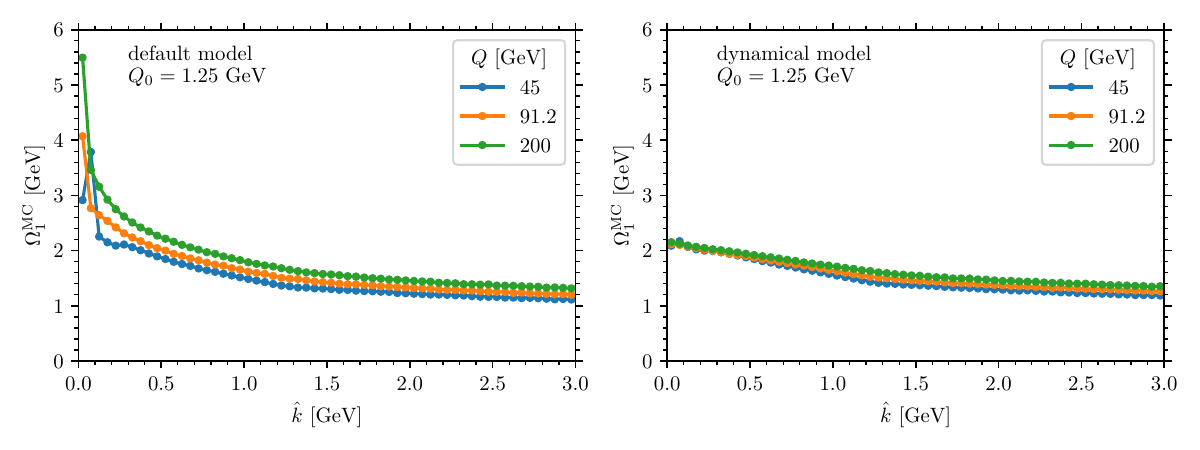}
		\caption{First moment $\Omega_1^{\rm MC}(\hat k,Q,Q_0)$ 
	of the rescaled parton-to-hadron migration function
    over $\hat k$ for the default (left panel) and the novel 
    dynamical hadronization model (right panel) extracted from
    the 2-jettiness distribution simulated for shower cutoff 
    $Q_0=1.25$~GeV at c.m.\ energies $Q=45$~GeV (blue),
    $91.2$ (orange) and $200$~GeV (green dots).
    }\label{fig:omega1Qs}
	\end{center}
\end{figure}

To gain a more quantitative insight into the $Q_0$-, $\hat k$- and
$Q$-dependence of the migration function
$S^{\rm MC}(\ell,\{\hat k,Q,Q_0\})$ it is instructive to have a close
look on its first moment defined by
\begin{align}
\label{eq:Omega1Q0MC}
	\Omega_1^{\rm MC}(\hat k,Q,Q_0) \, \equiv\,
	\frac{1}{2}	\int \!\mathrm{d}\ell\,\ell\, 
	S^{\rm MC}(\ell,\{\hat k,Q,Q_0\})\,,
\end{align}
in analogy to the QCD factorization's shape function first moment
$\Omega_1(Q_0)$ given in Eq.~(\ref{eq:Omega1Q0}). The latter only
depends on the shower cut $Q_0$, but is independent of $\hat k$ or
$Q$. In Fig.~\ref{fig:omega1Qs} the value of 
$\Omega_1^{\rm MC}(\hat k,Q,Q_0)$ is
displayed for $0<\hat k<3$~GeV and $Q=45$~GeV (blue), $91.2$~GeV
(orange) and $200$~GeV (green) for $Q_0=1.25$~GeV. The left panel
shows $\Omega_1^{\rm MC}$ for the default hadronization model and the
right panel that for the novel dynamical model. For both we only see a
mild $Q$- and $\hat k$-dependence for $\hat k\gtrsim 1.5$~GeV, where
the dependence is slightly smaller for the dynamical model. For small
$\hat k<1.5$~GeV, however, the dependence on $\hat k$ and $Q$ is
enormous for the default model. For the smallest $\hat k$ value we have
$\Omega_1^{\rm MC}\approx 2.1$~GeV for all $Q$ values for the dynamical
model. On the other hand, $\Omega_1^{\rm MC}$ varies wildly with $Q$ and 
even reaches $5.5$~GeV for $Q=200$~GeV. It is this uncontrolled behavior
for the default hadronization model that is responsible for the
stronger $Q_0$ dependence visible in
Fig.~\ref{fig:2jettihadronleveldefault} in comparison to
Fig.~\ref{fig:2jettihadronleveldynamic}.

Nevertheless, also for the dynamic model
$\Omega_1^{\rm MC}(\hat k,Q,Q_0)$ still exhibits a visible $\hat k$
dependence for $\hat k<1.5$~GeV which is in principle not compatible
with the shape functions first moment. This is again related to the
fact already mentioned at the end of Sec.~\ref{sec:observable}, namely
that for the \herwig{} MC setup the hadronization effects can
sometimes lead to a decrease of the hadron level 2-jettiness value
$\tau$ with respect to its parton level value $\hat{\tau}$.  For
values $\hat k \lesssim 1$~GeV this behavior cannot be realized any
longer because the physical 2-jettiness values cannot become
negative. Since the hadronization model's capability to increase the
hadron level 2-jettiness value remains unchanged for these low
$\hat k$ values, this feature leads to the more peaky shape of the
$S^{\rm MC}(\ell,\{\hat k,Q,Q_0\})$ just mentioned in the discussion
of Fig.~\ref{fig:2jettihadronleveldynamic} above, which unavoidably
leads to an increasing first moment for $\hat k \lesssim 1$~GeV.

\begin{figure}
	\begin{center}
		\includegraphics[width=1.0\textwidth]{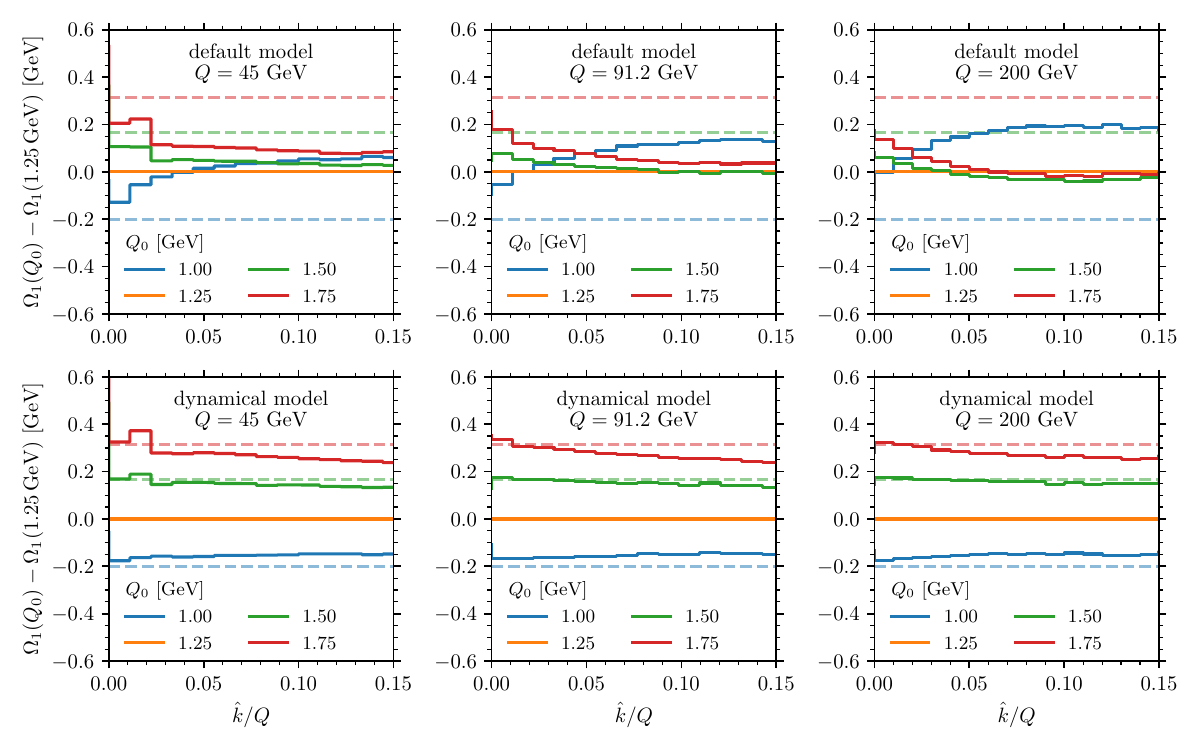}
		\caption{
	First moment difference 
	$\Omega_1^{\rm MC}(\hat k,Q,Q_0)-\Omega_1^{\rm MC}(\hat
	k,Q,Q_{0,\mathrm{ref}}=\SI{1.25}{\GeV})$ 
	of the rescaled parton-to-hadron migration matrix function
	over $\hat\tau=\hat k/Q$ for the default (upper panel) and 
	the novel 
	dynamical hadronization model (lower panels) extracted from
	the 2-jettiness distribution simulated for shower cutoff 
	$Q_0=1$ (blue), $1.25$ (orange), $1.5$ (green) and $1.75$~GeV 
	(red) at c.m.\ energies $Q=45$~GeV (left),
	$91.2$ (middle) and $200$~GeV (right panels). 
	The dashed lines show the results expected from
	QCD factorization employing the $\overline{\rm MS}$ strong
	coupling extracted from \herwig{}.
	}\label{fig:omegedifferences}
	\end{center}
\end{figure}

Finally, let us examine the dependence 
of $\Omega_1^{\rm MC}(\hat k,Q,Q_0)$ on the shower cutoff $Q_0$, which we 
have not yet covered in our analysis so far. 
Given the previous observations concerning
the $\hat k$ dependence of $\Omega_1^{\rm MC}$ for
$\hat k \lesssim 1.5$~GeV, the consistency of
$\Omega_1^{\rm MC}(\hat k,Q,Q_0)$ for the dynamic model with the 
$Q_0$-dependence of the shape function's first moment $\Omega_1$ in
Eq.~(\ref{eq:Omega1Q0}) should at least be realized for $\hat k > 1$~GeV,
which corresponds to $\hat{\tau}>(0.022,0.011,0.005)$ for
$Q=(45,91.2,200)$~GeV. In Fig.~\ref{fig:omegedifferences} the values
for
$\Omega_1^{\rm MC}(\hat k,Q,Q_0)-\Omega_1^{\rm MC}(\hat
k,Q,Q_{0,\mathrm{ref}})=\SI{1.25}{\GeV})$ are displayed for
$0<\hat{\tau}=\hat k/Q < 0.15$~GeV for $Q=45$~GeV (left panels),
$91.2$~GeV (middle panels) and $Q=200$~GeV (right panels), where we
have averaged the corresponding moment results in bins of width
$\Delta \hat\tau=0.01$~GeV. The upper panels show the results for the
default hadronization model and the lower panels for the novel dynamical
model. The results for the moment difference are shown for $Q_0=1.0$
(blue), $1.25$ (orange), $1.5$ (green) and $1.75$ (red). The
horizontal colored dashed lines represent the value of the
corresponding shape function moment differences from
Eq.~(\ref{eq:Omega1Q0primeO0}) as expected from QCD factorization:
\begin{align}
\label{eq:Omega1diffference}
\Omega_1(Q_0)-\Omega_1(Q_{0,\mathrm{ref}}) \, =\, \frac{1}{2}\,\Delta_{\rm soft}(Q_0,Q_{0,\mathrm{ref}})\,,
\end{align}
where $\Delta_{\rm soft}(Q_0,Q_{0,\mathrm{ref}})$ is determined from 
the $R$-evolution equation in Eq.~(\ref{eq:Deltasoftv2}) employing the 
strong coupling extracted from \herwig{} in the $\overline{\rm MS}$ 
scheme.

We see that for $\hat k > 1$~GeV the moment difference obtained for
the dynamical model (lower panels) are indeed nicely compatible with
Eq.~(\ref{eq:Omega1diffference}). The visible small discrepancies
are related to quadratic and higher order effects in $Q_0$ which 
the linear approximation of the evolution equation does not capture.
They are also compatible with the corresponding discrepancies 
concerning the $Q_0$ dependence of the partonic cumulant difference 
already discussed in Sec.~\ref{sec:numericalQ0} for 
Fig.~\ref{fig:cumulantdifference}. 
In stark contrast, for the default model results (upper panels) there is no 
sign of a similar agreement with Eq.~(\ref{eq:Omega1diffference}) 
or of any stability
with respect to the value of $\hat{\tau}$.  We see that for the default
model the moment differences are completely unrelated to the shower
cut dependence expected from QCD factorization, and thus essentially
uncontrolled from the QCD perspective. 
We stress, that this uncontrolled $Q_0$-dependence of the hadronization 
corrections from the default hadronization model takes place even 
for large $\hat k$ values where the migration function 
$S^{\rm MC}(\ell,\{\hat k,Q,Q_0\})$ appears stable (also with respect to 
changes of $Q$) as shown in the lower panels of 
Fig.~\ref{fig:softfunctions_default}.
There is furthermore no stability concerning
the $\hat{k}$ dependence of 
$\Omega_1^{\rm MC}(\hat k,Q,Q_0)-
\Omega_1^{\rm MC}(\hat k,Q,Q_{0,\mathrm{ref}})$
with respect the hard scattering scale $Q$. 

It is instructive to discuss this failure of the default
model from a the perspective of overall size of 
the hadronization corrections. As we can see from
Fig.~\ref{fig:omega1Qs}, for $Q_{0,\mathrm{ref}}=1.25$ 
we have $\Omega_1^{\rm MC}$  in the range between $1.2$ and $1.5$~GeV 
for partonic momenta $\hat k>2$~GeV. This should be compared to the
overall variation of $\Omega_1^{\rm MC}$ in the range 
between $Q_0=1.0$~GeV and $Q_0=1.75$~GeV predicted by 
$R$-evolution equation~(\ref{eq:Deltasoftv2}) obtained from 
QCD factorization. The latter amounts to around $0.5$~GeV, as we see
in the lower panels of Fig.~\ref{fig:omegedifferences}. However, 
because the choice of reference shower cut $Q_{0,\mathrm{ref}}$ is 
arbitrary, and we do not have any principle means to tell for which 
value of $Q_0$ the default model may have the best agreement with the expectations 
from QCD factorization (which can only be tested by a
$Q_0$-dependence consistent with QCD factorization), 
this implies that the size of the hadronization
corrections to the (true) partonic thrust distribution that is provided 
by the default model is inconsistent with QCD
factorization in the range of $40$\%. This is substantially
worse in comparison to the new default model, where we have
consistency at the level of better than $10$\%.  In the context of the
shower cut $Q_0$ adopting the role of an IR factorization scale, so 
that the hadronization effects have a well-defined and controlled 
scheme dependence so that they can be assigned field theoretic meaning, 
the default model thus fails quite badly. In this respect the novel
dynamical hadronization model performs substantially better.

Overall, the new dynamic hadronization model provides a significant
improvement concerning the control of the shower cut $Q_0$ as an IR
factorization scale. This feature is a prerequisite to combine the
hadronization corrections implemented and quantified in the
hadronization model with parton level theoretical calculations in a
meaningful and systematic manner.

\section{Conclusions and Outlook}
\label{sec:Conclusions}

In this article we promote the idea of the parton shower
cutoff $Q_0$ for MC simulations being an infrared factorization scale
that separates, in a controlled manner and compatible with QCD, 
perturbative and non-perturbative hadronization effects. 
In this context, features of the MC's hadronization model may be
given a more systematic field theoretic meaning and, at the same time,
QCD parameters appearing in the parton shower may be related to
their renormalization scheme dependent counter parts appearing in 
analytic QCD computations.  
An important prerequisite to achieve this is, to have a parton 
shower algorithm that has at least NLL precision for the observable
considered. However, the second essential and equally important 
prerequisite is to have a hadronization model
that can properly match the unavoidable infrared cutoff $Q_0$-dependence
of the parton level description that emerges from the parton 
shower. This is a highly nontrivial condition on the hadronization
model since it entails that the parton shower cutoff $Q_0$ is
not treated as a tuned hadronization parameter. Rather, the hadron
level MC observable description should be equivalent for different $Q_0$ 
values, at least within some low energy interval, where QCD perturbation
theory can still be trusted. 

In this work we have, for the first time in the literature, 
investigated the concept of the parton shower cutoff as a factorization scale
coherently for a multipurpose MC event generator from the perspective of the 
interplay between the parton shower and the hadronization model, in this case for the coherent branching algorithm and the cluster model as implemented in Herwig~7.
We have investigated these features using the angular
ordered parton shower and the cluster hadronization model implemented
in the \herwig~7.2 MC event generator focusing primarily on the
2-jettiness distribution in $e ^+e^-$ annihilation for which the
angular ordered parton shower is NLL precise. In the earlier
work~\cite{Hoang:2018zrp} some of us have analyzed the gluon
transverse momentum cutoff $Q_0$-dependence that emerges from the
angular ordered parton shower for the 2-jettiness distribution in detail
at the subleading ${\cal O}(\alpha_s)$ level using QCD factorization
in the dijet region.  It was also shown that the \herwig{} parton
shower implementation satisfies the $Q_0$ evolution equation from QCD
factorization very well. In this work we have now extended our
investigations concerning the parton shower infrared cutoff
being a factorization scale from the perspective of the \herwig{}
cluster hadronization model. 

The default cluster hadronization model has not been designed in a 
way so that it systematically matches to the parton shower for a 
range of $Q_0$ values. This is related to a number of  
features in the cluster formation and fission dynamics
that yield good data description through the tuning, but are otherwise 
ad-hoc and not compatible with the processes that happen in the parton 
shower evolution close to the cutoff $Q_0$. As a result, the
hadronization effects to the 2-jettiness distribution provided by 
the default \herwig{} cluster model do not satisfy the 
QCD constraints that emerge from $Q_0$ being promoted to a factorization
scale in a satisfactory manner. We have demonstrated this in a number
of tests based on tuning analyses for $Q_0$ shower cutoff 
values in the range between $1$ and $2$~GeV 
where $Q_0$ is treated as an external scale and not a tuned 
parameter. These tests involve analyses of
the $Q_0$-dependence of hadron level simulations for 2-jettiness 
and also other event-shapes and observables and of the parton-to-hadron
migration matrix for 2-jettiness for which QCD factorization provides
nontrivial constraints. 

To improve the cluster hadronization model we have added a number of
features to the cluster formation and cluster fission processes that
better mimic the gluon emission and splitting dynamics that takes place in 
the parton shower. These modifications provide a clearer separation of
model parameters that are expected to be correlated to $Q_0$, governing
the 'hard' aspects of the cluster hadronization dynamics, from those
which govern 'soft' hadron formation aspects that should be rather 
$Q_0$-independent. In our $Q_0$-dependent tuning studies we found
that this novel dynamical cluster hadronization model performs substantially
better concerning the $Q_0$-invariance of \herwig's hadron level 
predictions as well as concerning the QCD factorization constraints on
the 2-jettiness parton-to-hadron migration matrix. We emphasize that 
our analyses also involved hard scattering energies that are not 
accounted for in the reference data used for the tuning. This shows
that the novel dynamical hadronization model properly scales the
consistency to QCD factorization to other hard scattering energies.
We have also investigated the hadron level simulation of
many other $e^+e^-$ observables, including jet rates and multiplicities, 
and found globally a significant 
reduction of the $Q_0$-dependence in the dynamical model. However, since the concrete results 
for the parton level $Q_0$-dependence are currently only available for 2-jettiness and thrust related
event-shape observables~\cite{Hoang:2018zrp}, a separate test of the parton level
results and the hadronization corrections is not yet
possible for other types of observables. To acquire such results a systematic knowledge of
the infrared sensitivity of an observable (and thus the structure of its hadronization corrections) 
is mandatory. 

Even though the novel dynamical hadronization model we have designed
in this article is not perfect, its features provide an important step
forward in promoting the hadronization corrections encoded in MC
generators to have a well-defined scheme in the QCD context, as
also discussed previously in Ref.~\cite{Platzer:2022jny}. This is an essential
aspect that should be followed in parallel to the ongoing developments
of subleading order precise parton shower algorithms such that the
parameters of the parton shower and eventually even of the hadronization model may
acquire a systematic QCD field theoretic meaning. Beyond a more
precise and consistent description of experimental data, important
potential applications of such improvements are the determination of
QCD parameters directly from MC studies as well as a systematic
quantification of hadronization corrections to analytic QCD
calculations from MC simulations.  In an upcoming article we will
investigate the former application from the perspective of the MC top
quark mass parameter.

\section*{Acknowledgments}
\label{sec:acknowledgements}

We acknowledge support by the FWF Austrian Science Fund under the
Project No.~P32383-N27 and under the FWF Austrian Science Fund
Doctoral Program ``Particles and Interactions'' No.~W1252-N27 as well
as partial support by the COST Action No.\ CA16201 Particleface.  This
work has also been supported in part by the European Union’s Horizon
2020 research and innovation programme as part of the Marie
Skłodowska-Curie Innovative Training Network MCnetITN3 (grant
agreement No.~722104).  We are grateful the Erwin-Schr\"odinger
International Institute for Mathematics and Physics for partial
support during the Thematic Programme ``Quantum Field Theory at the
Frontiers of the Strong Interactions'', July 31 - September 1, 2023.

\appendix

\section{Comparison of Simulations with LEP Data}
\label{app:otherfigures}

\begin{figure}
	\centering
	\begin{minipage}{0.28\linewidth}
		\includegraphics[width=\linewidth]{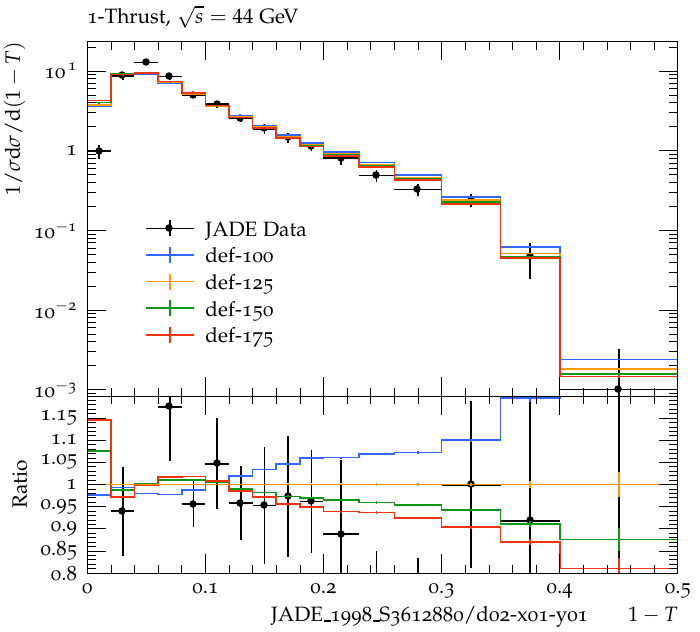}
	\end{minipage}
	\begin{minipage}{0.28\linewidth}
		\includegraphics[width=\linewidth]{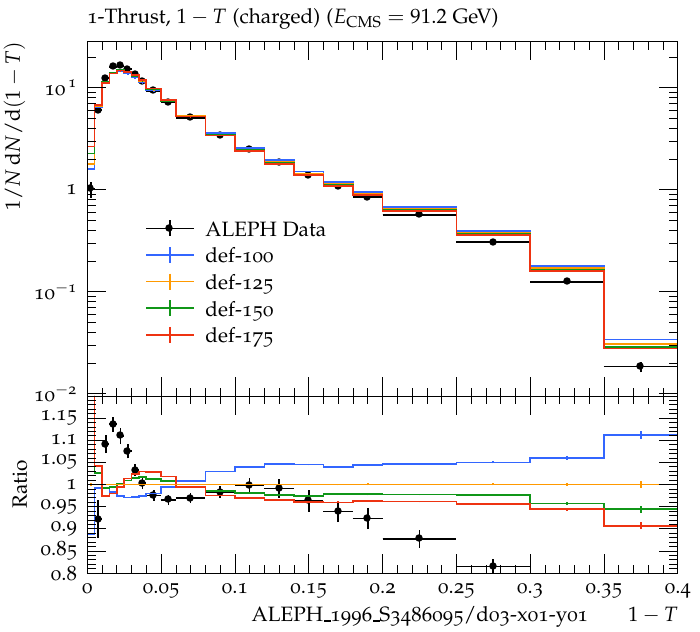}
	\end{minipage}
	\begin{minipage}{0.28\linewidth}
		\includegraphics[width=\linewidth]{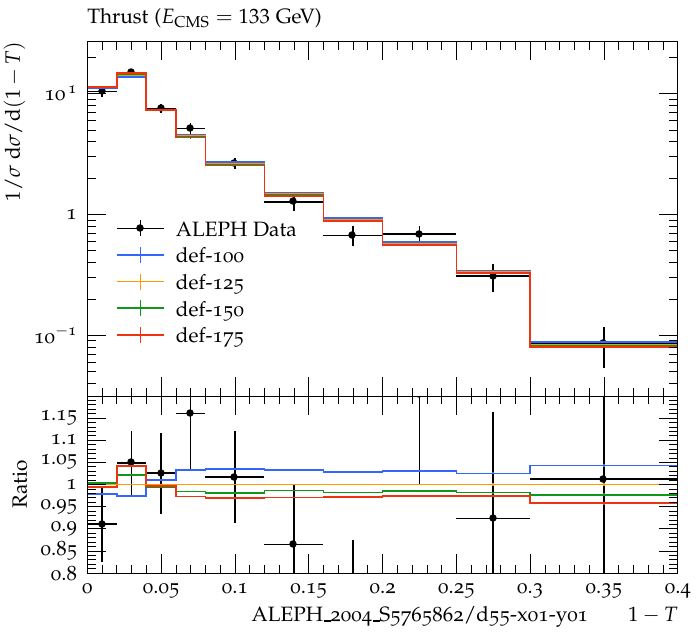}
	\end{minipage} \\	
	\begin{minipage}{0.28\linewidth}
		\includegraphics[width=\linewidth]{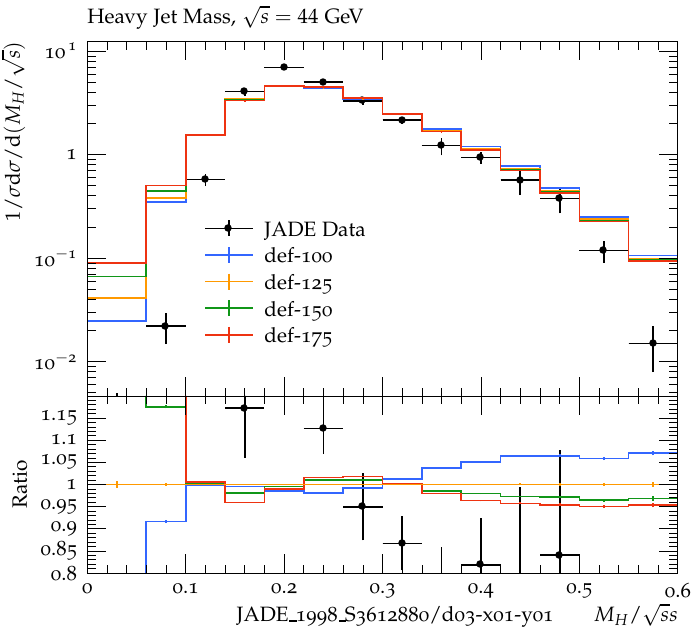}
	\end{minipage}
	\begin{minipage}{0.28\linewidth}
		\includegraphics[width=\linewidth]{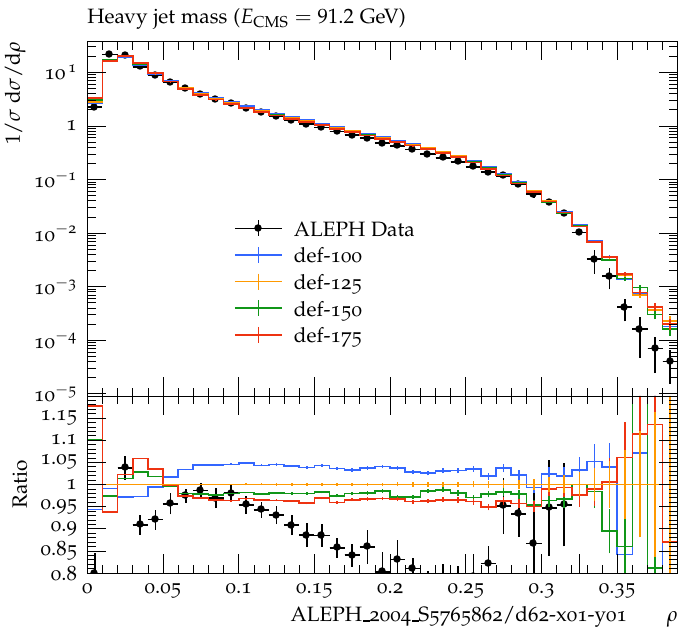}
	\end{minipage}
	\begin{minipage}{0.28\linewidth}
		\includegraphics[width=\linewidth]{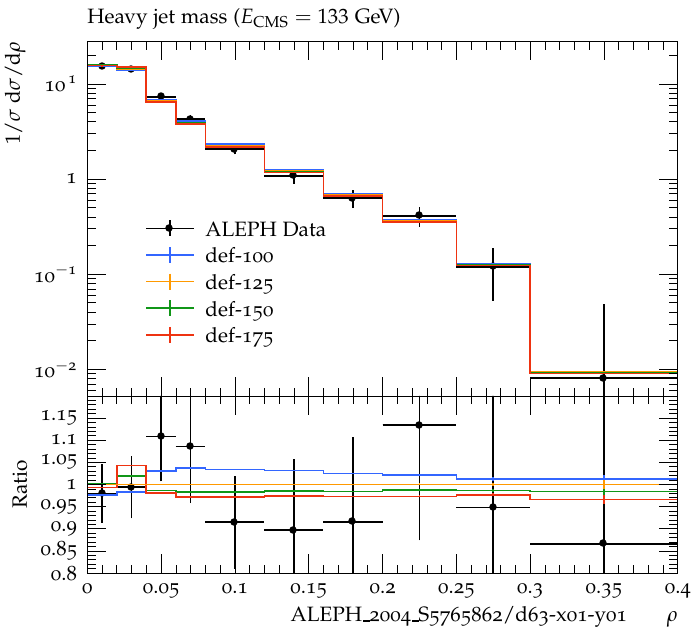}
	\end{minipage} \\
	\begin{minipage}{0.28\linewidth}
		\includegraphics[width=\linewidth]{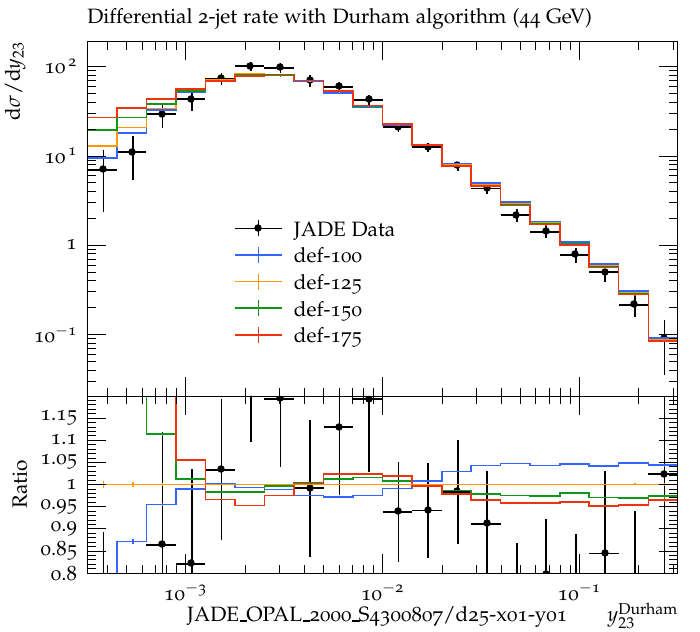}
	\end{minipage}
	\begin{minipage}{0.28\linewidth}
		\includegraphics[width=\linewidth]{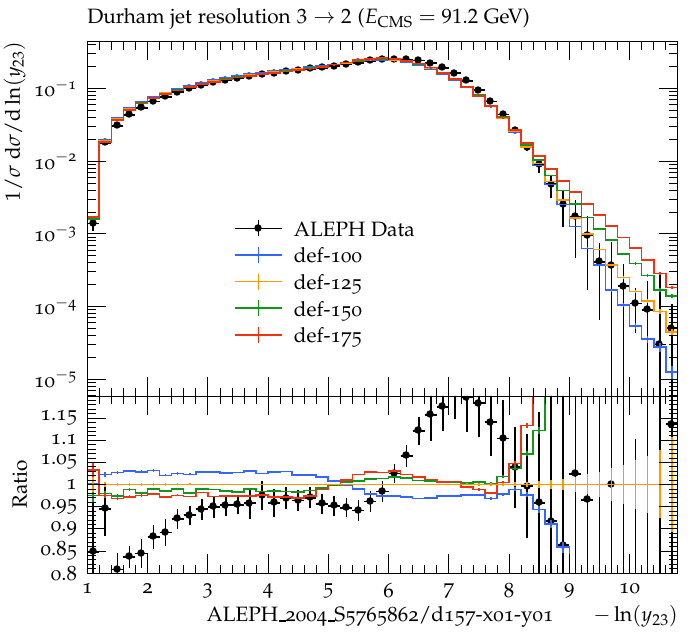}
	\end{minipage}
	\begin{minipage}{0.28\linewidth}
		\includegraphics[width=\linewidth]{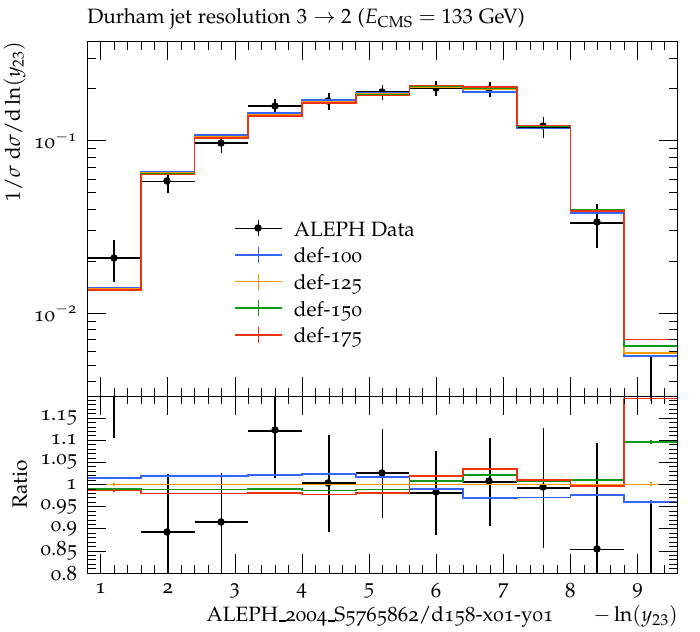}
	\end{minipage} \\
	\begin{minipage}{0.28\linewidth}
		\includegraphics[width=\linewidth]{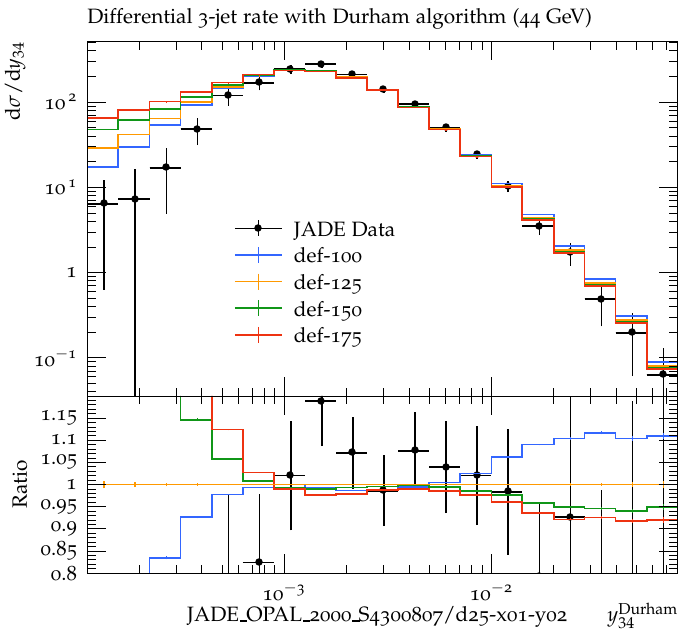}
	\end{minipage}
	\begin{minipage}{0.28\linewidth}
		\includegraphics[width=\linewidth]{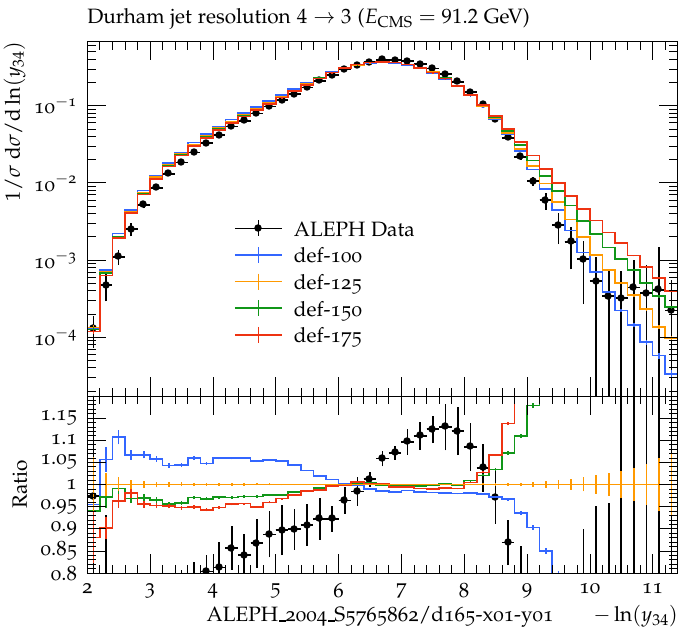}
	\end{minipage}
	\begin{minipage}{0.28\linewidth}
		\includegraphics[width=\linewidth]{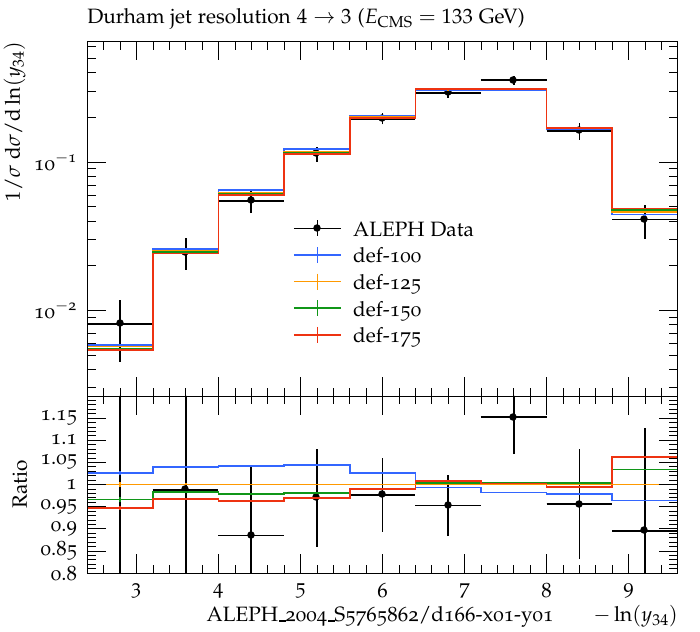}
	\end{minipage} \\
	\begin{minipage}{0.28\linewidth}
		\includegraphics[width=\linewidth]{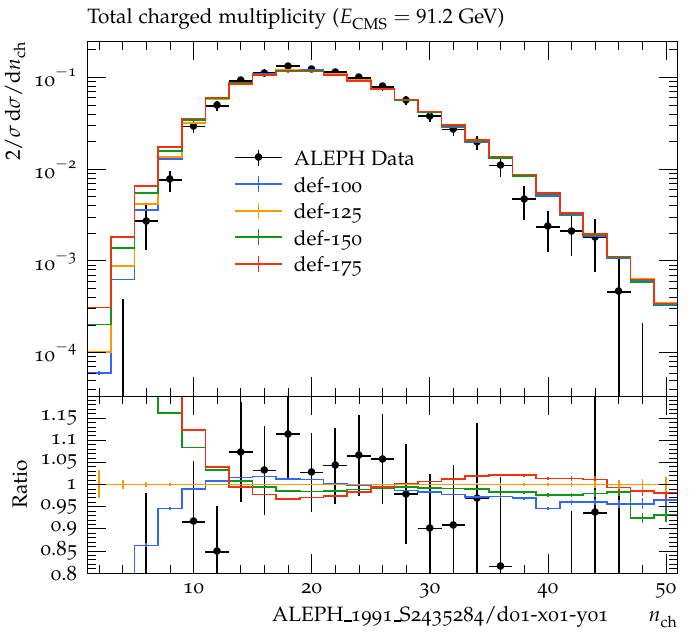}
	\end{minipage} 
	\begin{minipage}{0.28\linewidth}
		\includegraphics[width=\linewidth]{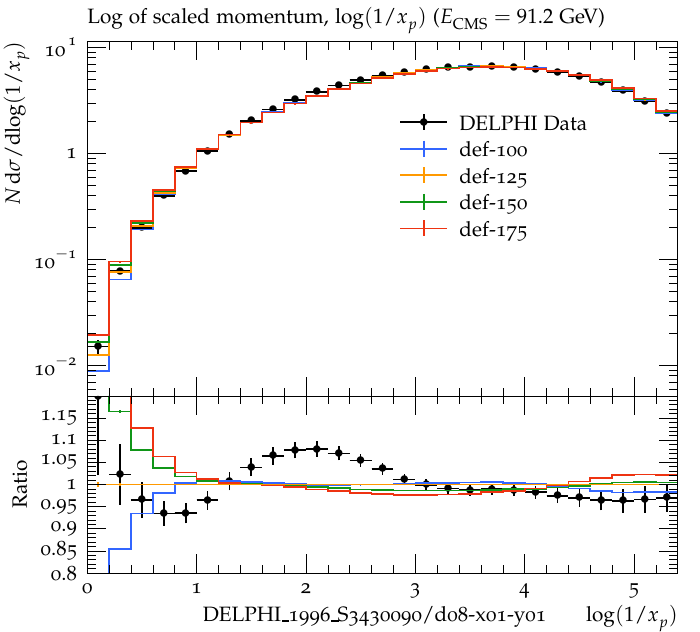}
	\end{minipage}
	\begin{minipage}{0.28\linewidth}
		\includegraphics[width=\linewidth]{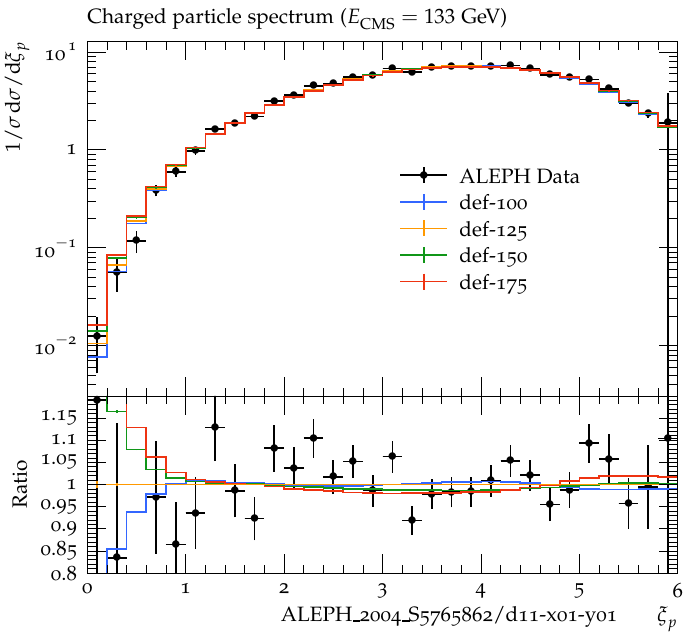}
	\end{minipage}
	\caption{Selected rivet analyses for $e^+e^-$ data (black) versus \herwig{} simulations for 
	the default hadronization model for shower cutoff values $Q_0=1$ (blue), $1.25$ (orange), $1.5$ 
	(green) and $1.75$~GeV (red). The ratios in the lower panel sections are shown w.r.\ to the
    simulations for $Q_0=1.25$~GeV.}
	\label{fig:Rivetdefaultotherdistributions}
\end{figure}

\begin{figure}
	\centering
	\begin{minipage}{0.28\linewidth}
		\includegraphics[width=\linewidth]{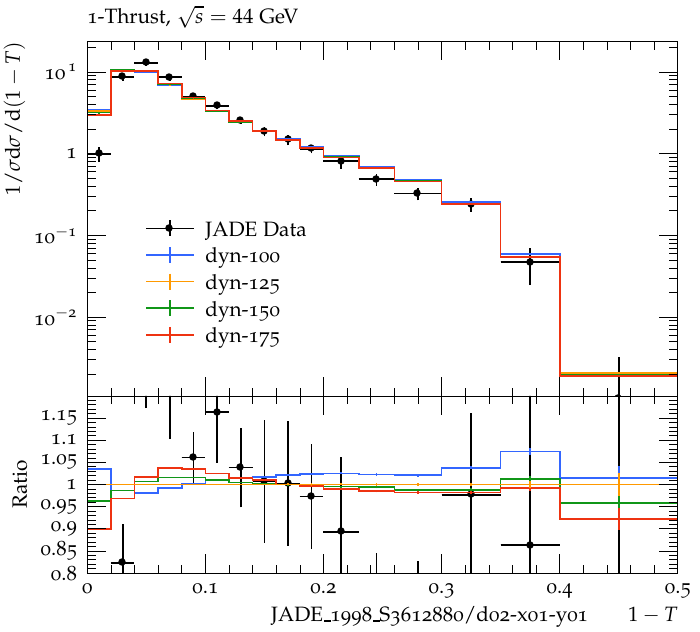}
	\end{minipage}
	\begin{minipage}{0.28\linewidth}
		\includegraphics[width=\linewidth]{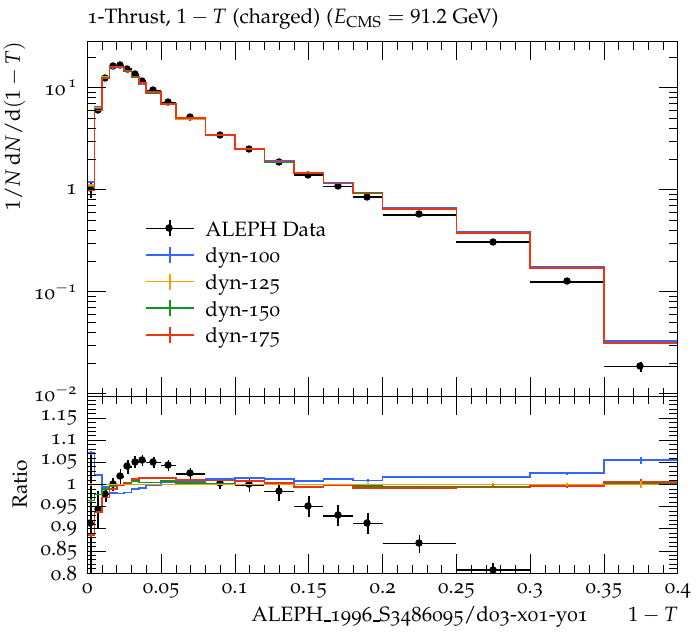}
	\end{minipage}
	\begin{minipage}{0.28\linewidth}
		\includegraphics[width=\linewidth]{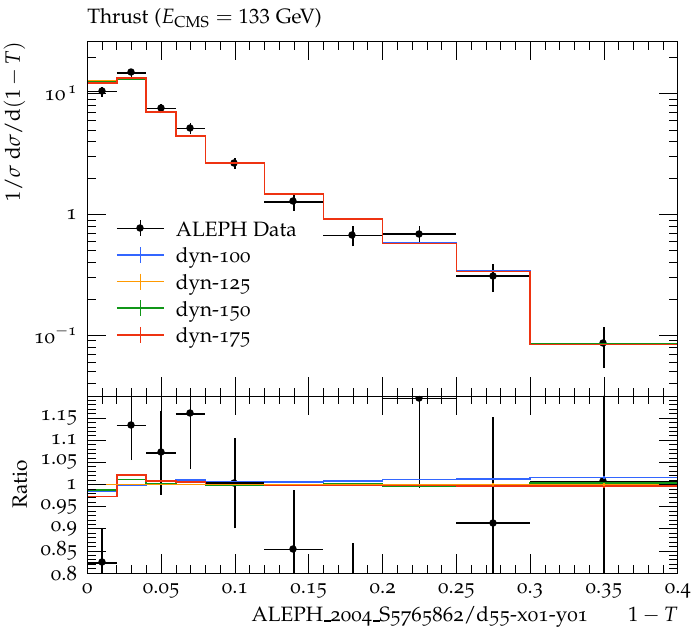}
	\end{minipage} \\	
	\begin{minipage}{0.28\linewidth}
		\includegraphics[width=\linewidth]{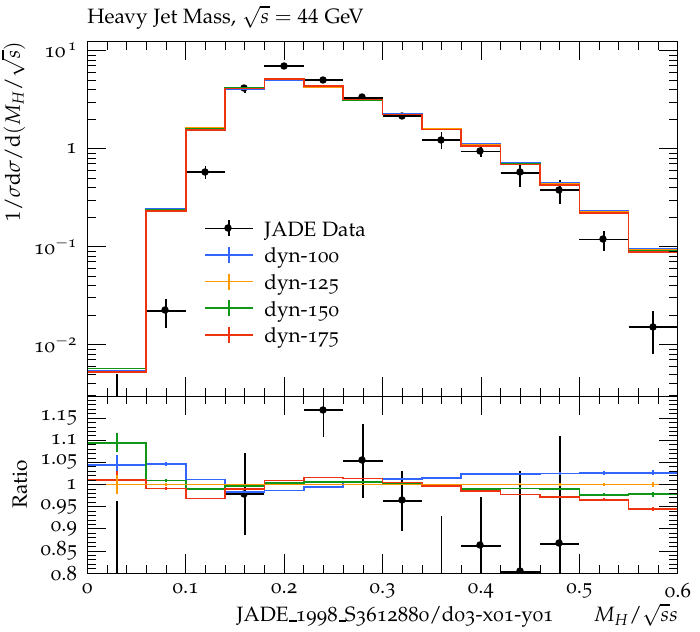}
	\end{minipage}
	\begin{minipage}{0.28\linewidth}
		\includegraphics[width=\linewidth]{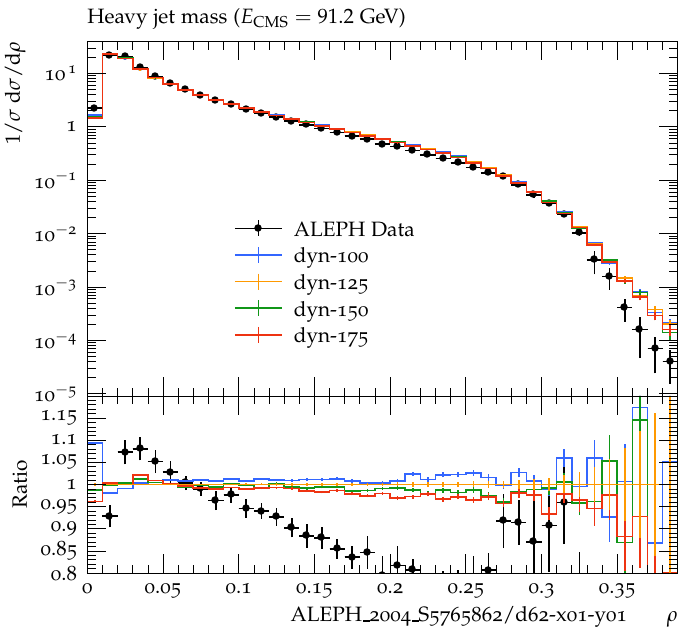}
	\end{minipage}
	\begin{minipage}{0.28\linewidth}
		\includegraphics[width=\linewidth]{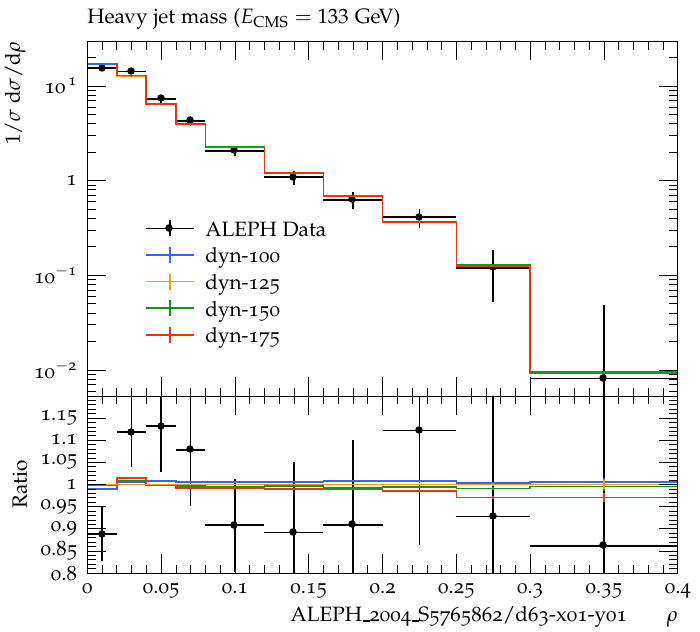}
	\end{minipage} \\
	\begin{minipage}{0.28\linewidth}
		\includegraphics[width=\linewidth]{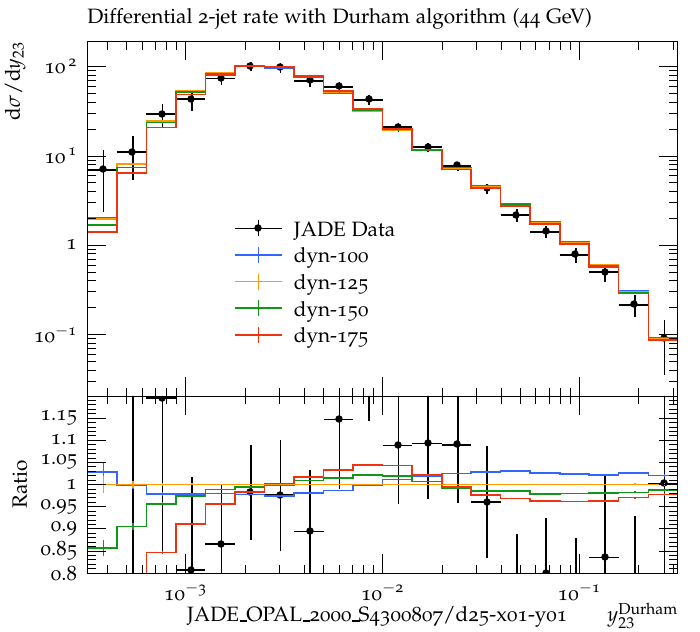}
	\end{minipage}
	\begin{minipage}{0.28\linewidth}
		\includegraphics[width=\linewidth]{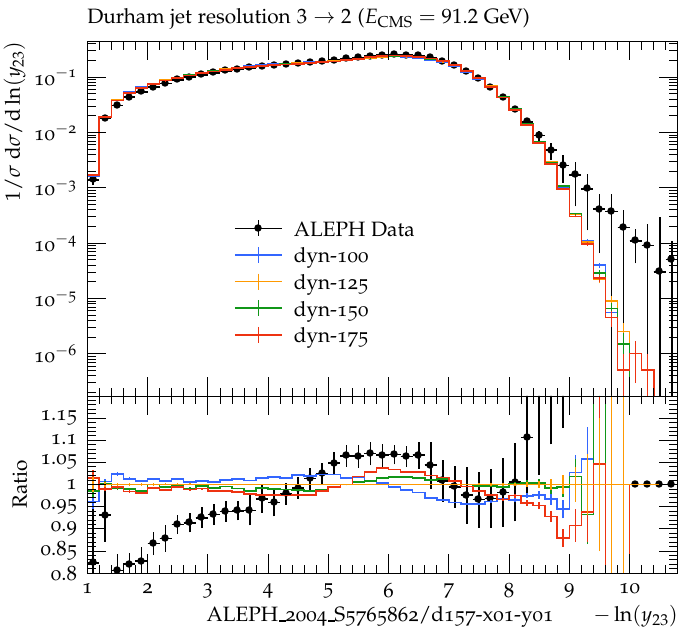}
	\end{minipage}
	\begin{minipage}{0.28\linewidth}
		\includegraphics[width=\linewidth]{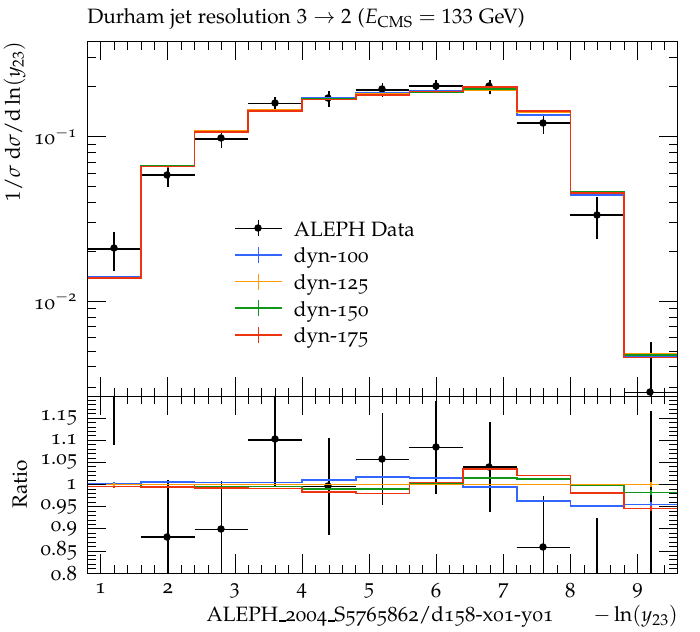}
	\end{minipage} \\
	\begin{minipage}{0.28\linewidth}
		\includegraphics[width=\linewidth]{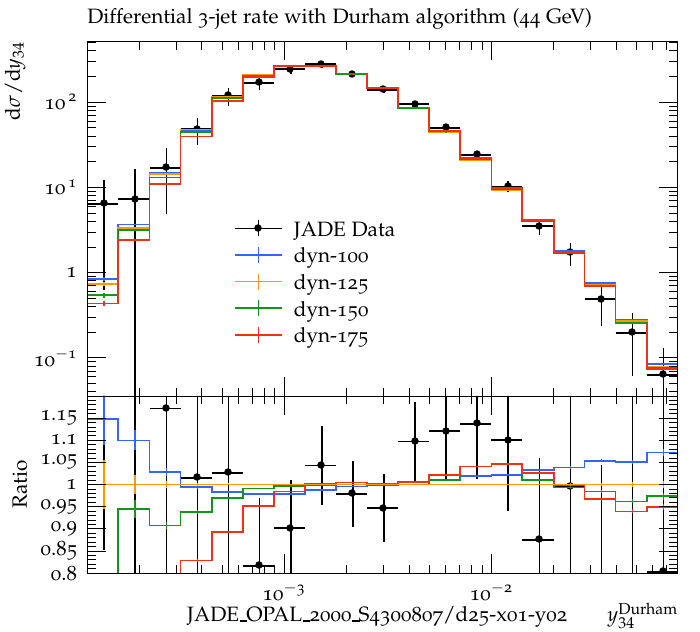}
	\end{minipage}
	\begin{minipage}{0.28\linewidth}
		\includegraphics[width=\linewidth]{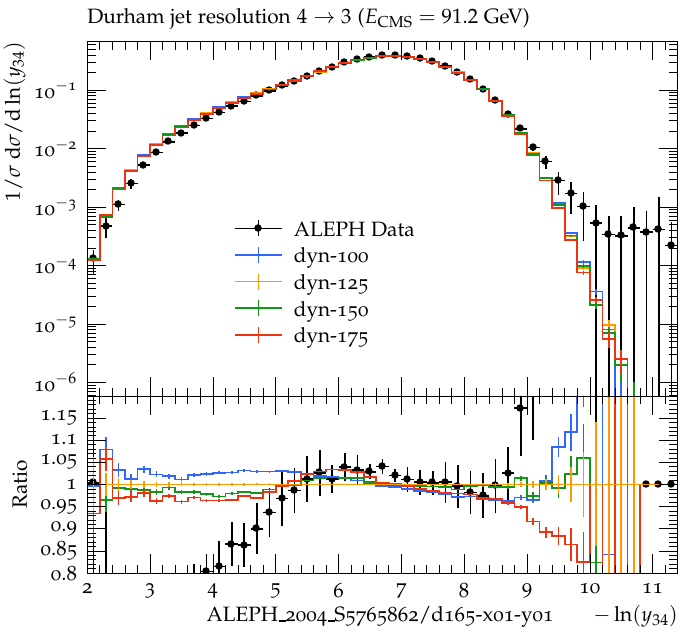}
	\end{minipage}
	\begin{minipage}{0.28\linewidth}
		\includegraphics[width=\linewidth]{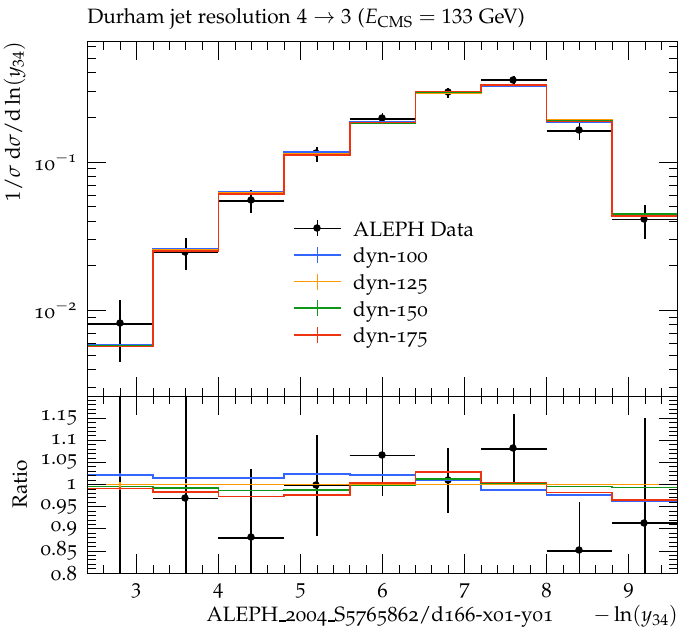}
	\end{minipage} \\
	\begin{minipage}{0.28\linewidth}
		\includegraphics[width=\linewidth]{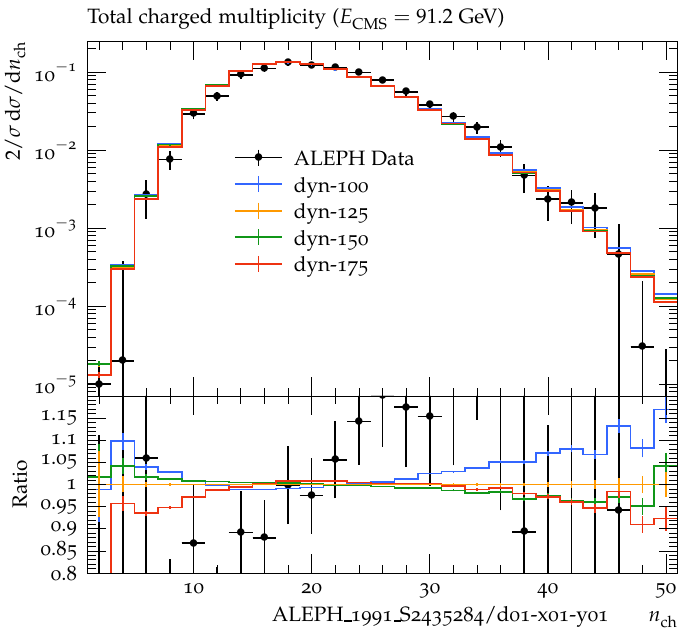}
	\end{minipage} 
	\begin{minipage}{0.28\linewidth}
		\includegraphics[width=\linewidth]{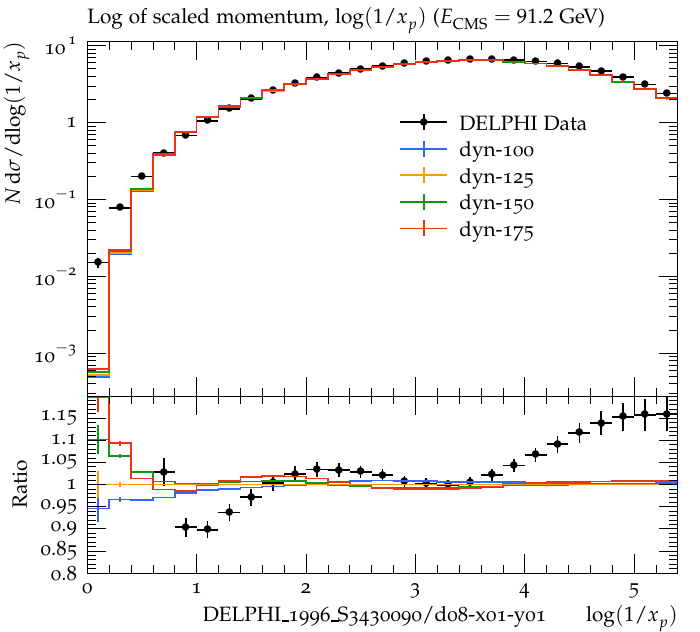}
	\end{minipage}
	\begin{minipage}{0.28\linewidth}
		\includegraphics[width=\linewidth]{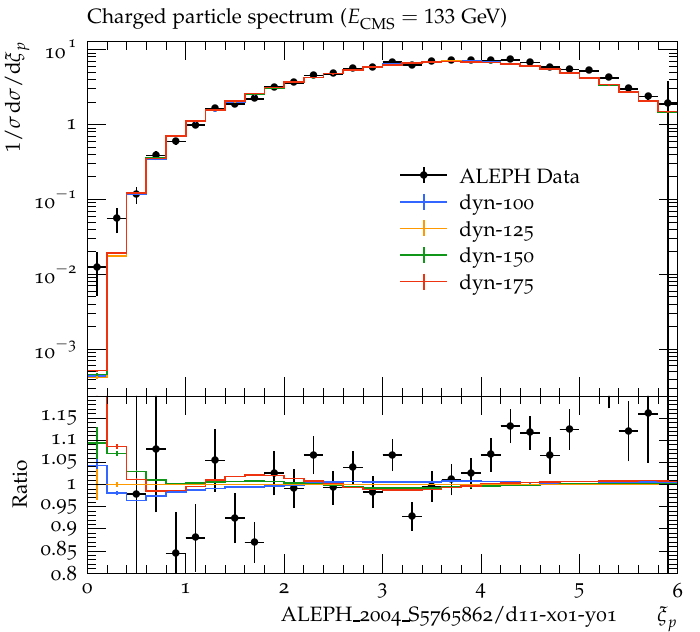}
	\end{minipage}
	\caption{Selected rivet analyses for $e^+e^-$ data (black) versus \herwig{} simulations for 
		the novel dynamical hadronization model for shower cutoff values $Q_0=1$ (blue), $1.25$ 
		(orange), $1.5$ (green) and $1.75$~GeV (red). The ratios in the lower panel sections 
		are shown w.r.\ to the simulations for $Q_0=1.25$~GeV.}
	\label{fig:Rivetdynamicotherdistributions}
\end{figure}

\newpage

\bibliography{./sources}	
\bibliographystyle{JHEP}

\end{document}